\documentclass[twocolumn,twocolappendix,tighten]{aastex63}
\usepackage{amstext}
\usepackage{amsmath}
\usepackage{amssymb}
\usepackage{apjfonts}
\usepackage{xcolor}

\shorttitle{BHL Accretion onto Binaries}
\shortauthors{Antoni, MacLeod, \& Ramirez-Ruiz}

\newcommand{\beq}{\begin{equation}}
\newcommand{\eeq}{\end{equation}}
\newcommand{\beqar}{\begin{eqnarray}}
\newcommand{\eeqar}{\end{eqnarray}}

\newcommand{\rs}{R_{\rm s}}
\newcommand{\ra}{R_{\rm a}}
\newcommand{\rbh}{R_{\rm BH}}

\newcommand{\vinf}{v_\infty}
\newcommand{\vorb}{v_{\rm orb}}
\newcommand{\vcm}{{\bf V}_{\rm CM}}
\newcommand{\rcm}{{\bf R}_{\rm CM}}

\newcommand{\rhoinf}{\rho_\infty}
\newcommand{\ninf}{n_\infty}
\newcommand{\mach}{\mathcal M_\infty}
\newcommand{\machorb}{\mathcal M_{\rm orb}}
\newcommand{\cs}{c_{s,\infty}}
\newcommand{\g}{\gamma}

\newcommand{\tinf}{\,\ra /\vinf}
\newcommand{\mhl}{\dot M_{\rm HL}}
\newcommand{\mbh}{\dot M_{\rm BH}}
\newcommand{\fhl}{F_{\rm HL}}
\newcommand{\fbh}{F_{\rm BH}}
\newcommand{\ehl}{\dot E_{\rm HL}}
\newcommand{\ebh}{\dot E_{\rm BH}}

\newcommand{\mb}{\dot{M}}
\newcommand{\ms}{\dot{M}(a_0 = 0)}
\newcommand{\mo}{\dot{m}_1}
\newcommand{\mt}{\dot{m}_2}

\newcommand{\tinsp}{\tau_{\rm insp}}
\newcommand{\taccr}{\tau_{\rm accr}}
\newcommand{\tstop}{\tau_{\rm stop}}

\newcommand{\tinspfit}{\tau_{\rm insp,fit}}
\newcommand{\tgw}{\tau_{\rm insp, GW}}
\newcommand{\tgas}{\tau_{\rm insp, gas}} 

\newcommand{\tinspBH}{\tau_{\rm insp,BH}}
\newcommand{\tinspHL}{\tau_{\rm insp,HL}}
\newcommand{\tstopBH}{\tau_{\rm stop,BH}}
\newcommand{\tstopHL}{\tau_{\rm stop,HL}}
\newcommand{\taccrBH}{\tau_{\rm accr,BH}}
\newcommand{\taccrHL}{\tau_{\rm accr,HL}}

\newcommand{\tildetau}{\widetilde\tau_{\rm insp,BH}}

\newcommand*\ave[1]{\langle{#1}\rangle}

\newcommand{\psr}{PSR J0337+1715 }

\newcommand{\msun}{M_\sun}

\begin{document}

\title{The Evolution of Binaries in a Gaseous Medium: Three-Dimensional Simulations of Binary Bondi-Hoyle-Lyttleton Accretion}

\correspondingauthor{Andrea Antoni}
\email{aantoni@berkeley.edu}

\author[0000-0003-3062-4773]{Andrea Antoni}
\affiliation{Department of Astronomy \& Astrophysics, University of California, Santa Cruz, CA 95064, USA}
\affiliation{Department of Astronomy, University of California, Berkeley, CA 94720, USA}

\author[0000-0002-1417-8024]{Morgan MacLeod}
\altaffiliation{NASA Einstein Fellow}
\affiliation{Harvard-Smithsonian Center for Astrophysics, 60 Garden Street, Cambridge, MA, 02138, USA}

\author[0000-0003-2558-3102]{Enrico Ramirez-Ruiz}
\affiliation{Department of Astronomy \& Astrophysics, University of California, Santa Cruz, CA 95064, USA}
\affiliation{Niels Bohr Institute, University of Copenhagen, Blegdamsvej 17, DK-2100 Copenhagen, Denmark}

\begin{abstract} 
Binary stars are common. While only those with small separations may exchange gas with one another, even the widest binaries interact with their gaseous surroundings.  Drag forces and accretion rates dictate how these systems are transformed by these interactions. We perform three-dimensional hydrodynamic simulations of Bondi-Hoyle-Lyttleton flows, in which a binary moves supersonically relative to a homogeneous medium, using the adaptive mesh refinement code FLASH.  We simulate a range of values of the initial semi-major axis of the orbit relative to the gravitational focusing impact parameter of the pair. When the binary separation is less than the gravitational focusing impact parameter, the pair orbits within a shared bow shock. When the pair is wider, each object has an individual bow-shock structure. The long-term evolution of the binary is determined by the timescales for accretion, slowing of the center of mass, and orbital inspiral. We find a clear hierarchy of these timescales; a binary's center-of-mass motion is slowed over a shorter timescale than the pair inspirals or accretes. In contrast to previous analytic predictions, which assume an unperturbed background medium, we find that the timescale for orbital inspiral is proportional to the semi-major axis to the $0.19 \pm 0.01$ power. This positive scaling indicates that gaseous drag forces can drive binaries either to coalescence or to the critical separation at which gravitational radiation dominates their further evolution. We discuss the implications of our results for binaries embedded in the interstellar medium, active galactic nuclei disks, and common envelope phases. 
\end{abstract}

\section{Introduction}
Many objects in the universe are found in binary systems.  The components of these binaries range from substellar mass objects to supermassive black holes (SMBHs) in the centers of galaxies \citep{1992ARA&A..30..287C,2013ARA&A..51..269D,1063-7869-57-4-359,2014LRR....17....3P}. In fact, roughly one-third of all main-sequence (MS) stars are members of binary systems, while $\sim$ 10\% of solar-type MS stars and an overwhelming 73\% of O-type MS stars have 2 or more stellar companions \citep{2017ApJS..230...15M}.  

Multiple-star systems are found throughout the galaxy, traveling through gaseous environments with widely varying densities, temperatures, and compositions.  The range of gaseous environments include the interstellar medium \citep[ISM;][]{2004ApJ...601..289C}, accretion flows in active galactic nuclei \citep[AGN;][]{1993ApJ...409..592A}, the dense envelopes of evolved companion stars \citep{2015MNRAS.450.1716S,2018arXiv180509338D}, and the intracluster medium in young star clusters \citep{2011ApJ...735...25N,2018MNRAS.478.2794N,2018MNRAS.481..627A,2019A&A...621L...1R,2019ApJ...876..142K}.

A gravitating object moving through gas captures mass and momentum from its surroundings \citep{1939PCPS...35..405H,1944MNRAS.104..273B}.  The drag force it feels due to its gravitationally induced wake robs it of linear momentum \citep{1943ApJ....97..255C,1999ApJ...513..252O}.  For binary or multiple objects, drag forces could lead to a metamorphosis of the size and shape of their shared orbit.

The effects of accretion and drag can have profound consequences. For example, a metal poor star moving through the ISM could accrete, over its history, enough material to significantly enhance its birth metallicity \citep{2017MNRAS.469.4012S}.  More dramatically, the orbit of a stellar-mass binary black hole (BBH) embedded in an AGN disk may be tightened by gaseous interaction to the point that the binary  merges under the influence of gravitational wave (GW) radiation  \citep{2017MNRAS.464..946S}.  Whether such binaries form within AGN disks or are dynamically captured  \citep{2017ApJ...835..165B}, they may be the only case of BBH mergers that are accompanied by  an electromagnetic signature \citep[due to the surrounding gas;][]{2017MNRAS.464..946S}. 

Our goal in this paper is to examine the nature of accretion and drag experienced by binaries moving through gaseous environments and to learn the extent to which we can derive physical intuition from  the classical Bondi-Hoyle-Lyttleton (BHL) formalism \citep{1939PCPS...35..405H,1944MNRAS.104..273B,1952MNRAS.112..195B}.  To that end we perform 3D hydrodynamical studies of binary systems moving supersonically through a gaseous medium.  

In the next section, we introduce the BHL formalism for a single accretor, discuss how this formalism can be extended to binary systems, and consider the scales for  astrophysically typical encounters.  In Section \ref{sec:methods}, we describe our numerical method and present the results of our hydrodynamic simulations in Sections \ref{sec:results} and \ref{sec:timescales}. In Section \ref{sec:discussion}, we discuss the implications of our results for astrophysical systems.  Finally, in Section \ref{sec:summary}, we conclude. 

\section{Scales and Typical Encounters in Binary BHL Flows}
\label{sec:scales}

In this section we define the characteristic scales of BHL and their extension to binary systems. Finally, we briefly discuss typical astrophysical environments for binary BHL flows.

\subsection{Characteristic Scales: Single-object BHL}	
A massive object moving through an initially uniform background gas accelerates matter in its direction as it moves.  If the motion of the mass is supersonic relative to the background, a bow shock forms, and shocked gas piles up in a stagnation region downstream of the moving object. If it is absorbing, the object accretes gravitationally bound material from the wake, while the rest is eventually advected away.  Although the material in this wake is, therefore, transient, the overall structure of the wake persists, and it exerts a gravitational drag force on the point mass until there ceases to be relative motion between the object and the gas \citep{1943ApJ....97..255C}.

The BHL formalism provides analytical estimates for these interactions. Let $M$ be the mass of the object, $\vinf$ be the speed of the object relative to the gas, and $\cs$ be the sound speed of the gas. The Mach number of the flow is $\mach = \vinf / \cs$. The object's influence on the gas is characterized by its accretion radius. If the object moves highly supersonically, then we can use the Hoyle-Lyttleton accretion radius
\beq
\ra = \frac{2GM}{\vinf^2},
\eeq
where $G$ is the gravitational constant \citep{1939PCPS...35..405H}. For very subsonic motion, the accretion radius is given by the Bondi radius:
\beq
R_B = \frac{GM}{\cs^2} = \frac{1}{2}\mach^2\ra
\label{eq:rb}
\eeq
\citep{1952MNRAS.112..195B}.  

\subsubsection{Hoyle-Lyttleton (HL) Scalings}
In general, BHL predicts an accretion rate that is given by the flux of matter through a circular cross section far upstream from the object, where the radius of the cross section is the accretion radius of the object.  In the highly supersonic regime, this gives
\beq
\mhl = \pi \ra^2 \rhoinf \vinf = 4 \pi G^2 M^2 \rhoinf \vinf^{-3}
\label{eq:mhl}
\eeq
\citep{1939PCPS...35..405H}.  The drag force on the object due to the wake is
\beq
\fhl = \mhl \vinf = 4 \pi G^2 M^2 \rhoinf \vinf^{-2},
\label{eq:fhl}
\eeq
and the rate at which energy is dissipated is 
\beq
\ehl = \frac{1}{2}\mhl \vinf^2 = 2 \pi G^2 M^2 \rhoinf \vinf^{-1}.
\label{eq:ehl}
\eeq
Finally, the approximate timescale for the object to stop is 
\beq
\tstopHL \equiv \frac{\frac{1}{2} M \vinf^2}{\ehl} = \frac{\vinf^3}{4\pi G^2 M \rhoinf}.
\label{eq:tstopHL}
\eeq
In the HL formalism, the accretion timescale is identical to the stopping timescale because
\beq
\taccrHL = \frac{M}{\mhl} = \frac{\frac{1}{2}M \vinf^2}{\frac{1}{2}\mhl\vinf^2} = \frac{\frac{1}{2}M \vinf^2}{\ehl} = \tstopHL.
\label{eq:taccrHL}
\eeq

\subsubsection{Bondi-Hoyle (BH) Scalings}
For mild Mach numbers, the work of \citet{1944MNRAS.104..273B} and \citet{1952MNRAS.112..195B} led to an interpolation formula for the accretion rate
\beq
\mbh = \frac{4\pi G^2 M^2\rhoinf}{\big(\vinf^2 + \cs^2\big)^{3/2}} = \mhl \bigg(\frac{\mach^2}{1 + \mach^2}\bigg)^{3/2}
\label{eq:mbh}
\eeq
where the factor of four was found by \citet{1985MNRAS.217..367S}. It is helpful to define quantities analogous to those of the HL formalism.  If we factor $\mbh$ similarly to $\mhl$ in Equation~\eqref{eq:mhl}, then
\beq
\mbh = \pi \rbh^2 \rhoinf (\vinf^2 + \cs^2)^{1/2}
\eeq
where the characteristic velocity is  $(\vinf^2 + \cs^2)^{1/2}$ and the BH accretion radius is 
\beq
\rbh = \frac{2GM}{\vinf^2 + \cs^2} = \ra\bigg(\frac{\mach^2}{1 + \mach^2}\bigg). 
\label{eq:rbh}
\eeq
In keeping with Equation \eqref{eq:fhl}, we define the BH drag force as an accretion rate times a characteristic velocity, or
\beq
\fbh = \mbh (\vinf^2 + \cs^2)^{1/2} = \fhl\bigg(\frac{\mach^2}{1 + \mach^2}\bigg).
\label{eq:fbh}
\eeq
For $\mach \ll 1$, $\fbh \approx \mach^2\fhl$, whereas the piecewise analytical dynamical friction (DF) formula for a subsonic, nonaccreting perturber goes as $\mach^3\fhl$ \citep{1999ApJ...513..252O}. For a highly supersonic object ($\mach \gg 1$), $\fbh \to \fhl$.

Similarly, we can define the characteristic energy dissipation rate to be
\beq
\ebh = \onehalf\mbh (\vinf^2 + \cs^2) = \ehl\bigg(\frac{\mach^2}{1 + \mach^2}\bigg)^{1/2}.
\label{eq:ebh}
\eeq
The translational kinetic energy of the object is $\onehalf M\vinf^2$, 
which gives a BH stopping time of 
\beq
\tstopBH \equiv \frac{\frac{1}{2} M \vinf^2}{\ebh} = \tstopHL\bigg(\frac{1 + \mach^2}{\mach^2}\bigg)^{1/2}.
\label{eq:tstopBH}
\eeq
The BH accretion timescale is 
\beq
\taccrBH = \frac{M}{\mbh} = \tstopBH\bigg(\frac{1 + \mach^2}{\mach^2}\bigg).
\label{eq:taccrBH}
\eeq

While the above equations are only approximately correct,  coefficients for drag and accretion for the above canonical case have been obtained numerically by \citet{1985MNRAS.217..367S}, \citet{1994ApJ...427..351R}, \citet{1994A&AS..106..505R,1995A&AS..113..133R}, and \citet{2012ApJ...752...30B}. See \citet{2005A&A...434...41E} for a review of the BHL accretion formalism, including derivations of the above equations, and \citet{1999ApJ...513..252O} for the first analytical treatment of the gaseous DF drag approximated by the expressions for $\fhl$ and $\fbh$. 

\subsection{Characteristic Scales: Binary BHL}
Now consider what happens if we replace the single object of mass $M$ with two objects, $m_1$ and $m_2$, whose total mass is $M$, and whose center of mass (CM) moves with velocity ${\bf v}_\infty$ relative to the gas.  Free parameters now include the binary mass ratio, the initial eccentricity of the orbit, the inclination of the orbital plane relative to ${\bf v}_\infty$, and the semi-major axis of the orbit, $a$. We will examine the simplifying case of an equal-mass binary in an initially circular orbit with inclination of $\pi/2$ (that is, the incoming gas intercepts the binary ``edge-on'' to the orbit). 

Now there are two key length scales in the problem: the gravitational focusing scale, $\rbh$, and the binary semi-major axis, $a$. Here, $\rbh$, defined by Equation \eqref{eq:rbh}, is the BH accretion radius of the CM.  We can re-express this ratio of length scales in terms of other parameters of the binary:
\beq
\frac{a}{\rbh} = \frac{GM/\vorb^2}{2GM/(\vinf^2 + \cs^2)} = \frac{\vinf^2 + \cs^2}{2\vorb^2} = \frac{\vinf^2(1 + \mach^{-2})}{2\vorb^2},
\eeq
where we have used the orbital speed, $\vorb = \sqrt{GM / a}$, for a binary in a circular orbit.  Therefore, in the limit of highly supersonic motion (large $\mach$), $a/\rbh \approx \vinf^2 /2 \vorb^2$. 

We can imagine two limiting cases for the ratio of $a/\rbh$. When $a \ll \rbh$, the binary appears essentially single as gas is focused toward the system's CM from the background medium. In this case, we might imagine that the derivations of the previous section provide a reasonable description of flow around, accretion toward, and drag forces on the system's CM. In the opposite limit, when $a \gg \rbh$, the binary is much wider than the gravitational focusing scale. In this case, we might expect that each object undergoes an essentially isolated BHL interaction with the gaseous background. That is, the system evolves according to the BHL prescription for two independent objects (each of mass $\frac{1}{2}M$). In this limit, we would expect the accretion rate of the system and the total DF force on the binary to be $\propto (m_1^2 + m_2^2) = M^2 / 2$.

If we allow $a \sim \rbh$, the two length scales are on the same order. Gas is focused toward both the individual objects and the system's CM. Orbital motion projects phase-dependent velocities for the individual objects with magnitude similar to that of the CM motion. Without a clear expectation for the flow structures in this intermediate regime, we turn to hydrodynamical simulations to provide a solution for the small- and large-scale structures that contribute to the evolution of the orbital and CM motion of the binary.  The purpose of the remainder of this paper is to explore binary BHL flow with varying $a/\rbh$ and to characterize the evolution of a binary as a function of this ratio.  

\subsection{Typical Encounters in Binary BHL}

Astrophysical binary systems span many orders  of magnitude in mass and semimajor axis and these systems interact with a wide variety of ambient gases.  In Table \ref{table:landscape}, we calculate $a / R_{\rm BH}$ for a sampling of binary systems moving through the ISM. For each gaseous medium in Table \ref{table:landscape}, we assume an isothermal ideal gas and calculate the sound speed as $\cs = \sqrt{k_BT / m_p}$.   We see from the table that a stellar-mass binary with a separation of 1 AU moving through the warm ISM may have $a/\rbh$ values from several hundredths to a few times 10.  Increasing the mass of the binary by a factor of 10 gives results of  $0.3 \lesssim a/ \rbh \lesssim 3$ when crossing the disk.

We discuss ambient gases of much greater density in Section \ref{sec:discussion}.  For example, a binary embedded in the model AGN disk presented in Section \ref{sec:agn} may encounter number densities, $\ninf$, of $10^8 - 10^{14}$cm$^{-3}$, depending on its radial location in the disk.  A stellar-mass binary with a separation of 1 AU located at a radius of 2 pc in this disk will have $a /\rbh \approx 0.6$.  The same binary embedded at a radius of 0.5 pc will have $a/\rbh \approx 2.5$.  As another example, we discuss in Section \ref{sec:ce} a common envelope (CE) interaction in the triple system \psr.  We find that the models of \citet{2015MNRAS.450.1716S} imply envelope densities of  $10^{16} \lesssim \ninf \lesssim 10^{19}$ and the embedded binary would have $0.8 < a/\rbh < 2.7$.  The scaling with $\rhoinf$ in Equation \eqref{eq:tstopBH} indicates that the binary in one of these environments will transform 10 orders of magnitude more quickly than in the ISM.

\begin{deluxetable*}{l  |l  | c | c c | c }
\tablecaption{$a / R_{\rm BH}$ in a Variety of Encounters. \label{table:landscape}}
\tablehead{
Binary Type  & Mass $[M_\odot]$   & $a$            & \multicolumn{3}{c }{$a / \rbh$}  }
\startdata
             &         &                & \multicolumn{2}{c |}{Warm ISM}  & Hot ISM          \\            
             &         &                & \textit{Motion in Plane of Disk}  & \textit{Motion Perpendicular to Disk} &          \\            
\hline
Stellar Binary 
             & 1 + 1   & 2 $R_\odot$   & $5 \times 10^{-3}$  & 0.03 - 0.24     &  0.05 - 0.24      \\
             &         & 1 AU		   & 0.05                & 2.8  - 25       & 5 - 26              \\
		   	 &   	   & 0.05 pc	   & 531                 & $10^4-10^5$     & $10^4 - 10^5$   \\
\hline
Massive Stellar Binary 
             & 10 + 10 & 30 $R_\odot$ & $7 \times 10^{-4}$  &  0.04 - 0.35    & 0.07 - 0.4        \\ 
             &         & 1 AU	      & 0.005               &  0.3 - 2.5      & 0.5  - 2.6        \\
\hline
Binary Black Holes
             & 30 + 30 & 10 $R_\odot$  & $8 \times 10^{-5}$ & 0.004 - 0.04    & 0.008 - 0.04     \\
             &         & 100 $R_\odot$ & $8 \times 10^{-4}$ & 0.04  - 0.4     & 0.08  - 0.4       \\
\hline
SMBHs        & $10^6$ - $10^8$	& 1 pc          & $10^{-4} - 10^{-2}$ & 0.01 - 10      & 0.02  - 11   \\
\hline
Double Planet &   $3.7 M_{\rm Jup} + 3.7 M_{\rm Jup}$  & 3.9 AU        & $56$                & $10^3-10^4$    & $10^3-10^4$   \\
\enddata
\tablecomments{The warm ISM consists of neutral and ionized gas and constitutes more than half of the volume of the ISM. We calculate the sound speed using $T = 10^4$ K \citep{2017ApJ...846..133K}.  For the warm ISM with motion in the plane of the disk, we assume the CM moves at $\vinf = 10$ km s$^{-1}$, which would be the case for newly formed binaries in the plane of the disk \citep[where the dispersion velocity is $\approx$ 10 - 50 km/s,][]{2012ApJ...755..115B}.  For the warm ISM with motion perpendicular to the disk, we use $\vinf = 100 - 300$ km s$^{-1}$.  This represents halo or bulge binaries crossing the disk plane. For the hot ISM, we use a temperature range of $T = 10^6 - 10^7$ K \citep{2017ApJ...846..133K} and a dispersion velocity of $\vinf = 100$  km s$^{-1}$ \citep{2010AJ....139...59B,2014ApJ...794...59K}.}
 \vspace{-14pt}
\end{deluxetable*}
\section{Numerical Approach}   \label{sec:methods}
We perform simulations of supersonic BHL flows around an embedded binary system. The traditional approach to numerical studies of BHL is to perform simulations in the frame of a point mass by constructing a wind tunnel that sends gas flowing past a single gravitating sink particle that is fixed at the coordinate origin. The sink particle accretes gas from the domain, and the forces experienced by the sink particle due to the gas can be recorded at each time step.  We modify this approach by instead placing two sink particles in orbit within a wind tunnel. The sink particles are free to move about the domain, and we start the simulations in the initial frame of the CM of the binary. The sink particles accrete gas from the domain and actively respond to the forces they experience due to the gas. In the following subsections, we give a brief history of numerical studies of BHL that are most relevant to the current work, we describe our numerical method, and we outline the parameters chosen for the suite of simulations that we run.

\subsection{Previous Numerical Approaches to BHL}

BHL flows have a long history of hydrodynamical study, beginning with \citet{1971MNRAS.154..141H}, which has been reviewed by \citet{2004NewAR..48..843E} and \citet{2005A&A...435..397F}.  Of particular importance to our work are the studies of \citet{1994ApJ...427..351R} and \citet{1995A&AS..113..133R,1996A&A...311..817R,1999A&A...346..861R}, who performed Eulerian hydrodynamic simulations of BHL. They model the central, accreting boundary condition with a low-pressure sink that effectively absorbs all incoming material and removes it from the domain. These papers examine the effects of gas adiabatic index, Mach number, sink size, resolution, and the size of the computational domain on the rates of accretion and drag.   They showed the qualitative features of BHL flows in 3D: a symmetric bow shock sets up in front of the accretor with an opening angle and standoff distance that are set by $\gamma$  and $\mach$. High-density material collects in a downstream wake from which material is accreted or is advected downstream.  Quantitatively, they found higher accretion rates with smaller adiabatic index and with lower Mach number. 

Ruffert's results, along with the more recent, higher resolution simulations of \citet{2009ApJ...700...95B} and \citet{2012ApJ...752...30B}, have shown that accretion rates are also sensitive to the size of the computational domain, the size of the sink, and the resolution of the grid.  We therefore studied those three aspects of our numerical setup before proceeding with our production runs.  The results of those studies are presented in Appendix \ref{sec:validation}.

In this paper, we use a very similar numerical approach to extend the efforts of \citet{1994ApJ...427..351R}, \citet{1995A&AS..113..133R,1996A&A...311..817R,1999A&A...346..861R}, \citet{2012ApJ...752...30B}, \citet{2015ApJ...803...41M}, \citet{2017ApJ...838...56M}, and \citet{2017ApJ...845..173M} to binary systems by simulating two sink particles orbiting within a supersonic BHL flow. While methodologically distinct, we note that a similar scenario has been simulated in 3D by \citet{2010PhRvD..81h4008F} in the ultrarelativistic regime in which a pair of black holes decays rapidly to merger while traversing through a surrounding gaseous medium.

\subsection{Notation for Two-body Orbits} \label{sec:two_body_orbits}
Before we describe our numerical method in further detail, it is necessary to define the notation we will use to describe the orbit of the binary. We follow the notation of \citet[][Chapter 2]{2001icbs.book.....H} with the exception that we use $M$ to represent the total mass of the binary.  Let us summarize the quantities relevant to this paper. If ${\bf r}_1$ and ${\bf r}_2$ are the positions of $m_1$ and $m_2$, respectively, relative to the coordinate origin, then the CM is located at
\beq
\rcm = \frac{m_1}{M}{\bf r}_1 + \frac{m_2}{M}{\bf r}_2.
\eeq 
Similarly, the velocity of the CM is 
\beq
\vcm = \frac{m_1}{M}{\bf v}_1 + \frac{m_2}{M}{\bf v}_2
\eeq 
where ${\bf v}_1$ and ${\bf v}_2$ are the velocities of $m_1$ and $m_2$ relative to the origin. Consistent with \citet{2001icbs.book.....H}, we use capital letters with subscripts for the barycentric quantities. The barycentric position and velocity of particle $m_i$ are, respectively,
\beq
\begin{aligned}
{\bf R}_i &= {\bf r}_i - \rcm \\
\end{aligned}
\eeq
and
\beq
\begin{aligned}
{\bf V}_i &= {\bf v}_i - \vcm.
\end{aligned}
\label{eq:barycentric_velocities}
\eeq
The position of $m_1$ relative to $m_2$ is ${\bf r} = {\bf r}_1 - {\bf r}_2$ and the semi-major axis of this ``relative'' orbit is denoted by $a$.  The semi-major axis of the barycentric orbit of $m_1$ is
\beq
a_1 = \frac{m_2}{M}a
\eeq  
and similarly for $m_2$.
For a circular orbit, the orbital speed is 
\beq
\begin{aligned}
\vorb &= \bigg(\frac{GM}{a}\bigg)^{1/2}.
\end{aligned}
\label{eq:linearspeeds}
\eeq

\subsection{Hydrodynamical Setup}
We model BHL accretion onto a binary system moving relative to a uniform gas using FLASH \citep{2000ApJS..131..273F}. FLASH is an Eulerian hydrodynamic code with adaptive mesh refinement (AMR). We use FLASH's directionally split Piecewise Parabolic Method Riemann solver \citep{1984JCoPh..54..174C}. The gas pressure is defined by an ideal gas, gamma-law equation of state,
\beq
P = (\gamma -1) \rho \epsilon.
\eeq
where $\rho$, $\epsilon$, and $\gamma$ are the density, internal energy, and adiabatic index of the gas, respectively. We initialize a 3D Cartesian grid with the coordinate origin at the center of the grid. The size of the computational domain in $x$, $y$, and $z$ is $L$, giving a domain volume of $L^3$. 

We use the active sink particles (ASP) unit in FLASH \citep{2010ApJ...713..269F} to model the binary system.  We represent the binary with two equal point masses, $m_1$ and $m_2$, placed at positions ${\bf r}_1$ and ${\bf r}_2$ relative to the grid.  We begin the simulations with the CM of the binary at rest ($\vcm = {\bf 0}$) at the origin of the coordinate system ($\rcm = {\bf 0}$). The two masses are initialized in a circular orbit of semi-major axis $a_0$ in the $x-y$ plane with the orbital angular momentum vector pointing in the $+z$ direction. For comparison to single-particle BHL, we also include the option to place a single particle of mass $M$ on the grid instead of a binary.

To model the supersonic motion of the binary relative to a uniform gas, we fill the domain with a wind of density $\rhoinf$ that flows in the $+x$ direction at speed $\vinf$.  The wind speed is related to the sound speed, $\cs$, through the Mach number $\mach = \vinf / \cs$.  The initial gas pressure is set according to 
\beq
P_\infty = \frac{\rhoinf \cs^2}{\gamma} = \frac{\rhoinf}{\gamma \mach^2}.
\eeq
An inflow boundary condition at the $-x$ boundary of the computational domain replenishes this homogeneous wind throughout the simulation.  All other boundaries of our computational domain have outflow conditions, which extrapolate the thermodynamic quantities and allow gas to freely leave and enter the domain. In our code units, $\vinf = 1$.  With this choice and $M = (2G)^{-1}$, $R_a = 2GM/ \vinf^2 = 1$.  The result is a dimensionless setup with length unit $R_a$ and time unit $\tinf$.  

\subsection{ASP Dynamics and Diagnostics}
We utilize many features built in to the ASP unit of FLASH with some additional diagnostics. See \citet{2010ApJ...713..269F} for a complete description; we briefly describe our implementation here.  The two massive particles that represent our binary system are gravitationally softened point masses surrounded by a spherical inflow boundary of radius $\rs$. The softening radius of each point mass is $0.4\rs$.  Once initialized, each particle moves in response to gravitational accelerations from the other particle and the gas in the domain. The inflow boundary, or sink, that surrounds each particle removes gas from the domain. Once material falls into the sink, it is added to the sink particle properties and then it is deleted.  This boundary condition represents perfectly efficient accretion without feedback. In our implementation of the ASP, we turn off the creation of new sink particles and we turn off self-gravity of the gas.  

\subsubsection{Accreted Quantities}
Gas absorbed by sink particles carries mass and momentum, which are acquired by the particle.  The accretion step is performed by integrating the quantity over all cells within the volume of the sink and by adding the summed values to the particle's properties.  To compute an accretion rate, we divide each accreted quantity by the time step. The accretion rates are recorded, and then the accreted gas is deleted by setting the density within the sink cells to $\rho_{\rm sink} = 10^{-2}\rhoinf$, and setting all components of the gas velocity to zero. The mass accretion rate for particle $m_i$ is calculated as a volume integral over the sink
\beq
\dot{m}_i = \frac{1}{\Delta t} \int_{\rm sink_i}  \big (\rho - \rho_{\rm sink}\big) dV.
\label{eq:mdoti}
\eeq
The total mass accretion rate for the binary is $\mb \equiv \mo + \mt$.  

The accretion of linear momentum is integrated in each Cartesian coordinate. For particle $i$, this is given by
\beq
\dot{{\bf p}}_i = \frac{1}{\Delta t} \int_{\rm sink_i} {\bf v}\big(\rho - \rho_{\rm sink}\big)  dV.
\label{eq:ftransi}
\eeq 
The accretion of linear momentum represents a force on each particle.  We will denote these ``momentum transport forces'' on $m_1$ and $m_2$, respectively, as ${\bf F}_{\dot{p}_1}$ and ${\bf F}_{\dot{p}_2}$.  Finally, 
\beq
{\bf F}_{\dot{p}} = {\bf F}_{\dot{p}_1} + {\bf F}_{\dot{p}_2}
\eeq
is the total momentum transport force on the binary CM.

\subsubsection{Dynamical Friction}
Each sink particle exerts a gravitational force on the gas, which restructures the gas in the domain.  In turn, the redistributed gas exerts a gravitational force on each particle. This gravitational drag force, or gaseous DF, will occur whenever the gas distribution is not spherically symmetric about the particle.  The total (vector) DF force acting on each particle is found by summing up the gravitational force on the particle that is due to each cell in the domain.  Recalling that ${\bf r}_i$ is the location of $m_i$ in the grid and letting ${\bf r}'$ denote the location of a particular gas cell, we find the force on $m_i$ due to the cell is 
\beq
d{\bf F}_{\rm DF_i} = -\frac{G m_i \rho({\bf r}') dV}{|{\bf r}_i - {\bf r}'|^3} \big({\bf r}_i - {\bf r}'\big)
\eeq
and the total DF force on $m_i$ is the integral over the domain
\beq
{\bf F}_{\rm DF_i} = -\int_{\rm domain} \frac{G m_i \rho({\bf r}') dV}{|{\bf r}_i - {\bf r}'|^3} \big({\bf r}_i - {\bf r}'\big).
\label{eq:fdfi}
\eeq
The total DF force on the CM is
\beq
{\bf F}_{\rm DF} = {\bf F}_{\rm DF_1} + {\bf F}_{\rm DF_2}.
\eeq

\subsubsection{Particle Advancement}
The CM position and CM velocity of each particle are updated to account for the accreted gas.  Before the gas is evolved further, each particle's motion is advanced using ASP's leapfrog integrator.  The advancement step applies the momentum transport, DF, and sink-on-sink accelerations to each sink particle.  At each time step, we record the accretion rates and forces as well as the total mass, position, and velocity of each particle.  

\subsection{Simulation Parameters}
To summarize, the free physical parameters of our numerical setup are the background density $\rhoinf$, the flow Mach number $\mach$, the adiabatic index $\gamma$, and the initial semi-major axis of the binary $a_0$.  In order to study the effects of changing binary separation relative to the gravitational focusing scale, we vary $a_0$ while holding $\mach$ and $\g$ constant. We simulate seven values of $a_0 / R_a$: 0.0, 0.16, 0.3, 0.41, 0.55, 0.74, and 1.0.  Each of these simulations adopts $\rhoinf = 10$, $\g = 5/3$, and $\mach =2$. 

We have studied the effects of domain size, spatial resolution, sink radius, and background density on our results.  We present these validations of our method in Appendix \ref{sec:validation}. As a result of these studies, we adopt the following in our production runs. We use a domain of volume $L^3 = (32 R_a)^3$, with the CM of the binary initialized at the center.  The base-level mesh consists of four blocks of $8^3$ cells in the $x$, $y$, and $z$ directions.  We adaptively apply between one and seven levels of refinement.  The finest grid cells are $\delta_{\rm min} = 0.016 R_a$ wide in $x$, $y$, and $z$.  The ASP unit maximally refines the region of each sink particle. We further instruct the AMR unit to refine on the second derivative of the internal energy of the gas.   The radius of each sink particle is set to $\rs = 0.05 \ra$ so that each sink has $\rs/\delta_{\rm min} = 3.125$ grid cells across its radius.  We run the simulations to $t = 50\tinf$, which is 18$\tinf$ beyond the box-crossing time of the wind.  We note that the $a_0 = 1.0 \ra$ simulation falls in the $a \gtrsim \rbh$ regime because $1.0 R_a = 1.25 \rbh$.

\section{Numerical Results}
\label{sec:results}

In this section, we describe the results of our gas dynamical simulations of binary BHL flows. We analyze the morphology of these flows, the resultant accretion onto the individual members, and the forces on the binary CM and orbit. 

\subsection{Flow}
\label{sec:flow}

\begin{figure*}
    \includegraphics[width=\textwidth]{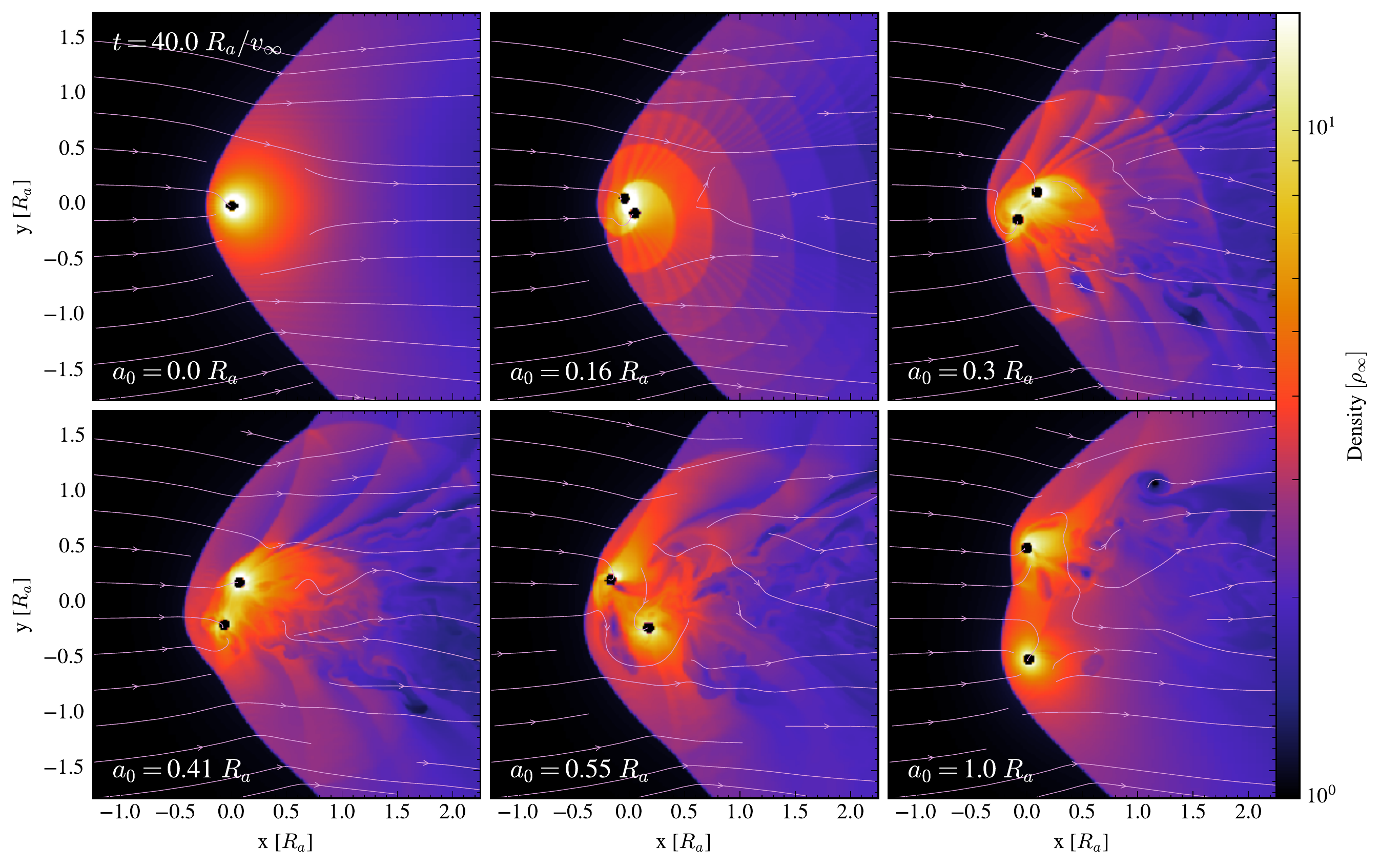}
    \caption{Comparison of binary flow morphology to single-particle BHL flow.  Shown are density slices through the orbital $(z=0)$ plane in units of $\rho_\infty$ at time $t= 40.0 \tinf$ with streamlines following the velocity field. The wind flows from left to right in these images. The axes are in units of $R_a = 2GM\vinf^{-2}$, the accretion radius of the CM of the particle(s). These slices show the region of the orbiting pair; the full computational domain extends to $\pm 16 R_a$ in $x$, $y$, and $z$. The upper left snapshot shows BHL flow for a single particle of mass $M = (2G)^{-1}$. The remaining snapshots show the binary systems simulated, each with total mass $M = (2G)^{-1}$ but a different value of the initial semi-major axis, $a_0$.  More compact binaries impart spiral shocks into the post-bow-shock region, but the effect of the binary on the shape of the overall bow shock increases with $a_0$.}
    \label{fig:sma_series_density}
\end{figure*}

The presence of a binary significantly alters the morphology of BHL accretion flows. At a given time, the manifestation of the binary's presence depends strongly on $a_0$ as compared to $\ra$, and on orbital phase.  We explore these two aspects of BHL flows in the following subsections and in Figures \ref{fig:sma_series_density} -  \ref{fig:flow-timeseries-041}.  These figures  show slices through the orbital ($z=0$) plane. 
 
\subsubsection{Binary Separation and Binary BHL Flows}
 
Figure \ref{fig:sma_series_density} illustrates the effect of semi-major axis on flow morphology. Here we plot density (in units of $\rhoinf$) for our single-particle simulation (upper left) and for binary simulations with five different values of semi-major axis. The single particle, with the same total mass as the binaries, represents the limit of zero binary separation ($a_0 \rightarrow 0.0\ra$). 

The $a_0 = 0.0 \ra$ snapshot of Figure \ref{fig:sma_series_density} illustrates key features of canonical BHL flow. Supersonic motion relative to the background gas sets up an upstream bow shock front that funnels material into a downstream wake. The shock front, which has an opening angle  proportional to $\cs/\vinf$, is smooth and axisymmetric about the line of motion of the point mass.

With the introduction of a binary, rather than a single, central object, some aspects of the flow remain similar while others diverge substantially. With $a_0 = 0.16\ra$, the size and shape of the primary shock remain  very similar to that of $a_0 = 0.0\ra$. As $a_0$ increases, the effect of the binary on the overall bow shock structure becomes more apparent. By $a_0 = 1.0\ra$, each particle sets up its own primary shock (during parts of the orbit), and the overall structure is the superposition of the two shock cones.

Figure \ref{fig:sma_series_density} also shows that the postshock density and velocity structures are also modified by the presence of a binary. In the single-object case, the highest densities surround the accreting object. Flow converges in the wake and compresses as it falls inward toward the accretor. With increasing binary separation relative to $\ra$, we see that this density enhancement divides into separate convergences near $m_1$ and $m_2$. This is particularly apparent in the $a_0=1.0\ra$ snapshot.  We can understand this transition by considering that, with the division of the central object into a binary, the formally divergent potential at the origin (with $a_0 = 0.0\ra$) has been replaced by a potential that is softened on the scale of the binary separation. This restructuring of material in the postshock region has implications for the rates of accretion and drag, which we discuss in detail in Sections \ref{sec:accretion} and \ref{sec:forces}. 

\begin{figure*}
    \includegraphics[width=\textwidth]{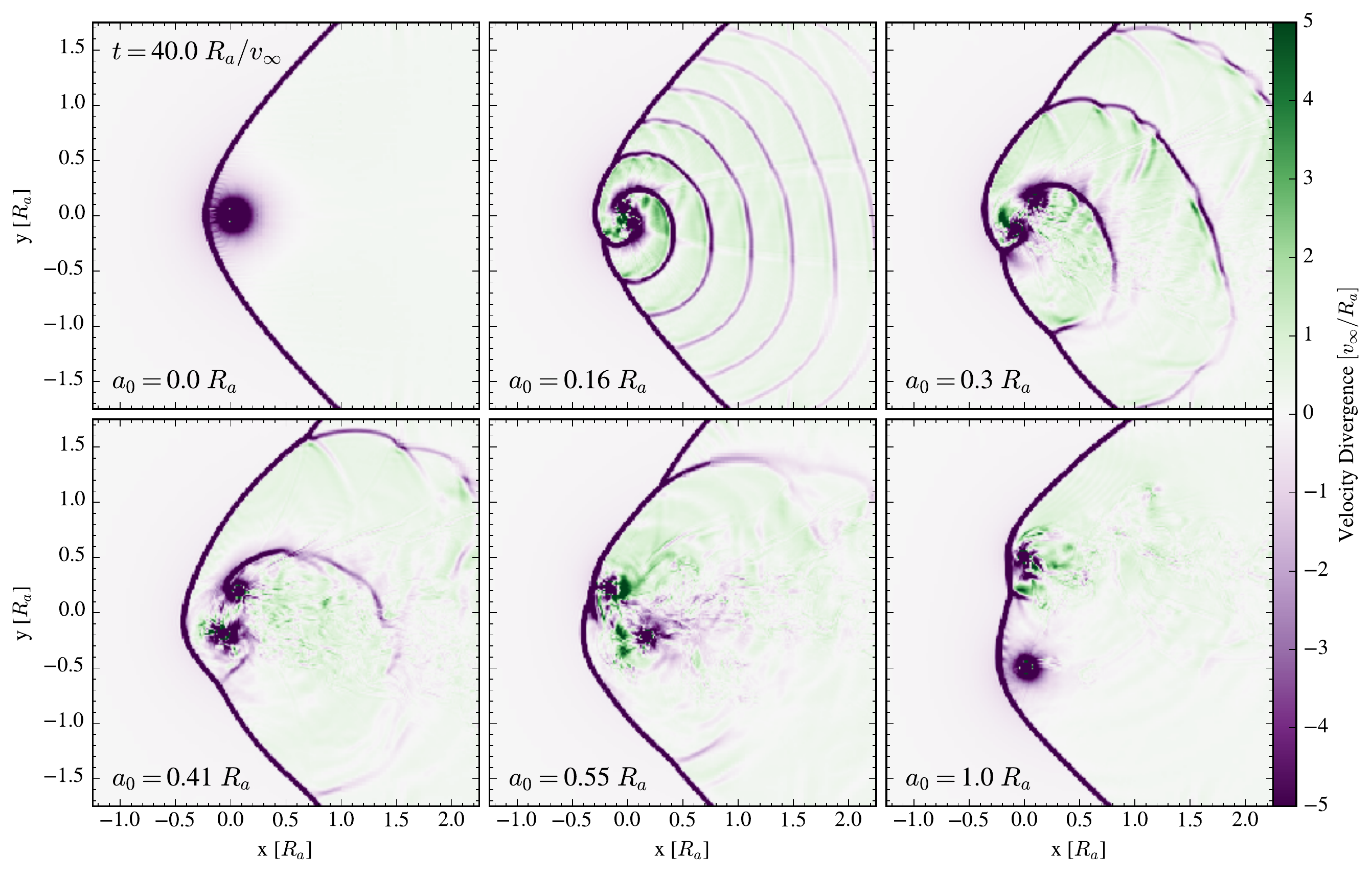}
    \caption{Same as in Figure \ref{fig:sma_series_density}, except that here we plot velocity divergence, $\nabla \cdot {\bf V}_{\rm gas}$. Shades of purple highlight shocked material, where $\nabla \cdot {\bf V}_{\rm gas} < 0$. Secondary shocks form in the vicinity of particles in supersonic relative motion.  The increasing orbital period with increasing $a$ means that the secondary shocks are absent by $a_0 = 1.0 R_a$. }
    \label{fig:sma_series_divergence}
\end{figure*}

Trailing the bow shock, flow is relatively smooth in the case of a single accretor. With the introduction of a binary, orbital motion imparts new features on the postshock flow. For compact binary separations (e.g. $a_0 = 0.16\ra$), some of the primary features are spiral shocks launched because the orbital motion of the binary is supersonic relative to the surrounding gas. These features are most apparent in Figure \ref{fig:sma_series_divergence}, in which we plot the velocity divergence, $\nabla \cdot {\bf V}_{\rm gas}$. Shocked material is revealed in purple in the figure, where $\nabla \cdot {\bf V}_{\rm gas} < 0$ indicates abrupt convergence of gas at a shock front. Especially for the binaries with $a_0 = 0.16\ra$ and $a_0 = 0.3\ra$, spiral shocks propagate outward away from the binary and are advected downstream in the wake. 

As the binary separation increases, the relative velocity between the binary object's orbital motion and the gas decreases. The particle moving through $-y$ positions transitions to subsonic relative motion first, because the particle in the lower half of the orbit moves in the same direction as the gas.  This transition to subsonic relative motion is visible in the bottom row of Figure \ref{fig:sma_series_divergence} through the disappearance of the leading shock front and in Figure~\ref{fig:sma_series_density} through the symmetric density distribution around the object.  In the $a_0 = 0.41\ra$ slice of Figure \ref{fig:sma_series_divergence}, the secondary shock ahead of the lower particle is barely discernible.  By $a_0 = 1.0\ra$, both secondary shocks are absent, and the only inflection of the binary envelope is from the superposition of two primary shock fronts. Here, all orbital motion of the binary is subsonic relative to the postshock sound speed. 

\subsubsection{Time-dependent Properties of the Flow}
\label{sec:time_dependent_flow}

\begin{figure*}[htbp]
\begin{center}
      	\includegraphics[width=\textwidth]{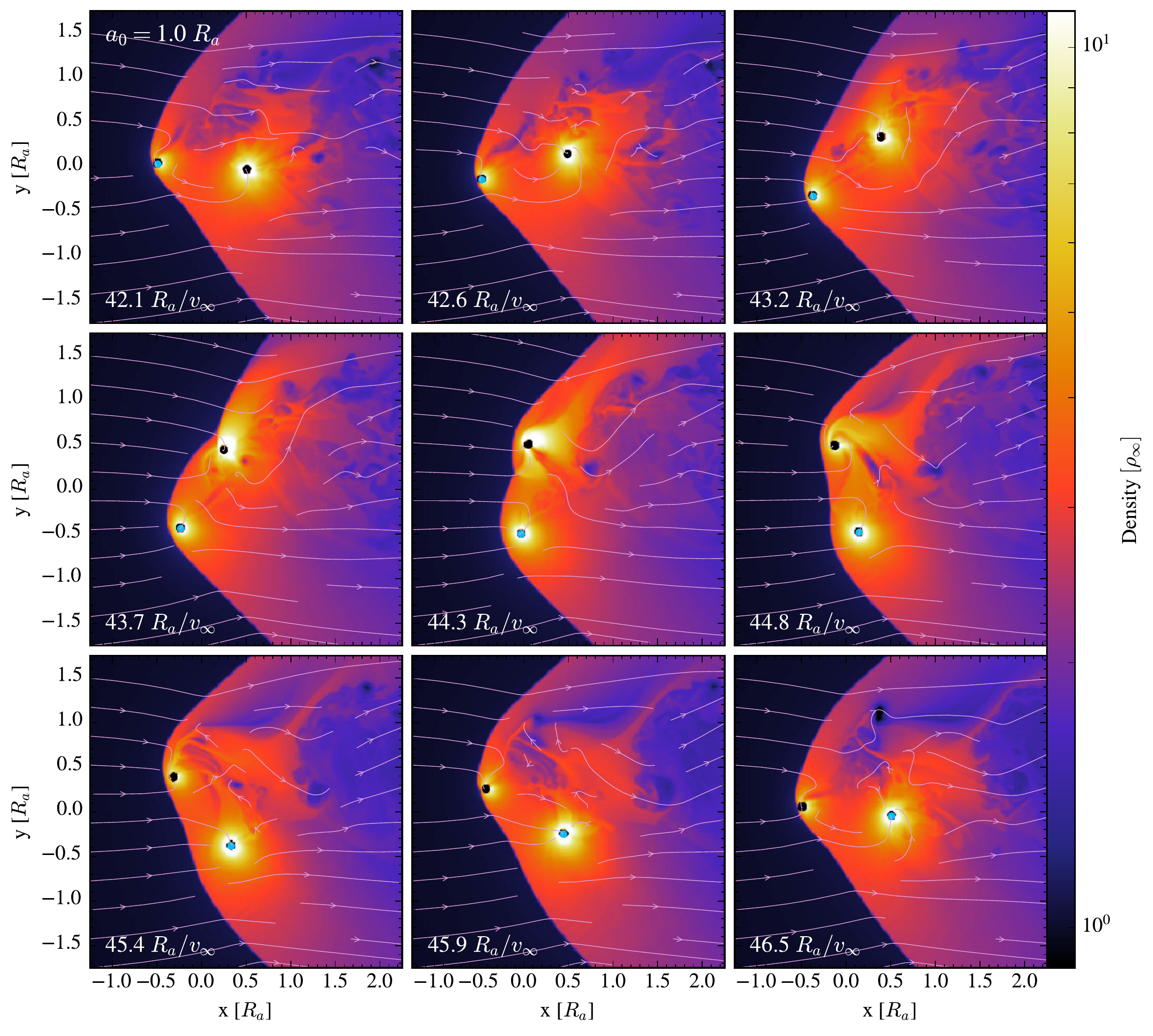}
        \caption{Flow morphology over half of an orbit for the $a_0 = 1.0\ra$ simulation.  As in Figure \ref{fig:sma_series_density}, we show density slices of a $3.5 R_a \times 3.5 R_a$ section of the orbital plane.  The simulation time is shown in the lower left corner of each snapshot, and we have marked $m_1$ with a blue dot and $m_2$ with a black dot.  The overall shape of the primary shock front responds to the rotation of the oblique potential of two particles with finite separation.  The density distribution about each particle is heavily influenced by the motion of the particle relative to the wind.  In this series, $m_1$ moves in the $+x$ direction, as does the wind, so it has a lower speed relative to the gas.  Also, $m_2$ moves in the $-x$ direction, head on into the wind.  The downstream particle interacts with material that has already interacted with its binary companion (see $m_1$ in the last two snapshots).}
        \label{fig:flow-timeseries-1}
\end{center}
\end{figure*}

\begin{figure*}[htbp]
    \vspace{0pt}
    \begin{center}
    \includegraphics[width=\textwidth]{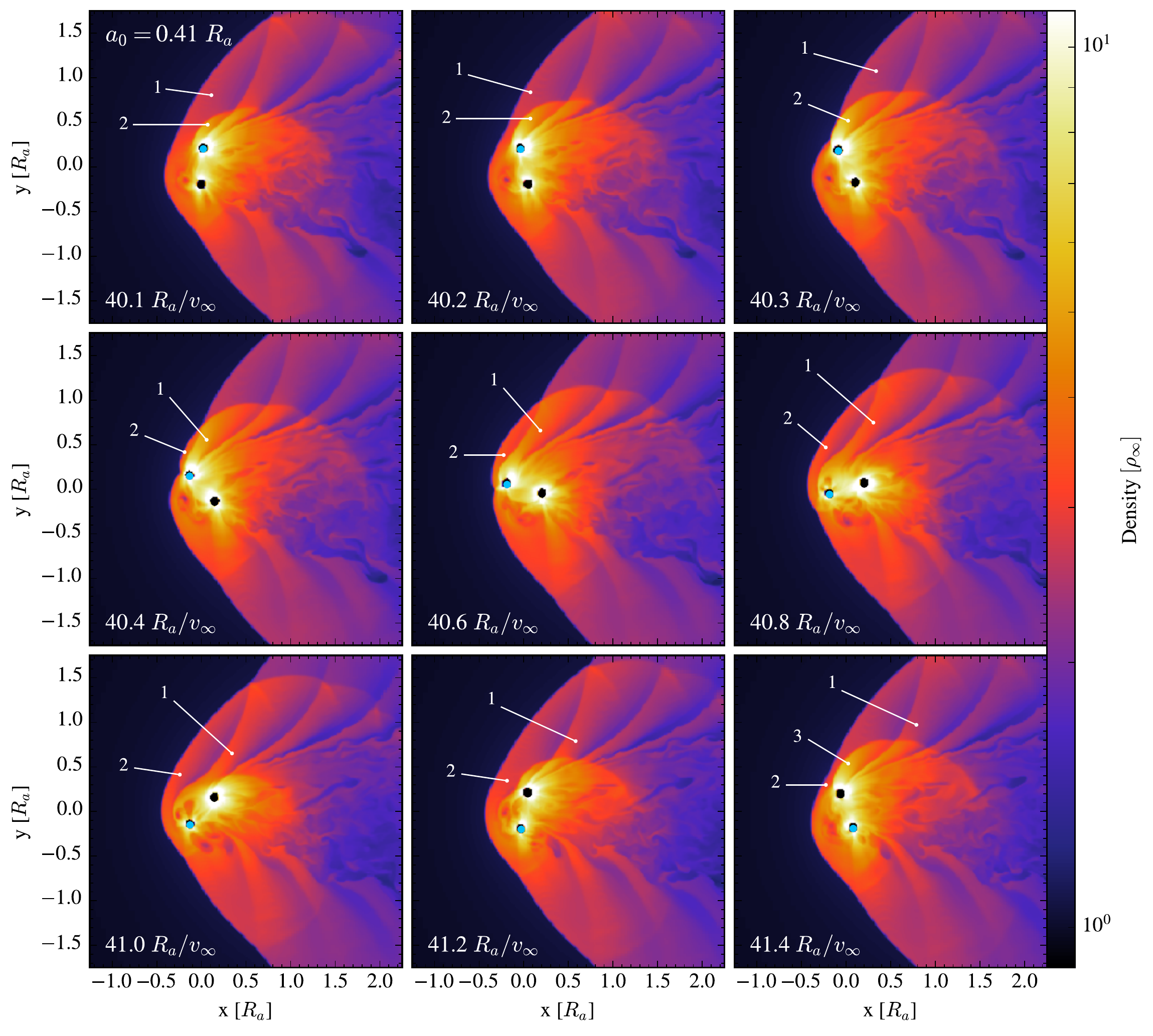}
    \caption{Same as in Figure \ref{fig:flow-timeseries-1}, but for the $a_0 = 0.41 R_a$ simulation.  We observe curved fan structures that arise four times each orbit, which are discussed in detail in the text.}
    \label{fig:flow-timeseries-041}
    \end{center}
    \vspace{22.5pt}
\end{figure*}

In the simulations involving a binary central object, orbital motion imparts time variability and orbital phase dependence to the flow structure. We examine those features in detail here.  Figures \ref{fig:flow-timeseries-1} and \ref{fig:flow-timeseries-041} show the time series of the flow for $a_0 = 1.0 R_a$ and $a_0 = 0.41 R_a$, respectively.  In both of these time series, we show the binary over half of an orbit only.  The second half of each orbit is qualitatively similar to the first, except that the roles of $m_1$ and $m_2$ are reversed.   

Figure \ref{fig:flow-timeseries-1} shows $a_0 = 1.0 R_a$ over the half-orbit that begins at $t = 42.1 \tinf$. In the first image, $m_1$ is fully upstream from $m_2$.  Although at this instant the particles lie along a line parallel to the wind direction (at $\infty$), the bow shock is not symmetric about this line but carries distortions that highlight the preceding orbital motion of the binary.  We will trace the origin of this distortion through  the remaining snapshots of Figure \ref{fig:flow-timeseries-1}.

Situated downstream, $m_2$ interacts with material that has already been influenced by $m_1$.  This dense, lower-velocity material piles up around $m_2$, resulting in a higher-than-average, and almost spherical, density distribution about $m_2$.   This density enhancement remains with $m_2$ as time advances across the upper row of Figure \ref{fig:flow-timeseries-1}, approaching the primary bow shock by the first image of the second row. 

At $t = 43.7 \tinf$, a second shock front is forming ahead of $m_2$ as the particle approaches its greatest velocity relative to the gas. Between this snapshot and the next, $m_2$ crosses the leading boundary of the bow shock of $m_1$.  The overdensity that has been carried along by $m_2$ remains with the particle during the crossing.  
In the next two images ($t = 44.3$ and $44.8$ $\tinf$), this dense material is impacted by the incoming wind.  By $t = 45.4 \tinf$ (first panel of third row), the over-density is largely detached from $m_2$ and is being pulled into the approaching potential well of $m_1$.  In the same panel, a bridge of material connects the two particles and, for the first time in these snapshots, the density enhancement about $m_1$ is greater than that of $m_2$.  

Two flow properties transition in the last row of Figure \ref{fig:flow-timeseries-1}. First, the dense material that is attached to $m_2$ at $t = 45.4 \tinf$ becomes fully detached from the particle as the downstream region is increasingly dominated by the gravitational pull of $m_1$.  Second, as $m_2$ moves down and to the left, the shock set up by the particle rotates with it, leading the $+y$ portion of the bow shock to protrude farther upstream than the $-y$ portion.		

Figure \ref{fig:flow-timeseries-041} shows the $a_0 = 0.41 R_a$  binary for the half-orbit beginning at $t = 40.1 \tinf$. We additionally annotate several features, which will be referenced here. As discussed in the context of Figure \ref{fig:sma_series_divergence}, the particle traveling upwind through the post-bow-shock material launches a secondary spiral shock wave behind the primary shock. 

The interaction of the binary with spiral shocks sets up the fan-like structures emanating from the binary in Figure~\ref{fig:flow-timeseries-041}. 
Material crossing the primary shock interacts with each particle twice, resulting in four fan structures per orbit.   In the first snapshot ($t = 40.1 \tinf$), the particles are located at $x \approx 0$.  There is a large standoff distance between the primary shock and the binary.  Behind the primary shock, a secondary shock is set up in front of $m_1$.  We label these structures ``1'' and ``2,'' respectively, in all of the snapshots shown, so that one can follow each structure over time.  From $t = 40.2$ to $40.3\tinf$, shock 1 is deflected by the gravitational pull of $m_1$.  By $t = 40.4\tinf$, shock 2 is pushed past shock 1 by $m_1$ as the particle advances in the $-x$ direction. 
From $t = 40.6$ to $t = 41.0\tinf$, $m_2$ advances toward $+y$ values, and shock 1 is defected by the gravitational pull of $m_2$. At $t = 41.2\tinf$, a new secondary shock (set up by $m_2$  and labeled ``3'' in Figure \ref{fig:flow-timeseries-041}) crosses shock 1 as shock 1 continues to advect downstream.  In the final image, shock 3 is advancing to meet shock 2 and will cross shock 2 as the orbit continues.

In addition to the fan structures, the density very close to the binary is of interest. In particular, it appears that the particle at $+y$ positions exhibits a higher local density than its counterpart at $-y$ values.  A persistent bridge of dense material always connects the two particles, giving a slightly higher density enclosed by the orbit than in the $a_0 = 1.0\ra$ simulation.

\subsection{Accretion} \label{sec:accretion}
In this section, we examine instantaneous and orbit-integrated accretion rates onto the objects. 

\subsubsection{Instantaneous Accretion Rates}
\begin{figure*}
\includegraphics[width=\textwidth]{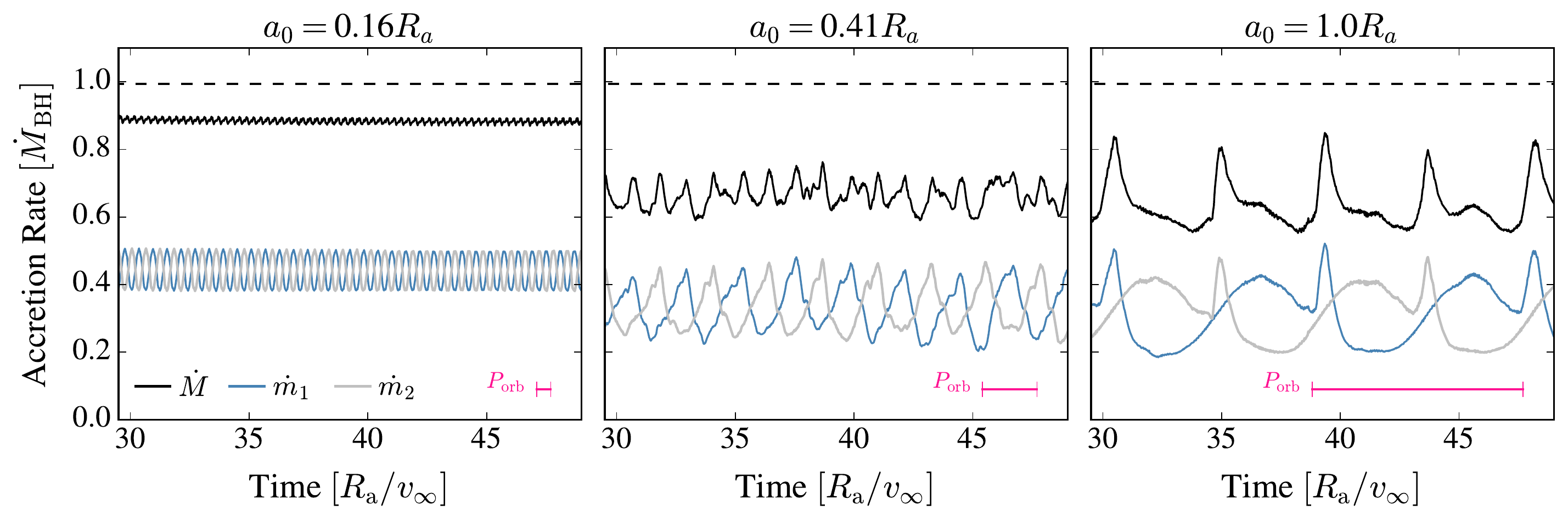}
    \caption{Instantaneous accretion rates over time for the binaries with $a_0 = 0.16 R_a$ (left panel),  $a_0 = 0.41 R_a$ (center panel), and $a_0 = 1.0 R_a$ (right panel).  The individual accretion rates,  $\mo$ and $\mt$, and the total rate of accretion, $\mb = \mo + \mt$, are shown.  The orbital period of each binary is included for reference.   On each panel, the dashed line indicates the steady-state accretion rate realized in the single-particle simulation, $a_0 = 0.0 R_a$.  The individual accretion rates for the  $a_0 = 0.16 R_a$ binary sum to an almost constant value, while the total accretion rates for $a_0 = 0.41 R_a$ and $a_0 =1.0 R_a$ exhibit significant time variability.  The wider binaries experience lower accretion rates than the tightest binary, whose total accretion rate approaches that of the single-particle simulation.}
    \label{fig:mdot-time-series}
\end{figure*}

In Figure \ref{fig:mdot-time-series}, we plot the instantaneous accretion rates versus time for $a_0 = 0.16, 0.41$, and $1.0 R_a$ for simulation times $\ge 30 \tinf$.   As expected from prior studies of canonical BHL accretion (with $\rs = 0.05 R_a$, $\mach \sim 2$, and at our resolution), $\ms \approx 0.7 \mhl \approx 1.0 \mbh$ \citep{1994A&AS..106..505R,2012ApJ...752...30B}.   

Although $\ms$ reaches a steady state, the accretion rates for the members of the binary systems are periodic.  For the tightest binary, $a_0 = 0.16 R_a$, and $\mo$ and $\mt$ are exactly out of phase, each with an accretion rate that follows the orbital period.  As such, $\mb$ is almost constant in time and approaches $\ms$.  For more widely separated binaries, the individual accretion rates of the two particles no longer sum to a constant value.  In the case of $a_0 = 1.0 R_a$, $\mo$ shows a sharp spike once per orbit before declining to a minimum value, and then the rate climbs to a local maximum before the next sharp spike in accretion. While the curve for  $\mt$ looks very similar, the variation in $\mb$  reveals that the individual accretion rates are not exactly out of phase.  In the case of $a_0 = 0.41 R_a$, the individual accretion rates reveal one peak and one valley per orbit with transitions between the two that are variable from one orbit to the next.  In addition, $\mo$, for example, reveals a slight overall increase and decrease over many orbits. The superposition of these trends in $\mo$ and $\mt$ yields an $\mb$ that has irregularly spaced peaks and valleys with variation from orbit to orbit.  

\begin{figure}
    \includegraphics[width=0.46\textwidth]{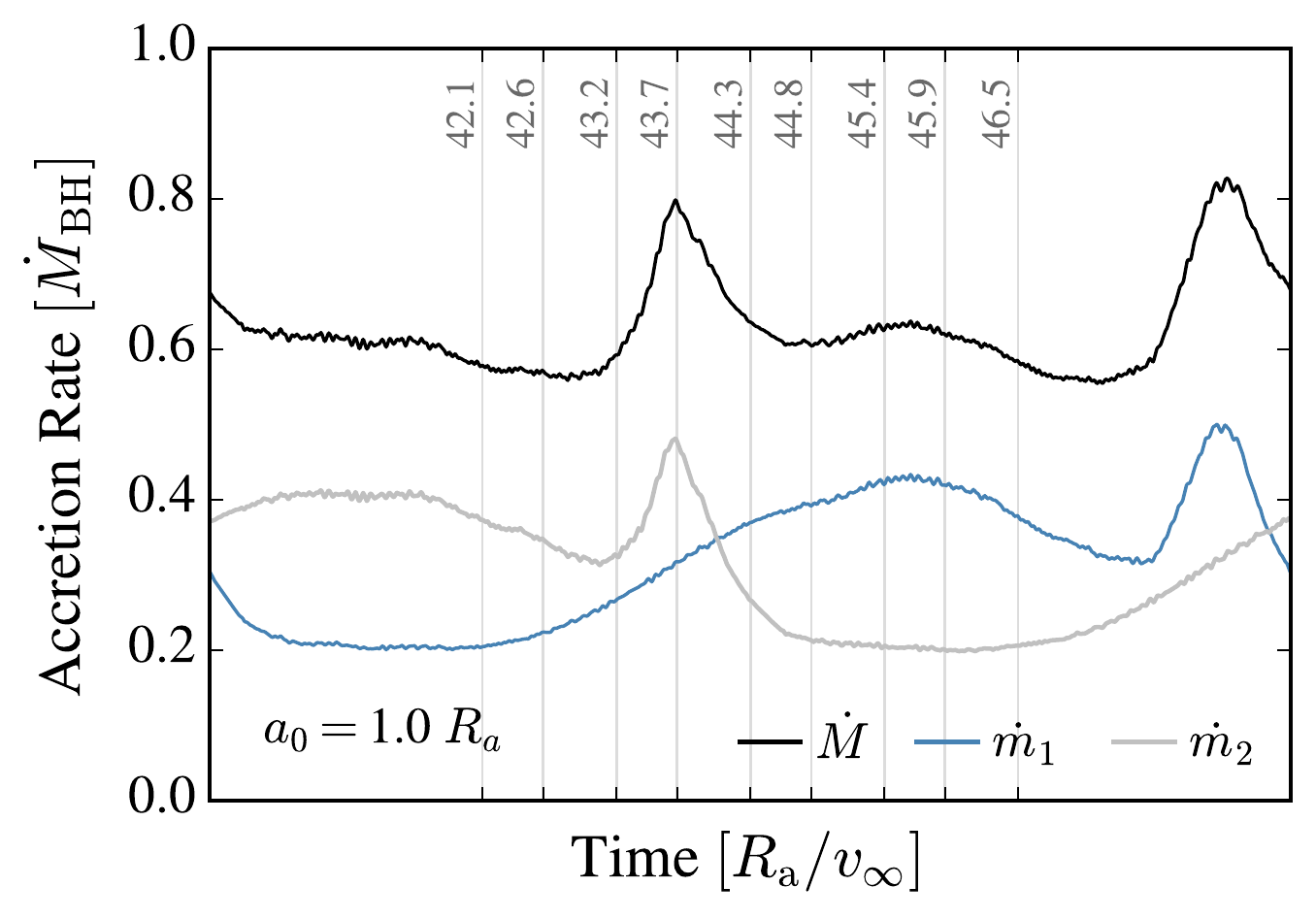}
    \caption{As in the right panel of Figure \ref{fig:mdot-time-series}, instantaneous accretion rates are shown for the $a_0 = 1.0 R_a$ binary. Here we focus on one orbit of the binary, which begins at $t = 39.9 \tinf$.  Each vertical line corresponds to one of the density snapshots of Figure \ref{fig:flow-timeseries-1}.}  
    \label{fig:mdot-zoom-a1}
\end{figure}

\begin{figure}
    \includegraphics[width=0.46\textwidth]{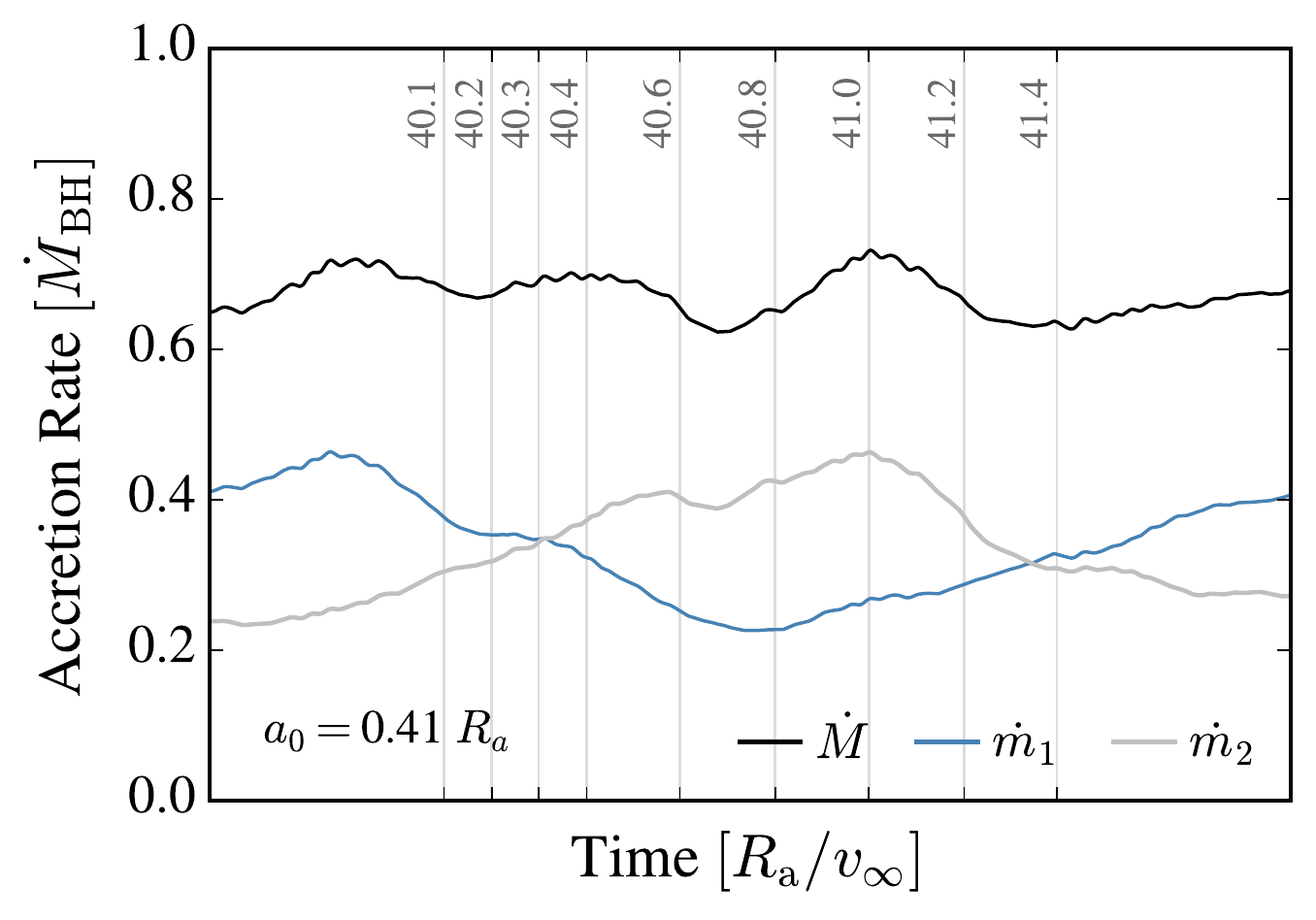}
    \caption{Instantaneous accretion rates for the $a_0 = 0.41 R_a$ binary for a single orbit, beginning at $t = 39.6 \tinf$.  Each vertical gray line corresponds to one of the density snapshots of Figure \ref{fig:flow-timeseries-041}.}
    \label{fig:mdot-zoom-a041}
\end{figure}

Let us now connect features of the accretion rates depicted in Figure \ref{fig:mdot-time-series} to the flow morphology discussed in Section  \ref{sec:time_dependent_flow}.  In Figure  \ref{fig:mdot-zoom-a1}, we show $\mo$, $\mt$, and $\mb$ for the $a_0 = 1.0 R_a$ simulation for simulation times $t = 39.9 - 48.7$ $\tinf$, as in the panels of Figure \ref{fig:flow-timeseries-1}. In the first row of Figure~\ref{fig:flow-timeseries-1}, $m_2$ moves from a position directly downstream from $m_1$ at $t = 42.1 \tinf$ into a lower-density pocket behind the primary shock by $t = 43.2 \tinf$. Figure \ref{fig:mdot-zoom-a1} reveals a decreasing accretion rate for $m_2$ during the same period of time.   Although $m_2$ has carried along higher-density material during this time interval, the material is able to expand into the low-density region behind the primary shock, giving rise to the local minimum in $\mt$ at $t \approx 43.2 \tinf$.  

Between $t = 43.2$ and $43.7 \tinf$, $m_2$ has turned into the oncoming wind and approaches the primary shock front; the particle nears its highest velocity relative to the gas at $t = 43.7 \tinf$ and a second shock front forms in front of $m_2$. Meanwhile, the high-density concentration that has followed $m_2$ is now stopped by this second shock, and a stream of material flows from $m_1$ to $m_2$.  This confluence of flow gives $\mt$ a sharp peak at $t = 43.7 \tinf$.  

By $t = 44.3 \tinf$, $m_2$ has crossed the primary bow shock of $m_1$, and now $\mt$ declines. From $t = 44.3$ to $44.8 \tinf$, $m_2$ begins to lose the high-density clump that has followed the particle since its time downstream from $m_1$.  Then $\mt$ bottoms out as the clump of material finally detaches from $m_2$ and is swept downstream ($t = 45.4$ to $46.5 \tinf$).  At the same time, $m_1$ is moving into a position downstream from $m_2$. There is an increase in $\mo$ from $t = 44.3-45.9 \tinf$ as $m_1$ moves with its slowest relative velocity compared to the surrounding gas and interacts with dense material stripped from the vicinity of $m_2$.
From $45.9$ to $46.5 \tinf$, $\mo$ declines slightly as the clump of material that has detached from $m_2$ crosses outside of the orbit of $m_1$ as it is advected downstream and out of reach.

Figure \ref{fig:mdot-zoom-a041} shows the accretion rates for the $a_0 = 0.41 \ra$ simulation for the time period $t = 39.6$ to $t = 41.9 \tinf$, also shown in the snapshots of Figure \ref{fig:flow-timeseries-041}.  From $t = 40.1$ to $40.8 \tinf$, $\mt$ steadily increases as $m_2$ moves in the same direction as the flow.  At $40.8 \tinf$, $m_2$ turns upstream. There is a steepening of $\mt$ to a maximum at $41.0 \tinf$ as $m_2$ deflects shock 1 while also stripping material from the region surrounding $m_1$.  Between $41.0$ and $41.2$ $\tinf$, shock 3 forms in front of $m_2$. Once this material crosses shock 1 (at $41.2 \tinf$), it expands into the lower-density region behind shock 2, reducing the rate of accretion onto $m_2$.

Meanwhile, $m_1$ experiences a decreasing accretion rate from $t = 40.1 \tinf$ to $t = 40.8 \tinf$ as the particle takes up a position in front of $m_2$.  There is a flattening  of $\mo$ from $t = 40.2$ to $40.3 \tinf$ when shock 1 is pulled into $m_1$ and shock 2 crosses shock 1.   The overall decrease in $\mo$ continues through $t = 40.8  \tinf$ while $m_1$ is in the upstream position and $m_2$ pulls material away from $m_1$.   Just before $t = 40.8 \tinf$, the slope of $\mo$ turns positive as $m_1$ moves to $-y$ values. There, $m_1$ moves in the same direction as the wind, reducing the relative motion between the particle and the gas.  From there the cycle depicted in Figure \ref{fig:flow-timeseries-041} begins again, but this time with the roles of $m_1$ and $m_2$ reversed.

\subsubsection{Orbit-averaged Accretion Rates}
Having examined the origins of time-variable accretion rates in binary BHL flows, we now compare time-integrated quantities for binaries of varying separation. 

It is apparent from Figure \ref{fig:mdot-time-series} that $\mo$, $\mt$, and $\mb$ for $a_0 = 0.41$ and $1.0 \ra$, are  lower than the corresponding accretion rates in the $a_0 = 0.16 R_a$ case.   To look for trends in the accretion rate as a function of $a_0$, let us compare the time-averaged total accretion rate  for each pair.  We use angled brackets, $\langle~\rangle$, to indicate quantities averaged over an integer number of orbits. Each time average begins at $t = 30 \tinf$ and terminates at the end of the last complete orbit before the end of the simulation ($t = 50 \tinf$).   The $a_0 = 1.0 R_a$ binary (with $P_{\rm orb} \approx 8.89 \tinf$) completes two full orbits during this time. The $a_0 = 0.16 R_a$ binary completes 33 full orbits.  

\begin{figure}
    \includegraphics[width=0.46\textwidth]{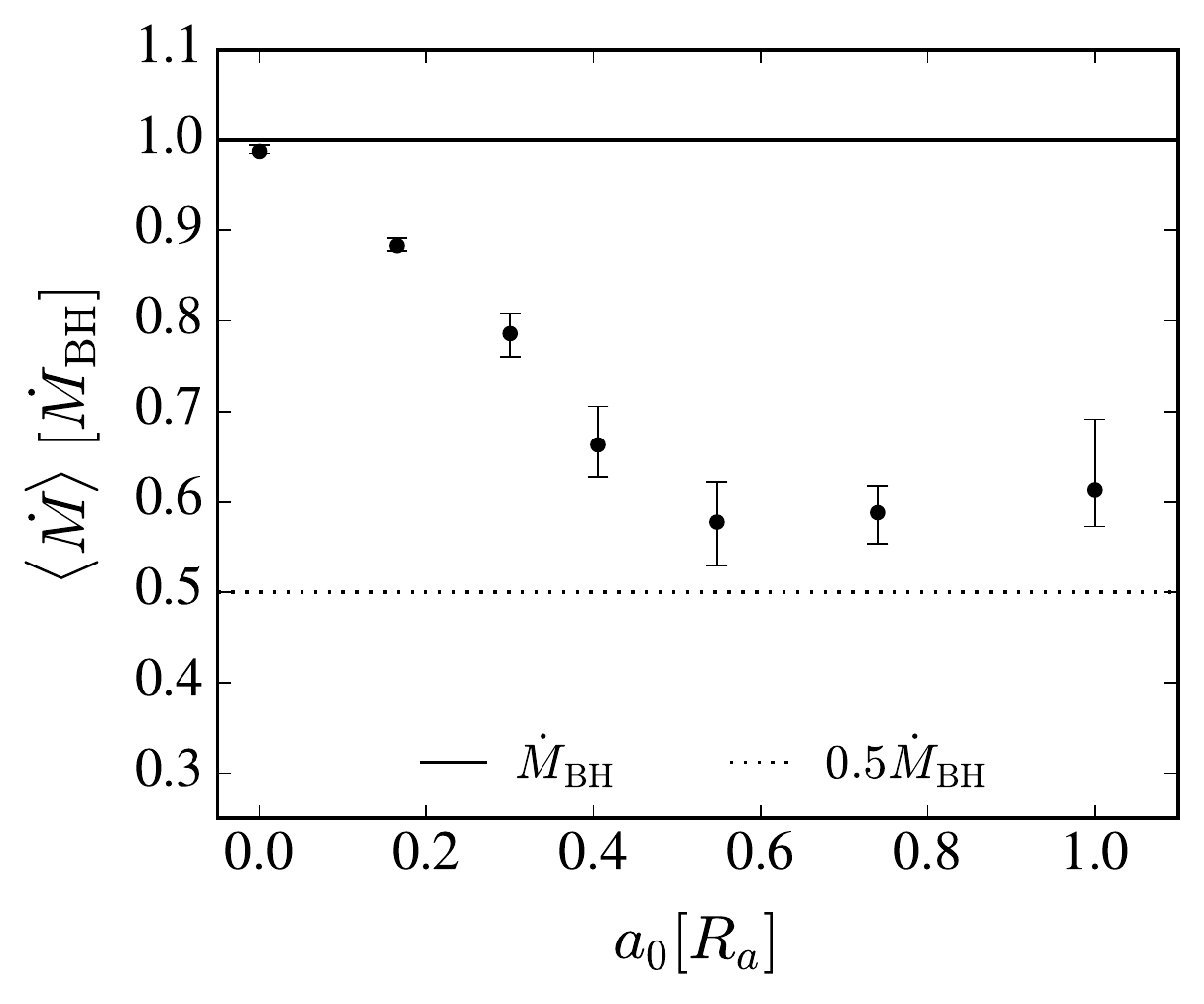}
    \caption{Orbit-averaged total accretion rate, $\ave\mb$, in units of $\mbh$ versus $a_0$, in units of $R_a$.  The black points show the median of $\mb(t)$ for each binary over an integer number of orbits between $t = 30 \tinf$ and $t = \, 50 \tinf$.  The error bars show the 15.87th to 85.13th percentile range about the median, driven by periodic variations with orbital phase. At smaller separations, the accretion rate approaches that of a single particle with $\ave\mb \approx \mbh \propto M^2$, while at larger separations, the accretion rate approaches that of two independent particles with $\ave\mb \approx \dot{M}_{\rm BH, independent}  \propto (m_1^2 + m_2^2)$.} 
    \label{fig:mdot-versus-a}
\end{figure}

In Figure \ref{fig:mdot-versus-a}, we plot the orbit-averaged total accretion rate versus $a_0$.  As the binary separation decreases, $\ave\mb$ approaches $\mbh  \propto M^2 \propto (m_1 + m_2)^2$.  
That is, binaries of smaller separations accrete like a single particle of mass $m_1 + m_2$.  Wider binaries, on the other hand, have a lower accretion rate and fall closer to $\ave\mb = 0.5 \mbh$, which assumes that $m_1$ and $m_2$ accrete independently of one another.  To see where the factor of $0.5$ comes from, let us write the BH accretion rate for $m_1 = M /2$:
\beq
\dot{M}_{\rm BH, 1} = \frac{4 \pi G^2 m_1^2 \rhoinf}{(\vinf^2 + \cs^2)^{3/2}} = \frac{4 \pi G^2 (M/2)^2 \rhoinf}{(\vinf^2 + \cs^2)^{3/2}} = \frac{1}{4} \mbh.
\eeq
Because $m_2 = M /2 = m_1$, $\dot{M}_{\rm BH, 2} = \dot{M}_{\rm BH, 1}$.  Then the total accretion rate for the pair (treating them as independent particles) is 
$\dot{M}_{\rm BH, independent} = 0.25 \mbh + 0.25 \mbh = 0.5 \mbh$.  For more general $m_1$ and $m_2$, we would say that because $\dot{M}_{\rm BH, 1} \propto m_1^2$ and $\dot{M}_{\rm BH, 2} \propto m_2^2$, $\dot{M}_{\rm BH, independent} \propto (m_1^2 + m^2_2)$.

\subsection{Forces}
\label{sec:forces}
When either a single object or binary moves through a gaseous medium, it restructures the distribution of the surrounding gas.  The restructured gas exerts forces on the CM and the components of the binary.  In this section, we describe how these forces are calculated from our numerical simulations, and we show an example calculation for our $a_0 = 0.41\ra$ simulation. 

\begin{figure}
    \includegraphics[width=0.46\textwidth]{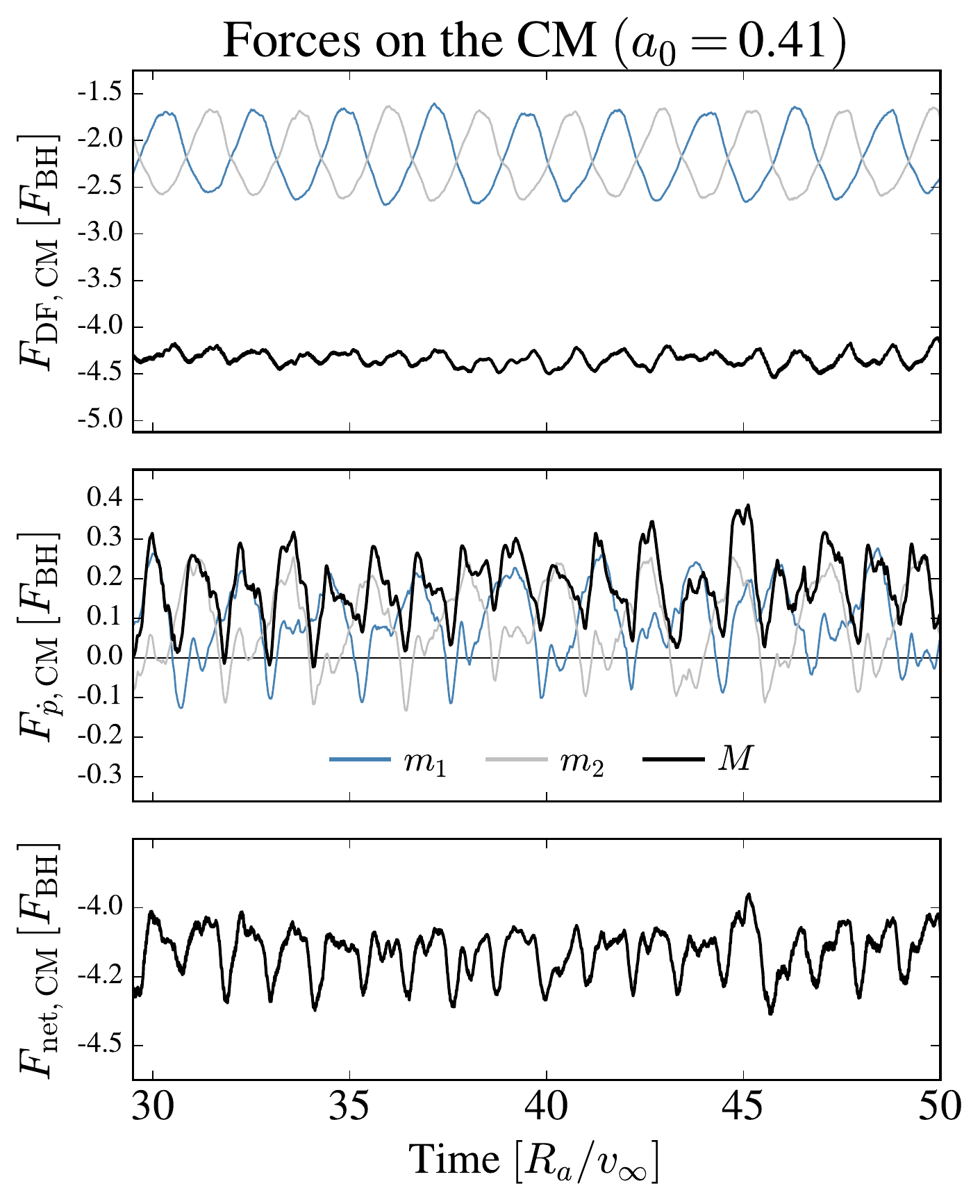}
    \caption{Center-of-mass forces over time for the $a_0= 0.41R_a$ binary. In the upper and middle panels, we show the DF and momentum transport forces, respectively, acting along the CM velocity vector relative to the wind, $\hat{V}_{\rm rel}$.  The blue and gray curves show the contributions from $m_1$ and $m_2$, respectively, to the total of each type of force (black curve).  The bottom panel is the total force on the CM due to both DF and momentum transport.  In each panel, a positive value indicates a net thrust of the CM while  a negative value indicates a net drag on the CM.  The DF force arising from the downstream wake dominates the total force on the CM of the binary. At all times, this net drag force acts to slow down the translational motion of the CM relative to the background gas.}  
    \label{fig:forces_a041_cm}
\end{figure}

\subsubsection{Forces on the Center of Mass}
First, let us consider how the forces change the translational motion of the binary CM. At each time step, we calculate the DF and momentum transport forces acting on each particle. The DF forces on $m_1$ and $m_2$ are calculated according to Equation \eqref{eq:fdfi}, while the momentum transport forces are calculated according to Equation \eqref{eq:ftransi}.  We can ask how these forces influence the motion of the CM by considering the component of each force along the CM velocity relative to the gas. 

The velocity of the CM relative to the grid is $\vcm$ and the velocity of the wind relative to the grid is ${\bf v}_\infty = \vinf\hat x$, so the velocity of the CM relative to the gas is ${\bf V}_{\rm rel} = \vcm - {\bf v}_\infty$.  Letting $\hat{V}_{\rm rel}$ be the unit vector in the direction of ${\bf V}_{\rm rel}$, we find the scalar component of the DF and momentum transport forces along ${\bf V}_{\rm rel}$ are, respectively,
\beq
F_{\rm DF, CM} \equiv \big({\bf F}_{\rm DF_1} + {\bf F}_{\rm DF_2} \big) \cdot \hat{V}_{\rm rel}
\label{eq:f_df_cm}
\eeq
and the momentum transport force on the CM is 
\beq
F_{\dot{p}\rm, CM} \equiv \big({\bf F}_{\dot{p}_1} + {\bf F}_{\dot{p}_2} \big) \cdot \hat{V}_{\rm rel}.
\label{eq:f_pdot_cm}
\eeq
Notice that $F_{\rm DF, CM}$ and $F_{\dot{p}\rm, CM}$ may be positive or negative.  A positive value indicates a net thrust, while a negative value exerts a net drag.  The net force on the CM 
\beq
F_{\rm net, CM} = F_{\rm DF, CM} + F_{\dot{p}\rm, CM},
\label{eq:f_net_cm}
\eeq
is the total drag (or thrust) that results from the sum of the two projections.

Figure \ref{fig:forces_a041_cm} shows the forces on the CM of the $a_0 = 0.41 R_a$ binary from $t = 30 \tinf$ until the end of the simulation. The top panel shows the DF forces on the CM (Equation \eqref{eq:f_df_cm}).  The contribution due to $m_1$ is ${\bf F}_{\rm DF_1}\cdot \hat{V}_{\rm rel}$, while the contribution from $m_2$ is ${\bf F}_{\rm DF_2}  \cdot \hat{V}_{\rm rel}$. The periodic modulation of the blue and gray curves is due to the orbital motion of each particle.  Even though the density structure at large scales is relatively stable, the position and velocity of each particle relative to the gas distribution change as the particles complete each orbit.  
This gives rise to a changing angle between each force vector and $\hat{V}_{\rm rel}$ as well as a changing magnitude of the force vectors themselves (${\bf F}_{\rm DF_1}$ and ${\bf F}_{\rm DF_2}$).  In the same panel, $F_{\rm DF, CM}$ (in black) also shows variation due to the orbital motion about a nearly constant value of -4.2 $\fbh$, which is less than the single-particle value of -5.1 $\fbh$.  Overall, the negative value of $F_{\rm DF, CM}$ indicates that the large-scale structure always exerts a drag on the CM motion of the pair.  

The center panel of Figure \ref{fig:forces_a041_cm} shows the momentum transport forces acting along the CM motion of the binary, which are calculated according to Equation \eqref{eq:f_pdot_cm}.  The $m_1$ and $m_2$ curves show ${\bf F}_{\dot{p}_1} \cdot \hat{V}_{\rm rel}$ and ${\bf F}_{\dot{p}_2} \cdot \hat{V}_{\rm rel}$, respectively.  The blue and gray curves dip to negative values when accreted material is moving opposite the CM motion of the particle. Their sum is always positive, though, so the momentum transport force always exerts a net thrust on the CM.  This is similar to the single-particle case, although the median value of $|F_{\dot{p}, \rm CM}| \approx$ 0.18 $\fbh$ is $74\%$ less than the magnitude of the single-particle value of 0.69 $\fbh$.  Finally, the bottom panel of Figure \ref{fig:forces_a041_cm} shows the net force on the CM, given by Equation \eqref{eq:f_net_cm}.  The combination of DF and momentum transport forces always exerts a drag on the CM. The orbit-averaged net force on the CM for the $a_0 = 0.41 R_a$ binary is -4.06 $\fbh$, a value that is only $7\%$ lower than in the single-particle case.  

\subsubsection{Forces on the Orbital Motion}
\begin{figure}
    \includegraphics[width=0.46\textwidth]{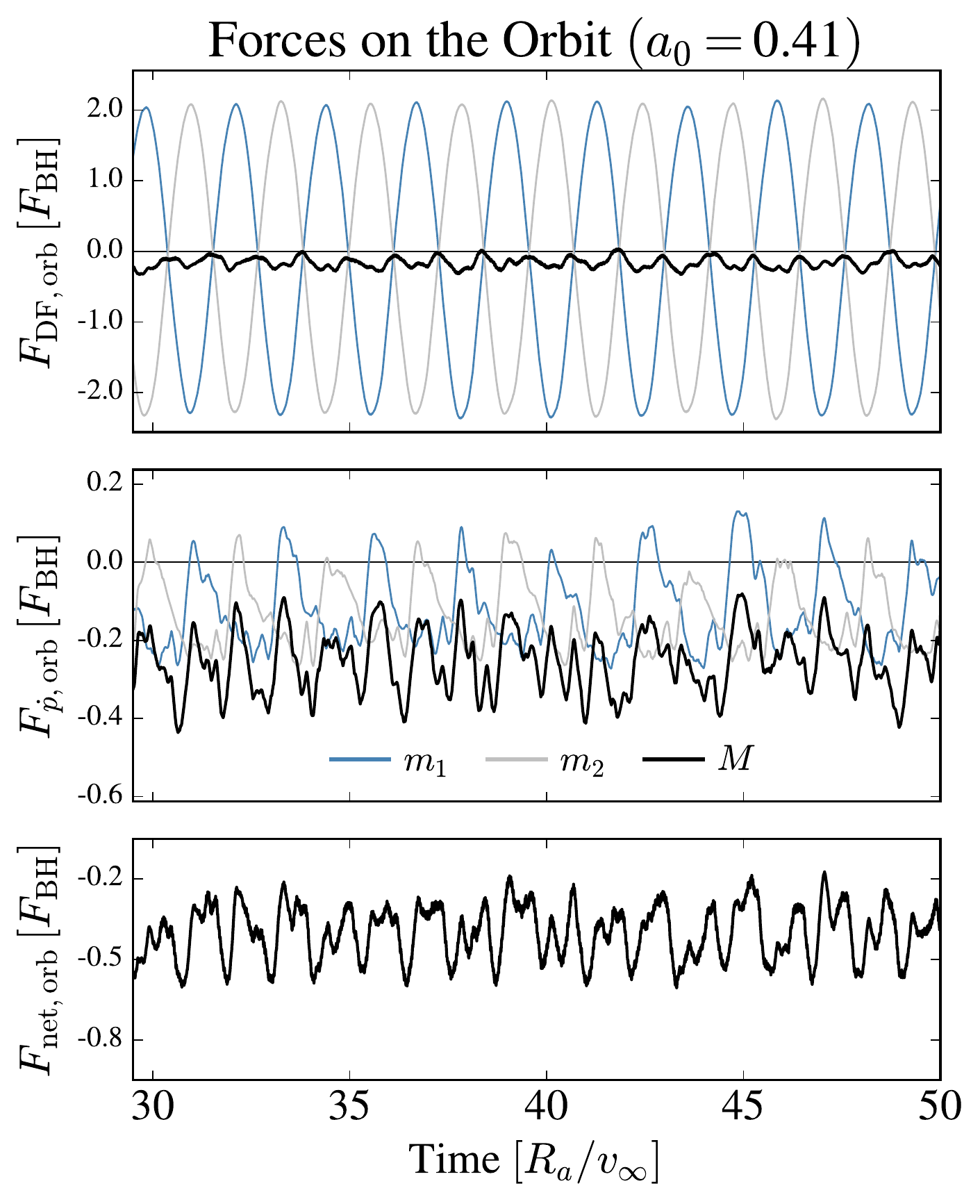}
    \caption{Orbital forces over time for the $a_0= 0.41R_a$ binary.  Same as in Figure \ref{fig:forces_a041_cm}, except that we plot the projections of the forces along the barycentric velocities of $m_1$ and $m_2$.  Both DF and momentum transport cause a net drag on the orbital motion at all times.  Momentum transport and DF contribute almost equally to the overall drag on the motion. The flow at small scales in the vicinity of the orbiting binary cannot be neglected, as momentum accretion contributes nearly half of the overall drag on the orbital kinetic energy of the binary.}
    \label{fig:forces_a041_orb}
\end{figure}
 We now consider the forces affecting the motion of the binary about the CM.  To do this, we find the projections of the forces along the barycentric velocity vectors of the two particles.  The barycentric velocities of $m_1$ and $m_2$ are ${\bf V}_1$ and ${\bf V}_2$, respectively (Equation \eqref{eq:barycentric_velocities}).  The unit vectors in the directions of these velocities are then $\hat{V}_1 = {\bf V}_1 / V_1$ and $\hat{V}_2 = {\bf V}_2 / V_2$.  The DF force on the orbital motion of the binary is 
\beq
F_{\rm DF, orb} \equiv {\bf F}_{\rm DF_1}  \cdot \hat{V}_1 + {\bf F}_{\rm DF_2}  \cdot \hat{V}_2
\label{eq:f_df_orb}
\eeq
and the momentum transport force on the orbit is 
\beq
F_{\dot{p}\rm, orb} \equiv {\bf F}_{\dot{p}_1} \cdot \hat{V}_1 + {\bf F}_{\dot{p}_2} \cdot \hat{V}_2,
\label{eq:f_pdot_orb}
\eeq
which gives a net ``orbital'' force of
\beq
F_{\rm net, orb} = F_{\rm DF, orb} + F_{\dot{p}\rm, orb},
\eeq
which is the force projected along the objects' respective directions of orbital motion. 

In Figure \ref{fig:forces_a041_orb}, we show these orbital forces calculated for $a_0 = 0.41 R_a$.   The DF forces, ${\bf F}_{\rm DF_1}  \cdot \hat{V}_1$ and ${\bf F}_{\rm DF_2}  \cdot \hat{V}_2$, shown in the upper panel of Figure \ref{fig:forces_a041_orb},  show a sinusoidal fluctuation between large negative and positive values.  Because the overdense wake located at positive $x$ values remains the primary driver of DF forces, this periodic change in the sign of $F_{\rm DF, orb}$ is dominated by the changing direction of the barycentric velocities over each single orbit. The net force on the orbit due to DF (black curve in the upper panel) is small and negative, always exerting a small drag on the orbit. 

The momentum accretion forces, ${\bf F}_{\dot{p}_1} \cdot \hat{V}_1$ and ${\bf F}_{\dot{p}_2} \cdot \hat{V}_2$, shown in the center panel of Figure \ref{fig:forces_a041_orb}, exhibit more complex behavior throughout the orbit.  Consider the blue curve for $m_1$.  The momentum accretion often causes a drag on the orbital motion of $m_1$. The most negative value (largest drag force) occurs while $m_1$ is moving in the same direction as the wind (for example, the bottom three panels of Figure \ref{fig:flow-timeseries-041}).  At some instances, the blue curve turns positive. This occurs when $m_1$ is traveling head on into the wind. At this time, $m_1$ is moving supersonically relative to the material behind the primary shock.  The sharp (positive) peaks in the blue curve occur when a secondary shock forms ahead of $m_1$ (as in the first snapshot in Figure \ref{fig:flow-timeseries-041}).  In these cases, $m_1$ shocks and focuses material locally, accreting some of this material from downstream relative to its instantaneous motion. Because this material is moving in the same direction as $m_1$ as it is accreted, it imparts a thrust. 

Although the momentum transport forces on $m_1$ and $m_2$ are sometimes positive, these forces always sum to a negative value (black curve in the middle panel of Figure \ref{fig:forces_a041_orb}). This reflects the fact that, in the case of $a_0 = 0.41 R_a$, the highest accretion rates happen during the subsonic portions of the orbit of each particle.  Material accreted from upstream exerts a drag force on each particle's orbital motion, and the overall result of momentum accretion is a net drag on the orbit. 

The net force on the orbit due to both momentum accretion and DF (bottom panel of Figure \ref{fig:forces_a041_orb}) is always negative (a drag).  Momentum accretion contributes nearly half of the drag force on the orbit.  We note that momentum accretion is a local phenomenon that depends on the details of the flow in the region of the orbiting binary.  An understanding of the long-term evolution of the orbit of an astrophysical binary, therefore, requires an understanding of the dynamics and microphysics at small scales. This conclusion can be contrasted to the CM forces shown in Figure \ref{fig:forces_a041_cm}, in which the forces are dominated by the DF force, which results from the large-scale structure of the gas.

\section{Analysis: Timescales for Binary Transformation} 
\label{sec:timescales}
A binary interacting with an ambient gas will transform as a result of drag and accretion. The mass of the binary grows by accretion over a timescale $M/\dot M$. Forces on the CM reduce the CM kinetic energy, $E_{\rm trans} =  M \vcm^2/2$, over a timescale $E_{\rm trans}/\dot E_{\rm trans}$. Forces on the binary orbit tighten the separation over a timescale $a/\dot a$. In this section we explore these timescales and their hierarchy. We first derive analytic results from BHL theory as context and then we discuss the numerical results from our simulation models.

\subsection{Analytical Timescales}
\label{sec:analytic}

Before turning to our numerical simulations, let us see what the BHL formalism tells us about these timescales. The BHL stopping timescales are $\tstopHL$ and $\tstopBH$ (equations \eqref{eq:tstopHL} and \eqref{eq:tstopBH}, respectively) with $M = m_1 + m_2$.   Similarly, the accretion timescales are $\taccrHL$ and $\taccrBH$, given by equations \eqref{eq:taccrHL} and \eqref{eq:taccrBH}, respectively, with $M = m_1 + m_2$.

We derive expressions for the BHL inspiral timescales in Appendix \ref{sec:timescale_derive}. For large Mach numbers, the HL characteristic scalings give an orbital inspiral time that can be expressed in terms of the stopping time: 
\beq
\tinspHL = \frac{\tstopHL}{8\sqrt{2}}\bigg(\frac{a}{R_a}\bigg)^{-3/2}.
\label{eq:tinspHL}
\eeq
In the regime of mild Mach numbers, the BH formalism gives an orbital inspiral time of
\beq
\tinspBH = 
\frac{\tstopHL}{8\sqrt{2}}\bigg(\frac{a}{R_a}\bigg)^{-3/2}\bigg[1 + \frac{8a}{\mach^2 R_a}\bigg]^{1/2}.
\label{eq:tinspBH}
\eeq
Because $\tstopHL \propto \rhoinf^{-1}$, the inspiral timescales are inversely proportional to the ambient density.  If the density of the medium is increased, then the binary evolves more quickly.  

A less intuitive feature of the two inspiral timescales is the leading-order dependence on $a^{-3/2}$.  This dependence arises from the fact that the accretion radius of each particle is inversely proportional to the square of its speed relative to the gas.  As the binary separation decreases, the orbital speed of each object increases, so the BHL force decreases with decreasing $a$. The lengthening of $\tinspBH$ with decreasing $a$, for fixed $\rhoinf$, implies that the binary will never merge. 

A natural question to ask is whether the stopping time is longer than the inspiral timescale for a given binary.  In the BH regime, the ratio of the timescales is
\beq
\frac{\tinspBH}{\tstopBH} = \frac{1}{8\sqrt{2}}\bigg(\frac{\mach^2}{1 + \mach^2}\bigg)^{1/2}\bigg(\frac{a}{R_a}\bigg)^{-3/2}\bigg[1 + \frac{8a}{\mach^2 R_a}\bigg]^{1/2}.
\eeq
For $\mach = 2$, this implies that $\tstopBH > \tinspBH$ where $a \gtrsim 0.21 R_a$ (this transition would occur at $a= 0.198 R_a$ if the HL timescales are used instead). 

\subsection{Numerical Timescales and Their Hierarchy} 
\label{sec:numtimescales}

\begin{figure}
    \begin{center}
    \includegraphics[width=0.45\textwidth]{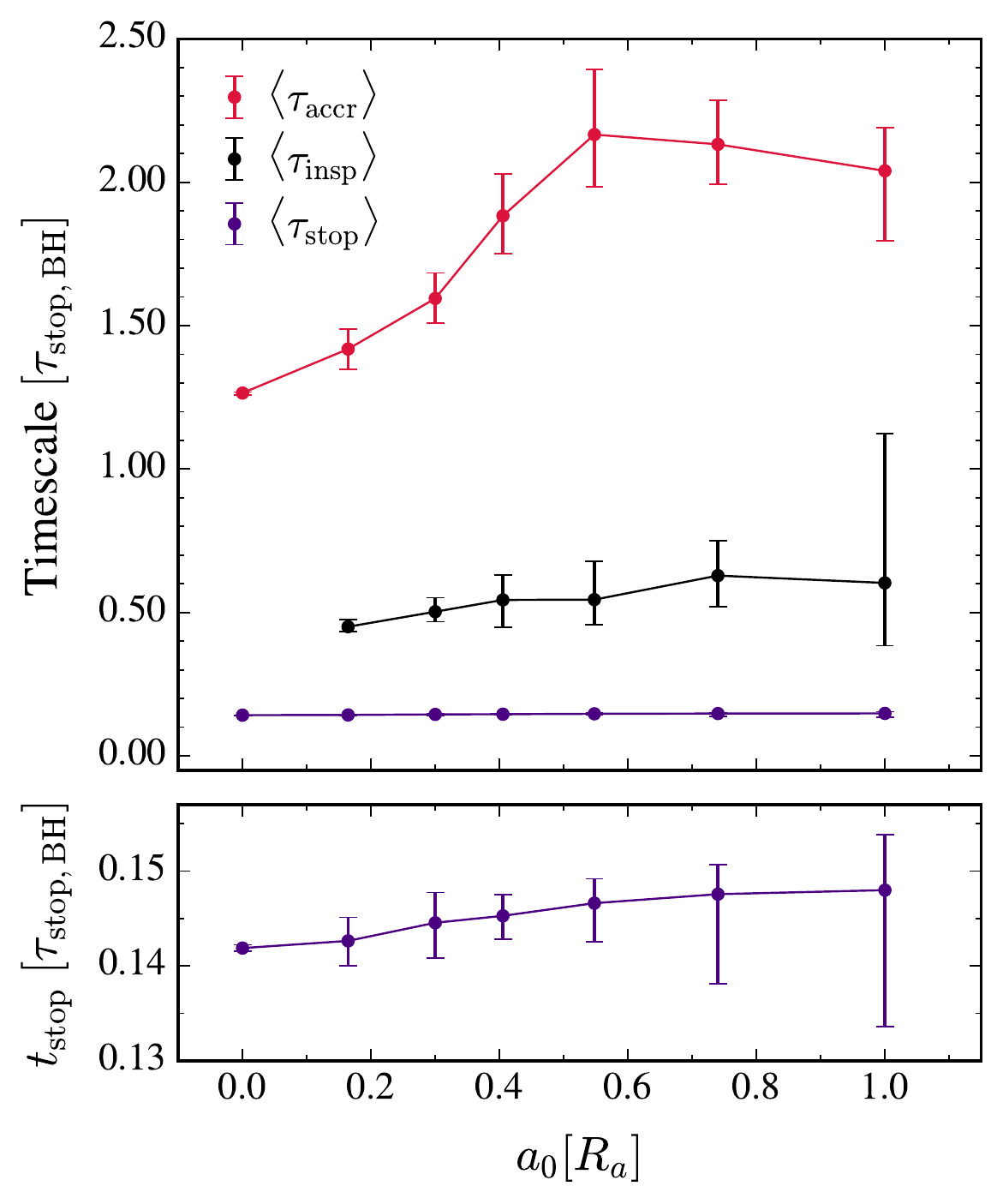}
        \end{center}
    \caption{Orbit-averaged timescales versus initial semi-major axis, $a_0$, calculated according to Equations \eqref{eq:tdouble_num}-\eqref{eq:tinsp_num}. The error bars show the  15.87th to 87.13th percentile range of instantaneous timescales, where the majority of this modulation occurs periodically, with varying binary orbital phase (see Figures \ref{fig:mdot-time-series}, \ref{fig:forces_a041_cm}, and \ref{fig:forces_a041_orb}).  All values are normalized to the BH stopping time, given by Equation \eqref{eq:tstopBH}. Unlike the prediction of BHL, we find that for all of the simulated binaries, the CM motion slows faster than the orbit decays. The difference between $\ave\taccr$ and $\ave\tstop$ for the $a_0 = 0.0 R_a$ simulation (single particle) indicates the difference in efficiency between accretion and drag on the CM.  The shape of $\ave\taccr$ with increasing $a_0$, though, shows the truncation of the accretion rate as the binary separation grows.}
    \label{fig:timescales}
\end{figure}

Having established the analytic predictions for timescales of accretion, stopping, and inspiral, we now proceed to measure these same quantities in our numerical simulations. 
For our simulated binaries, we measure orbit-averaged timescales of mass accretion
\beq
\ave\taccr \equiv \frac{M_0}{\ave{\dot M}},
\label{eq:tdouble_num}
\eeq
slowdown of the CM
\beq
\ave\tstop \equiv \frac{E_{\rm trans}}{\ave{{\dot E}_{\rm trans}}},
\label{eq:tstop_num}
\eeq
and orbital inspiral
\beq
\ave\tinsp \equiv \frac{a_0}{\ave{\dot a}},
\label{eq:tinsp_num}
\eeq
by taking the median over an integer number of orbits from  $t = 30 \tinf$ until the end of each simulation ($t = 50 \, R_a/\vinf$).  For $a_0 =0.0\ra$, the median and percentile ranges of $\ave\taccr$ and $\ave\tstop$ are computed over $30 \le t  \le 50 \tinf$.  

Numerically derived orbit-averaged timescales are shown in  Figure \ref{fig:timescales}.  The difference in $\ave\taccr$ and $\ave\tstop$ at $a_0 = 0.0\ra$ conveys the difference in efficiency of accretion and drag. The fact that $\ave\tstop \ll \ave\taccr$ implies that drag coefficients, compared to the BH predictions, are generally larger than unity, while accretion coefficients are of order unity or lower. This hierarchy of timescales has been consistently observed in previous simulations \citep[e.g.][]{2012ApJ...752...30B,2016A&A...589A..10T} and arises from large-scale gravitational interactions contributing to the DF drag \citep{1999ApJ...513..252O}, while accretion is a local property. 

As we introduce binaries of increasing separation, the timescales change, but their basic hierarchy stays the same.  The lengthening accretion timescale with larger binary separation arises from a drop in $\ave\mb$ with larger $a_0$, as shown in Figure \ref{fig:mdot-versus-a}.    We also see that $\ave\tstop$ changes by only 5\% as the binary separation increases. This reflects the fact that the slowdown of the CM is dominated by the DF force that is due to the downstream wake.  The wider binaries show larger time-variability because the primary shock changes shape as the particles move in and out of each other's wake (see Figure \ref{fig:sma_series_density}), but the overall structure downstream from the binary is similar in each simulation.  Critically, the overall size of the wake and the standoff distance of the primary shock are similar in all simulations, implying the Coulomb logarithm, $\ln(r_{\rm max}/r_{\rm min})$ in the DF force is essentially unchanged \citep[][Section 5.3]{1999ApJ...513..252O,2016A&A...589A..10T}. 

Figure \ref{fig:timescales} shows that the timescales fall into a clear hierarchy with $\ave\taccr > \ave\tinsp > \ave\tstop$. The most striking aspect of this hierarchy is that $\ave\tinsp > \ave\tstop$ for all of the binary systems we simulated. In Section \ref{sec:analytic}, we found that a simple application of BHL predicts that $\ave\tstop$ should be longer than $\ave\tinsp$ for all binaries with $a_0 \ge 0.21 R_a$.   Instead, the stopping time is always shorter, which means that the CM motion will come to a stop before the orbit decays.

\begin{figure}
    \begin{center}
    \includegraphics[width=0.47\textwidth]{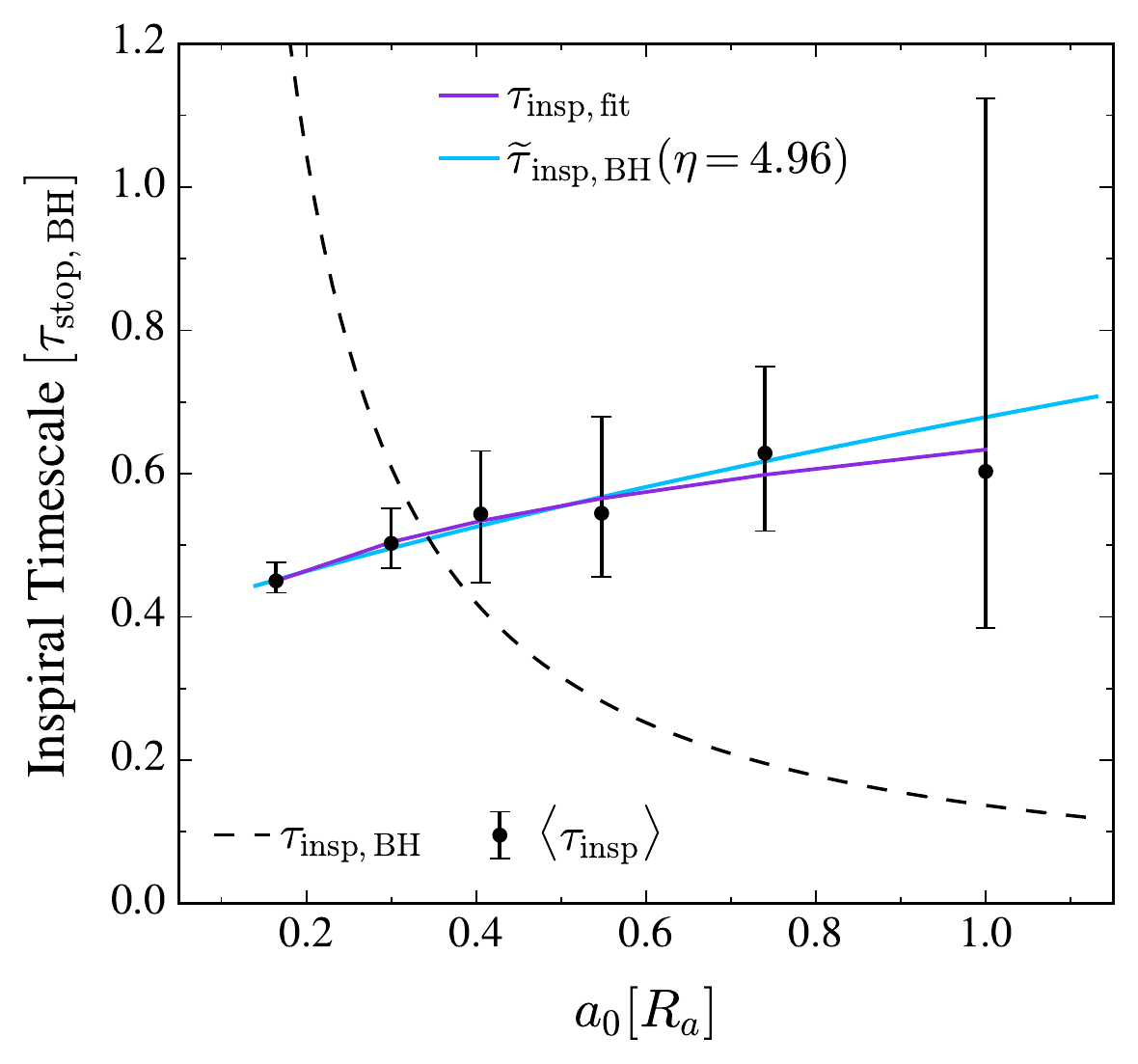}
    \caption{Numerical and analytical inspiral timescales.  As in Figure \ref{fig:timescales}, the black points with error bars show $\ave\tinsp$ for our simulated binaries.  A power-law fit to the data gives $\tinspfit = 0.64(a_0 / R_a)^{0.19}$.  This increasing inspiral time with $a_0$ is in stark contrast to the timescale predicted by BHL theory, $\tinspBH \propto a^{-3/2}$, which says the binary will never merge.  Instead, our simulations show that the inspiral timescale decreases as the binary spirals in.  Replacement of $\rhoinf$ with the Bondi density profile of Equation \eqref{eq:tilderho} in $\tinsp$ gives the modified analytical timescale, $\tildetau$, given by Equation \eqref{eq:tildeinsp}. With a normalization factor $\eta = 4.96$, $\tildetau$ captures our simulated data well, indicating that the positive scaling with $a_0$ is a result of the density enhancement in the region roughly enclosed by the binary orbit due to the gravitational influence of the CM.}
    \label{fig:decay_fit}
    \end{center}
\vspace{-5pt}
\end{figure}

Another prediction of the BHL formalism is the shape of $\ave\tinsp$ as a function of $a_0$. We plot $\tinspBH$ versus $a_0$ in Figure \ref{fig:decay_fit} (dashed line).  The plot reveals the $a^{-3/2}$ leading-order shape of $\tinspBH$.  We include in Figure \ref{fig:decay_fit} the calculated values of $\ave\tinsp$ of Figure \ref{fig:timescales}.  Instead of following $a^{-3/2}$, $\ave\tinsp$ decreases with decreasing $a_0$.  

To obtain the $a_0$-dependence of $\ave\tinsp$, we perform a least-squares fit to a power law of the form 
\beq
f(a_0) = A(a_0 / R_a)^B.
\eeq
We fit to the median values (the black points in Figure \ref{fig:decay_fit}) and use one standard deviation (approximately the average of the upper and lower error bars) as the uncertainty on each point when performing the fit.  We find $A = 0.64 \pm 0.01$ and $B = 0.19 \pm 0.01$, giving the functional form
\beq
\tinspfit = 0.64(a_0 / R_a)^{0.19}\tstopBH.
\label{eq:tinspfit}
\eeq
This fit is shown in Figure \ref{fig:decay_fit}. 

\subsection{Physical Interpretation of $\ave\tinsp$} 
\label{sec:physinterp}
The sharp difference in scaling between $\ave\tinsp \propto a_0^{0.19\pm 0.01}$, computed from the hydrodynamic simulations, and $\tinspBH \propto a^{-3/2}$, from BHL theory, is apparent in Figure \ref{fig:decay_fit}.  While BHL theory predicts that the binary will take an infinite time to merge, $\tinspfit$ grows shorter as the orbit decays.  
 
By assuming a background density of $\rhoinf$, $\tinspBH$ fails to capture the enhancement of the local density that is due to the gravitational pull of the binary CM.  If the CM of the binary were at rest relative to the gas, we would expect the density as a function of distance from the CM to be similar to a Bondi profile \citep[][albeit truncated inside the orbit of the binary since the potential is smoothed to a distance of approximately $a/2$]{1952MNRAS.112..195B}.  We use a Bondi density profile to express the density of the gas at a distance of $a/2$ from the CM of the binary: 
\beq
\tilde\rho \equiv \frac{\mach^3\rhoinf}{8} \bigg(\frac{R_a}{a}\bigg)^{3/2}
\label{eq:tilderho}
\eeq
\citep[for $\gamma = 5/3$ and $a/2 \le R_B$;][]{1983bhwd.book.....S}. Replacing $\rhoinf$ with $\tilde\rho$ in Equation \eqref{eq:tinspBH} cancels the leading $a^{-3/2}$ dependence of $\tinspBH$.   The modified timescale is then
\beq
\tildetau = \frac{\eta}{2\pi\sqrt{2}}\frac{\vinf}{G\mach^3\rhoinf \ra}\bigg[1 + \frac{8a}{\mach^2\ra}\bigg]^{1/2}
\label{eq:tildeinsp}
\eeq
where the coefficient $\eta$ is an overall normalization factor. We choose the normalization such that $\tildetau = \ave\tinsp$ at $a_0 = 0.16 \ra$, which yields  $\eta = 4.96$. 

We plot $\tildetau(\eta = 4.96)$  in Figure \ref{fig:decay_fit}. With this normalization, $\tildetau$ nearly replicates the power-law fit and the numerical data. The normalization factor of $\eta \approx 5$ indicates that Equation \eqref{eq:tilderho} overestimates the density by a factor of approximately 5.  This may be because the softening of the potential to the size of the orbit and the pressure support provided by the orbital motion of the binary both work to reduce the local density that can be realized versus the pure Bondi case. In addition, the incoming wind sweeps material from the region of the binary, constantly modifying the density structure of the local region.  

\subsection{Comparison to Previous Work}
\label{sec:previouswork}
In previous analytical and semianalytical work, there has been considerable disagreement of how the inspiral timescale and forces scale with binary separation. Here we briefly discuss these results in light of our semianalytical interpretation of the inspiral timescale of Section \ref{sec:physinterp}, noting the different physical scenarios and assumptions that led previous authors to varying results. 

\begin{figure}
    \includegraphics[width=0.47\textwidth]{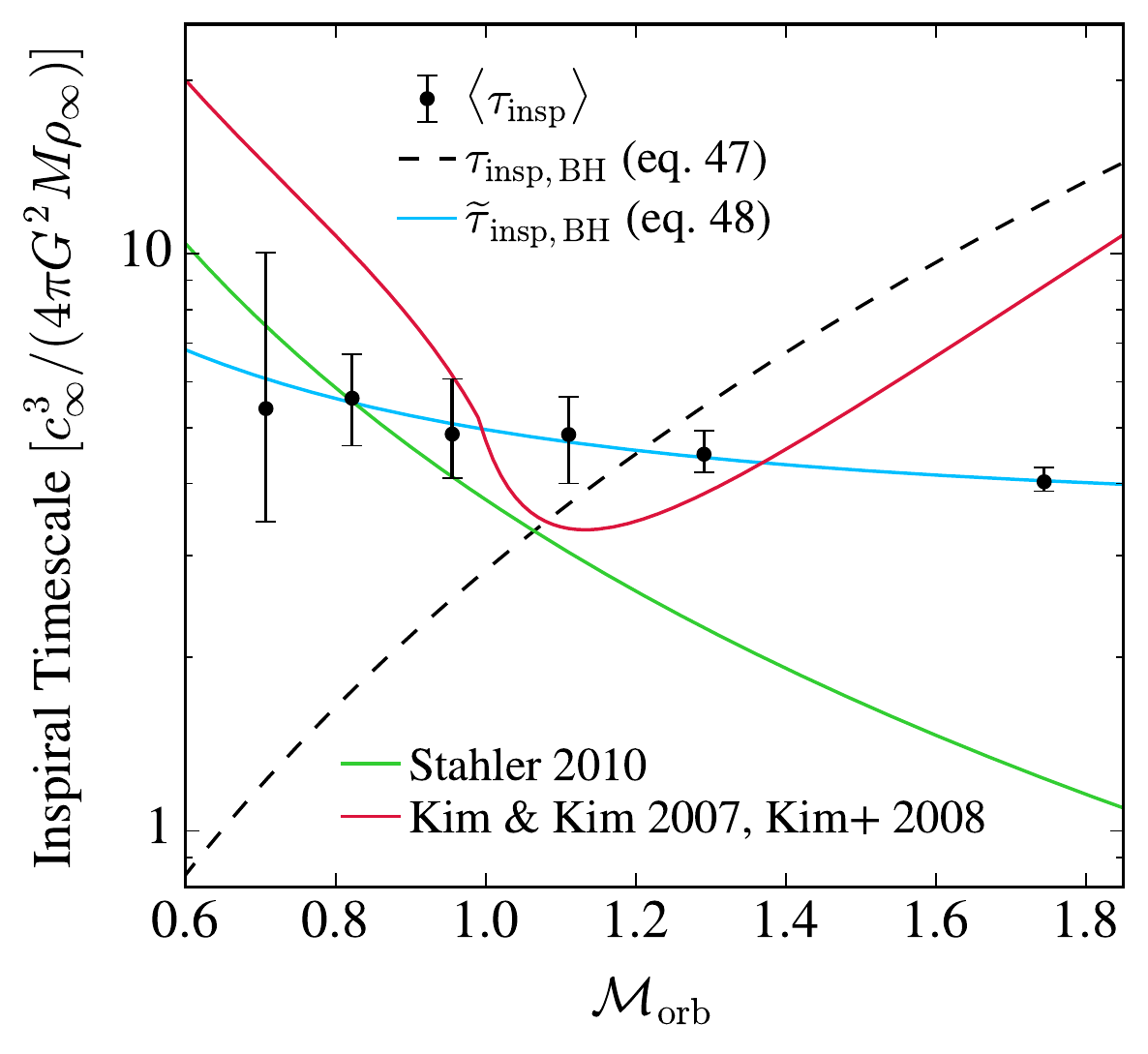}
    \caption{Comparison of our semianalytical inspiral timescale to previous studies of dynamical friction forces on circular-orbit perturbers in the linear regime. \citet{2008ApJ...679L..33K} finds a lengthening inspiral timescale with increasing orbital Mach number (decreasing separation), reminiscent of $\tinspBH$. In comparison, the calculation of \citet{2010MNRAS.402.1758S} gives a rapidly shortening inspiral timescale with lengthening orbital Mach number (decaying binary separation).   The result of our nonlinear calculations,  $\ave\tinsp$,  disagrees with these results from linear theory, which assume an unperturbed background density.  In contrast, our semianalytical timescale ($\tildetau$ with $\eta = 4.96$), which includes an enhanced density profile resulting from the presence of the binary potential, captures  $\ave\tinsp$ well.}
    \label{fig:previous_work}
\end{figure}

To simplify the comparison to previous studies, we first write our inspiral timescales in terms of the barycentric orbital Mach number 
\begin{equation*}
    \machorb^2 \equiv \frac{V_1^2}{\cs^2} = \frac{\vorb^2}{4\cs^2} = \frac{\mach^2}{8}\frac{\ra}{a}
\end{equation*}
which says
\begin{equation*}
    \frac{a}{\ra} = \frac{1}{8} \bigg(\frac{\mach}{\machorb}\bigg)^2.
\end{equation*}
Substituting the last expression into equations \eqref{eq:tinspBH} and \eqref{eq:tildeinsp} yields the forms
\beq
    \tinspBH = 2 \machorb^3\bigg[1 + \machorb^{-2}\bigg]^{1/2}\frac{\cs^3}{4\pi G^2 M\rhoinf}
    \label{eq:tinspBH_machorb}
\eeq
and
\beq
    \tildetau  = \frac{\eta}{\sqrt{2}}\bigg[1 + \machorb^{-2}\bigg]^{1/2}\frac{\cs^3}{4\pi G^2 M\rhoinf},
    \label{eq:tildeinsp_machorb}
\eeq
respectively. We plot equations \eqref{eq:tinspBH_machorb} and \eqref{eq:tildeinsp_machorb} along with our simulation data, $\ave\tinsp$, in Figure \ref{fig:previous_work}.   In these units, the fit to our simulation data, Equation \eqref{eq:tinspfit}, becomes
\begin{equation*}
    \tinspfit = 5.02 \machorb^{-0.38}\frac{\cs^3}{4\pi G^2 M \rhoinf}
\end{equation*}
implying $$\ave\tinsp \propto a^{0.19} \propto \machorb^{-0.38}.$$

The gaseous DF force acting on a point mass in straight-line motion relative to a background gas was first calculated analytically by \citet{1999ApJ...513..252O}.  As in the case of a collisionless gas, the integral nature of the DF force gives rise to a Coulomb logarithm $\ln(r_{\rm max}/ r_{\rm min})$. The radii $r_{\rm max}$ and  $r_{\rm min}$ represent the range of size scales spanned by the asymmetric wake trailing the point mass. \citet{2007ApJ...665..432K} applied the methods of \citet{1999ApJ...513..252O} to a perturber moving in a circular orbit relative to an initially uniform gas, finding that the orbital motion truncates the maximum coherence length of the wake to $r_{\rm max}\approx a$. 

\citet{2008ApJ...679L..33K} extended these calculations to a symmetric binary pair of perturbers in a circular orbit with CM at rest ($\mach = 0$).  \citet{2007ApJ...665..432K} and \citet{2008ApJ...679L..33K} report their results with piecewise functions of $\machorb$\footnote{In their notation, the barycentric Mach number is $\mathcal{M} \equiv V_p /c_s$, where $V_p$ is the orbital speed of their perturber and $c_s$ is the unperturbed sound speed of the background gas.  Their quantity $\mathcal{M}$ is identical to our $\machorb$. The difference in notation is to avoid confusion with $\mach$, which is the CM Mach number of the binary.   We note for convenience that they label the mass and orbit radius of the perturber as $M_p$ and $R_p$, respectively, which are $m_1 = M / 2$ and $a/2$ in our notation, respectively. The linear momentum of the perturber is $M_p V_p$, which is $M a / 4$ in our notation.}. 
To compare their results to ours, we construct an inspiral timescale from their azimuthal drag forces as follows. The azimuthal DF force on one perturber due to its own wake is $\mathcal{F}\mathcal{I}_{\phi,1}$.  The azimuthal DF force on that same perturber due to the companion's wake is $\mathcal{F} \mathcal{I}_{\phi,2}$.  Here, $\mathcal{F}$ and $\mathcal{I}_{\phi,1}$ are given by equations 11 and 14 of \citet{2007ApJ...665..432K}, respectively, and  $\mathcal{I}_{\phi,2}$ by equation 5 of \citet{2008ApJ...679L..33K}.  When computing $\mathcal{I}_{\phi,1}$ ,we take $r_{\rm min} = 0.05 \ra$, which is the size of the sink in our simulations and, as such, is the minimum distance to which we integrate the DF force.  The inspiral timescale is the linear momentum of the perturber divided by the total azimuthal force on the perturber, which simplifies to
\begin{equation*}
    \tau_{\rm insp, K08} = \frac{\machorb^3}{\mathcal{I}_{\phi,1} + \mathcal{I}_{\phi,2}}\frac{\cs^3}{4\pi G^2 M\rhoinf}.
\end{equation*}
We plot $\tau_{\rm insp, K08}$ as a function of $\machorb$ in Figure \ref{fig:previous_work}.

Figure \ref{fig:previous_work} shows that  for $\machorb \lesssim 1.1$, $\tau_{\rm insp, K08}$ increases much more rapidly than $\ave\tinsp$. For $\machorb \gtrsim 1.1$, $\tau_{\rm insp, K08}$ traces $\tinspBH \propto \machorb^3$, which similarly assumes a uniform background density.  


\citet{2010MNRAS.402.1758S} also considered angular momentum loss by a binary orbiting in a uniform gas with CM at rest in the linear regime, but came to a very different conclusion. They computed the angular momentum and energy carried by waves induced by the time-varying quadrupole moment of the binary potential. Their equation 39 gives the total torque, $\Gamma$, on a binary with  moment of inertia $I$ and orbital frequency $\omega$. For comparison,\footnote{\citet{2010MNRAS.402.1758S} denotes the total mass and total semi-major axis of the binary as  $M_{\rm tot}$ and $a_{\rm tot}$, respectively, which are $M$ and $a$ in our notation, respectively. In their equation 39, $\rho_0$ and $c_s$ are the unperturbed density and sound speed, respectively.  In constructing $J / \Gamma$, we use the fact that, in our notation, $I \omega^2 = Ma^2 / 4 = M\machorb^2 \cs^2$ for an equal-mass binary.}, we construct an inspiral timescale by dividing the total angular momentum of the binary, $J = I \omega$, by their expression for $\Gamma$. The resulting timescale 
\begin{equation*}
    \tau_{\rm insp, S10} = \frac{15}{4}\machorb^{-2}\frac{\cs^3}{4\pi G^2 M\rhoinf}
\end{equation*}
is shown in green in Figure \ref{fig:previous_work}. It is apparent that the scaling with $\machorb$ for $\machorb \gtrsim 1.1$ is opposite in sign from both  $\tau_{\rm insp, K08}$ and $\tinspBH$. \citet{2014ApJ...794..167S} report on this discrepancy and demonstrate that it can be traced to the multipole expansion that underlies the solution from \citet{2010MNRAS.402.1758S}.

 Similar to \citet{2007ApJ...665..432K}, \citet{2014ApJ...794..167S} calculated the torque on a single circular-orbit perturber in the linear regime, but they allowed the orbit center to move at ${\bf V}_{\rm CM}$ relative to the initially uniform background gas. They considered the cases of face-on orbits, in which ${\bf V}_{\rm CM}$ is parallel to the orbital angular momentum vector, as well as edge-on cases (as in the present work), in which the orbital angular momentum vector is perpendicular to ${\bf V}_{\rm CM}$. Their Figures 9 (face-on cases) and 12 (edge-on cases) present computed values of the azimuthal DF force on a single perturber due to its own wake, denoted $F_{\rm DF, \phi}$, for different combinations of $\mach$ and $\machorb$.  Although their method is slightly different, the DF torque on their face-on perturber with subsonic CM motion ($\mach = 0.5$) tends to the $\mach = 0$ torque calculated by \citet{2007ApJ...665..432K}, with $\tau_{\rm insp} \sim \tinspBH$ for $\machorb \gtrsim 1.1$.  
 
 By allowing for CM motion, the calculations of \citet{2014ApJ...794..167S} are most applicable to our particular study of $\tinsp$.  To apply their values of $F_{\rm DF, \phi}$ to a perturber in an equal-mass binary system, the contribution of the companion's wake to the azimuthal force must be included.  The total azimuthal force on the perturber (due to both wakes) is $F^{(1)}_{\rm DF, \phi} = \mathcal{R}_\phi F_{\rm DF, \phi}$ where $0.1 \lesssim \mathcal{R}_\phi \lesssim 1$ (see their \S 5.1). Only a few values of $\mathcal{R}_\phi$ are given, but \cite{2008ApJ...679L..33K} showed that $\mathcal{R}_\phi$ is a function of $\machorb$. Without an understanding of how $\mathcal{R}_\phi$ scales with $\machorb$, we are unable to include results from \citet{2014ApJ...794..167S} in Figure \ref{fig:previous_work}, although such a comparison is highly desirable.


Our numerical result exhibits a key difference relative to each of these previous works, which all assume a constant-density background. We can improve the analytic representation to bring it into agreement with the numerical findings by considering the enhanced density that arises from flow convergence onto the binary pair. Modifying $\tinspBH$ by adopting a local density, $\tilde \rho \propto r^{-n}$ (where $n=1.5$ in the BHL case; Section \ref{sec:physinterp}), changes the scaling of the analytic predictions and can reconcile the results of the analytic models with our simulations, as shown in Figure~\ref{fig:decay_fit}.

\section{Discussion}
\label{sec:discussion}

In this section, we discuss some implications of our simulation results for astrophysical binaries. 

\subsection{Implications of Timescale Hierarchy and Scaling}

In our models, we focused on the ratio of the initial separation of the binary, $a_0$, relative to the accretion radius of its center of mass, $R_a$, as an essential parameter in describing the flow. This ratio informs the rate of mass accretion, and it determines the nature of the drag forces acting on both the orbital and translational motion of the pair.  The timescales for accretion, $\taccr$, orbital inspiral, $\tinsp$, and slowdown of the center of mass, $\tstop$ inherit this dependence on $a_0 / R_a$.  The timescales follow a hierarchy, with $ \taccr > \tinsp > \tstop$ for all $a_0/\ra$ that we simulated.  In contrast to the expectation from BHL,  $\tinsp$ is a monotonically increasing function of $a_0$. 

This hierarchy of timescales implies that astrophysical binaries moving through gaseous surroundings will slow relative to the surrounding gas before they either inspiral or accrete significant quantities of gas. Binaries will be slowed when their gravitational cross section has intercepted a gas mass of about 15\% of the binary mass (because $\tstop \approx 0.15 \tstopBH$). To be driven toward merger, they would need to traverse a larger column of approximately 50\% their mass, and to double in mass, they need to intercept nearly twice their own mass.\footnote{We note, however, that the numerically derived accretion rate (and therefore timescale) is sensitive to the size of the absorbing sink boundary condition, such that smaller objects would be expected to accrete less material (Appendix \ref{sec:validation}). Further, this mass-doubling timescale assumes that the object is able to accrete efficiently from any gas that falls through the sink boundary without, for example, forming a disk or feeding back substantially into its surroundings. } This suggests that a binary in motion through a uniform gaseous environment would be slowed such that it would enter into a phase of binary Bondi accretion in which the binary CM was approximately at rest relative to the gas.

The scaling of $\tinsp$ with binary separation is also of great importance. Because this timescale has a positive slope, binaries with shrinking $a_0/\ra$ continue to be driven to merger on finite timescales. The ratio $a_0/\ra$ can shrink either because the binary separation is decreasing or as $\ra$ increases when the binary CM motion is slowed. Therefore, whereas the naive scaling $\tinspBH$ indicates that pairs of objects can rarely be driven fully to merger by gas -- instead being halted at finite separation by the increasing timescale as their orbits shrink -- our updated results indicate that a tightening orbit yields a trend toward ever-shorter $\tinsp$. 

\subsection{Critical Separation between Gas-dominated and GW-dominated Inspiral}
\label{sec:criticalsep} 

The fact that gaseous forces drive binaries to tighter separations leads to the conclusion that GWs may become important at some separation, especially if the binary is composed of compact objects. For a circular-orbit, equal-mass binary, \citet{1964PhRv..136.1224P} showed that the inspiral timescale is
\beqar
\tau_{\rm  insp, GW} &\equiv& \frac{a}{\dot{a}_{\rm GW}} =  \frac{5 a_0^4 c^5}{64 G^3 M^3}, \\
&=& 1.3 \times 10^{18} {\rm yr} \bigg[\frac{a_0}{\rm AU}\bigg]^4\bigg[\frac{M_\odot}{M}\bigg]^3.
\eeqar
 The inspiral timescale that we found numerically, $\tinspfit$ given in Equation \eqref{eq:tinspfit}, says that $\tgas \propto a_0^{0.19}$.  This mild, positive power law allows the gas to be the dominant driver of orbital inspiral at large separations.  At smaller separations, $\tgw < \tgas$ and GW radiation dominates the orbital decay until the binary finally merges.  Below we find the boundary between gas-dominated and GW-dominated inspiral, by computing the critical separation, $a_{0, \rm crit}$, at which $\tgas = \tgw$.

Our expression for $\tinspfit$ in Equation \eqref{eq:tinspfit} is given in units of $\tstopBH$.  To simplify the algebra, let us write our fit in terms of $\tstopHL$, instead. Let us define
\beq
\tgas = A\bigg(\frac{a_0}{R_a}\bigg)^B\frac{v_\infty^3}{4\pi G^2M\rho_\infty}, \\
\label{eq:tgas}
\eeq
where, in these units, $A = 0.71$ instead of $0.64$. The power $B = 0.19$ is unchanged. In astrophysical units,
\beq
\begin{split}
\tgas = 2.9\times 10^{10} {\rm yr}\bigg[\frac{a_0}{\rm AU}\bigg]^{0.19} \bigg[\frac{v_\infty}{\rm km/s}\bigg]^{3.38} \\
\times \bigg[\frac{M_\odot}{M}\bigg]^{1.19}\bigg[\frac{{\rm cm}^{-3}}{n_\infty}\bigg].
\end{split}
\label{eq:tinspgas}
\eeq
\begin{figure}
    \begin{center}
    \includegraphics[width=0.47\textwidth]{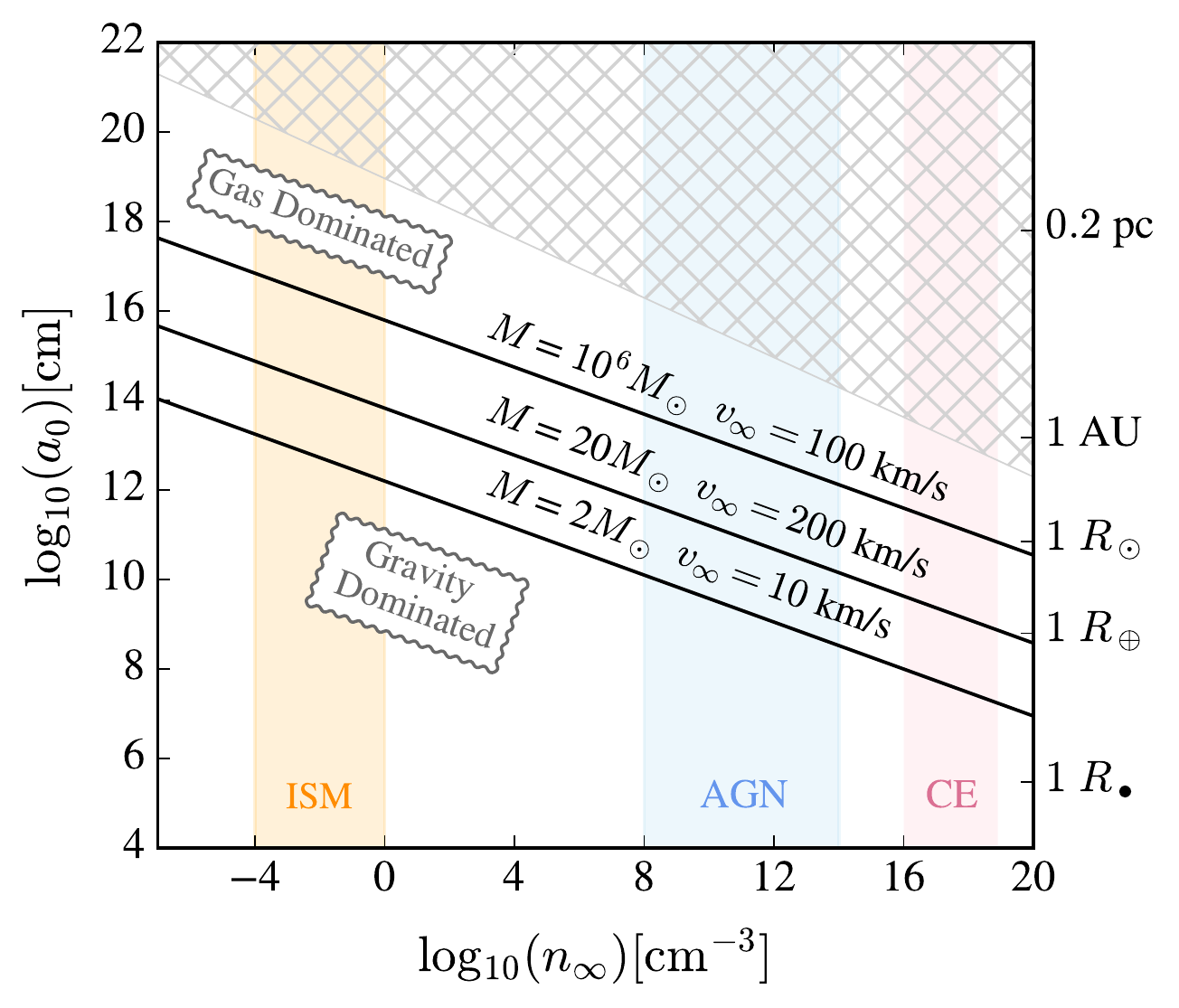}
    \caption{Critical separation between gas-dominated and GW-dominated inspiral as a function of background number density, $\ninf$.  The black curves show $a_{0,\rm crit}$ (Equation \eqref{eq:critical_a}) for three combinations of $M$ and $\vinf$.  Above each line, $\tgas < \tgw$.  Below each line, $\tgas > \tgw$. Vertical shaded regions correspond to the densities found in the warm/hot ISM, a model AGN disk, and a red giant envelope (CE).  In the gray hatched region, a $2\msun$ binary would enclose more than its own mass and, thus, would not remain bound. For higher-mass binaries, this ``forbidden'' region moves up and to the right. See the text for an application of this plot to a few astrophysical systems.}
    \label{fig:critical_a}
    \end{center}
   \vspace{-8pt}
\end{figure}
Equating $\tgas$ and $\tgw$ gives the critical separation 
\beq
a_{0, \rm crit} =2\bigg[\frac{A}{5 \pi}\frac{G^{1-B}}{c^5}M^{2-B} v_\infty^{2B+3}\rho_\infty^{-1} \bigg]^{1/(4-B)}
			\label{eq:critical_a}
\eeq
with $A = 0.71$ and $B = 0.19$. In astrophysical units, 
\beq
a_{0, \rm crit} = 0.03 {\rm AU} \bigg[\frac{M}{\msun}\bigg]^{\frac{2-B}{4-B}}\bigg[\frac{\vinf}{\rm km /s}\bigg]^{\frac{2B+3}{4-B}}\bigg[\frac{{\rm cm}^{-3}}{\ninf}\bigg]^{\frac{1}{4-B}}.
\eeq

In Figure \ref{fig:critical_a}, we plot $a_{0,\rm crit}$ versus number density, $\ninf$, for three sample binary systems (black lines), where we have converted mass density to number density through $\rhoinf = m_p\ninf $.  The vertical shaded regions give the typical densities encountered in several astrophysical environments.  The ISM densities use values from \citet{2017ApJ...846..133K}, the AGN conditions use the disk models of \citet{2017ApJ...835..165B}, and the CE density range is modeled after the red giant proposed by \citet{2015MNRAS.450.1716S} to have initiated a CE phase in the hierarchical triple  system \psr.  The gray hatched region shows where the enclosed gas mass becomes similar to the binary mass ($M_{\rm encl} > 2 \msun$). For higher-mass binaries, the lower boundary of the hatched region moves up and to the right. 

Each black line in the figure plots Equation \eqref{eq:critical_a} for a different combination of $M$ and $\vinf$.  Above each line, $\tgas$ is faster than $\tgw$ and we say that the binary is in the gas-dominated regime. Below each line, $\tgw < \tgas$, and the binary is in the GW-dominated regime. 
\begin{figure*}
    \begin{center}
    \includegraphics[width=0.95\textwidth]{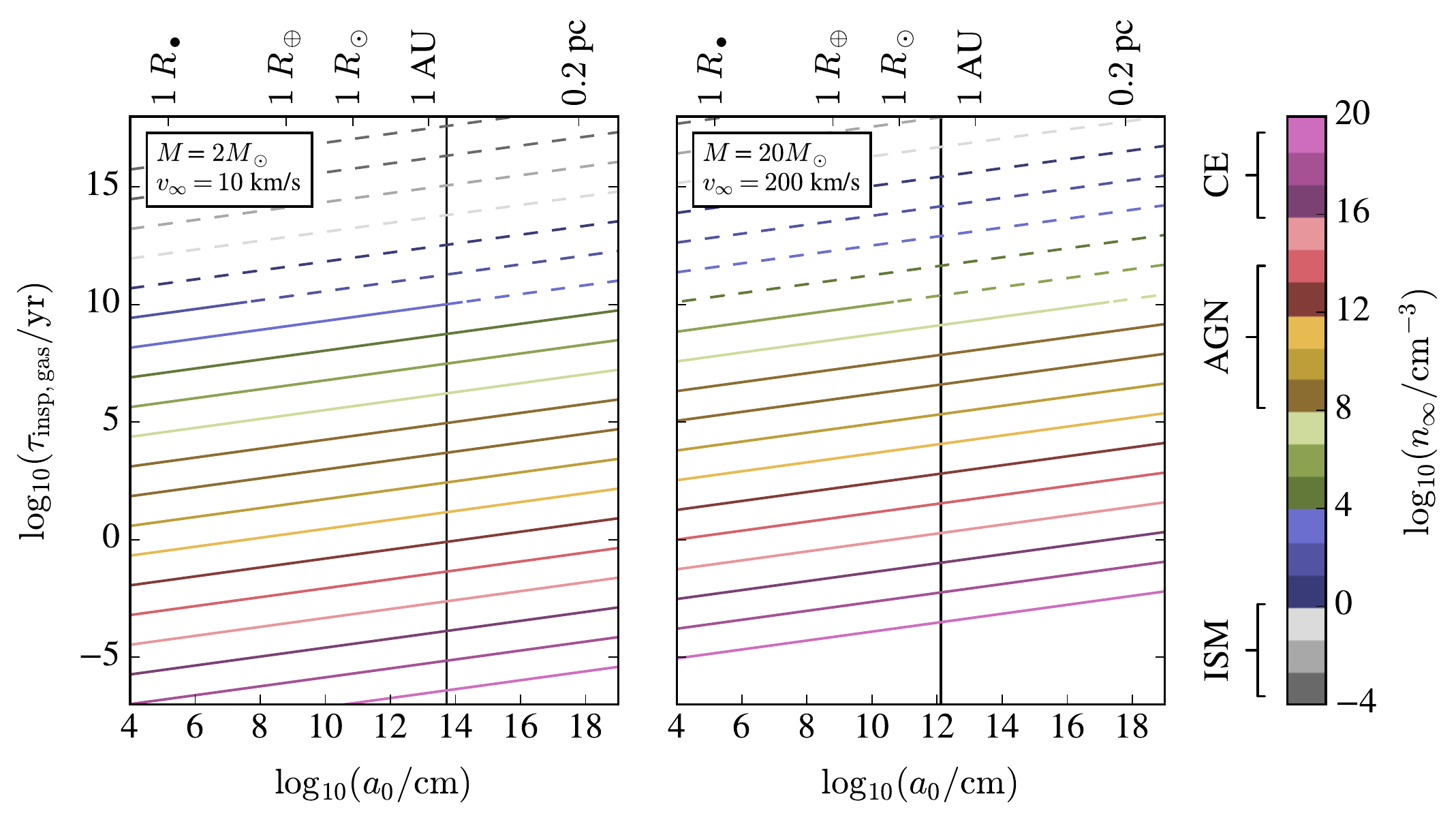}
    \caption{{Inspiral timescale for astrophysical binaries}. Each panel shows the gas-driven inspiral timescale in years versus the initial semi-major axis in cm, according to Equation \eqref{eq:tinspgas}, for fixed values of the binary's total mass, $M$, and CM speed, $\vinf$,  motivated by the examples discussed in Sections \ref{sec:binairesISM} and \ref{sec:binariesAGN}.  The individual curves in each panel represent different choices of $\ninf$ and the associated color bar is labeled to highlight the astrophysical environments discussed in the text.  The curves become dashed lines where the inspiral time is longer than a Hubble time.  In each panel, the vertical black line gives the minimum value of $a_0$ that we simulated.  Therefore, the range of validity of Equation \eqref{eq:tinspgas} is to the right of the black line, and everything to the left is an extrapolation of our results to smaller $a_0$.}
    \label{fig:tinspgas}
    \end{center}
\end{figure*} 
Below, we consider some example systems in the context of Figure \ref{fig:critical_a}.   For convenience, we also plot Equation \eqref{eq:tinspgas} for two different choices of $M$ and $\vinf$ in Figure \ref{fig:tinspgas}. 

\subsection{Binaries in the ISM}
\label{sec:binairesISM}
The multiphase ISM spans a large range of density and temperature conditions. Here, for the sake of a concrete example, we focus on the warm ISM, which occupies about one-half of the interstellar volume, has a number density of approximately 1 cm$^{-3}$, and has a temperature of the order of $10^4$K. A binary traveling at $10$~km~s$^{-1}$ relative to the gas has a Mach number of  1.1.  At this speed, a $2M_\odot$ binary with a separation of 1 au has $\tau_{\rm  insp, gas} < \tau_{\rm  insp, GW}$.  For this system, our numerical fits of $\ave\tinsp$ and $\ave\tstop$ (equations \eqref{eq:tinspfit} and \eqref{eq:tstopfit}, respectively) give a stopping timescale of $2\times 10^{14}$ yr and an orbital inspiral timescale of $5 \times 10^{13}$ yr (see Figure \ref{fig:tinspgas}). These timescales indicate that binaries traversing the warm ISM will suffer little dynamical effect from this interaction over the age of the universe. 

Nonetheless, this effect might potentially be measurable in the case of pulsar binaries with extremely precise timing.   Using $2\dot{P}_{\rm orb} / {P}_{\rm orb} = 3 \dot{a}/{a} = 3 / \tinsp$, the inspiral time gives a period derivative of $\dot{P}_{\rm orb} \approx 10^{-13}$.\footnote{In this estimate, we have neglected the pulsar's influence on its surroundings, which may not be justified because many pulsars are surrounded by hot, low-density wind nebulae. In cases where the standoff distance of the cavity is similar to the binary separation or bow shock size, drag forces on the binary orbit would likely be substantially reduced.}  By comparison, for the binary radio pulsar PSR J0751+1807, \citet{2008AIPC..983..453N} found an orbital period derivative of $(-3.1\pm0.5) \times 10^{-14}$, a precision that is two orders of magnitude better than our sample case presented here.

\subsection{Binaries Interacting with AGN Disks} \label{sec:agn}
\label{sec:binariesAGN}
The central regions of galaxies hosting AGNs experience an inflow of gas that settles into a cold accretion disk surrounding the central SMBH. Stars and binaries coexist with the SMBH and accretion disk in these central regions, forming a dense cluster around the black hole. Dynamical interactions between the stars and disk can trap stars within the disk plane \citep[e.g.][]{1991MNRAS.250..505S,1993ApJ...409..592A,1998ApJ...507..131I,2005A&A...433..405S,2005ApJ...619...30M,2007A&A...470...11K,2007ApJ...656..709P,2011ApJ...726...28B,2011MNRAS.417L.103M,2012MNRAS.425..460M,2012ApJ...758...51J,2014ApJ...780...84D,2016MNRAS.460..240K, 2018MNRAS.476.4224P}.

Our result that $\tstop < \tinsp$ for all of the binaries we simulated demonstrates that trapping by the AGN disk is not reserved for single stars. Instead, the CM motion of a binary can be slowed by the gaseous disk without causing a merger of the binary (we expect this conclusion to hold as long as $a_0<H$, the scale height of the disk).  That is, binaries formed outside of the disk can remain intact after undergoing capture into the disk plane.  When this process involves BBHs, the gaseous drag can aid the trend toward merger until the separation is reduced sufficiently that GWs take over, as discussed in Section \ref{sec:criticalsep}. These mergers are of additional interest because BBH mergers in dense, gaseous environments may be accompanied by detectable electromagnetic emission \citep[e.g.][]{2015MNRAS.446L..36F,2017MNRAS.464..946S,2017PhRvD..96l3003K}.

To give a concrete example, we use  the disk model of \citet[their Equations (1)-(4)]{2017ApJ...835..165B} with a central black hole mass of $M_\bullet = 10^6 \msun$. The central black hole dominates the gravitational force within a sphere of influence of $\approx 1$~pc, so we assume that binaries orbiting the black hole in this region do so in Keplerian orbits with a CM speed of $\vinf = \sqrt{GM_\bullet / R_{\rm disk}}$.   At $R_{\rm disk} = 0.1$ pc, $\ninf = 3 \times 10^{11}$cm$^{-3}$, and the Keplerian speed is $\vinf = 207$~km~s$^{-1}$. This speed gives a CM Mach number of 4.6 (neglecting the effect of rotation of the disk).  A $20 \msun$ BBH with $a_0 = 2$ AU falls in the gas-dominated regime (see Figure \ref{fig:critical_a}).   If inclined such that it is continuously within the disk plane, we find $\tinsp = 2 \times 10^5$~yr (see Figure \ref{fig:tinspgas}), $\tstop = 4 \times 10^4$~yr,  and $\taccr = 5 \times 10^5$ yr. This implies that the binary is entrained within the disk flow (and any eccentricity is damped) within approximately 14 orbits, and the BBH inspirals within approximately 70 orbits. Over this time, the BBH might accrete up to 40\% its own mass, growing to approximately 30$M_\odot$. The brevity of these timescales relative to a typical AGN activity cycle, on the order of $10^6$~yr, indicates that under certain conditions AGN disks very likely play a role in tightening BBHs. 

\subsection{Embedded Binaries in the CE} \label{sec:ce}

Triple-star systems are common. While about 10\% of solar-mass stars are in triple or higher-order systems, nearly 60\% of O-type stars are in similar hierarchical configurations \citep[e.g.][Figure 39]{2017ApJS..230...15M}. For long-term dynamical stability, these systems must be hierarchical \citep{2014ApJ...793..137N}. Many such triple systems containing compact objects must have experienced significant mass-transfer events as their stellar progenitors evolved \citep{2014ApJ...781L..13T,2015MNRAS.450.1716S,2016ComAC...3....6T,2018arXiv180509338D}. In some of these hierarchical triple systems, the outer star evolves to transfer mass toward an inner, more compact binary. Depending on the stability of this mass transfer, the donor star's mass transfer could either proceed at a low rate or lead to runaway orbital tightening and the engulfment of the more compact pair within the envelope of the donor star. 
What happens when a close binary becomes embedded in a CE has been a topic of previous uncertainty \citep[e.g.][]{2015MNRAS.450.1716S}. In particular, would the binary spiral inward within the stellar envelope, or be driven to merger? 

Unraveling the outcomes  of these triple-star CE encounters requires an understanding of gas accretion and orbital evolution as studied in this paper.  If one assumes a simple formalism, such as the timescales derived from BHL, it would appear that a binary embedded in a CE should merge faster than its CM spirals into the donor star's envelope. We have shown here that this is not necessarily the case. Instead, we have found that the slowdown of the binary's CM (which is the driver of the CE inspiral) occurs more quickly than the embedded binary's own separation decays. Therefore, the embedded binary can spiral into the envelope gas, and that process alone will not cause the binary to merge. 

Since the binary may be able to complete part of the CE inspiral intact, there are many possible outcomes for the interaction.   The binary may survive the CE if the envelope can be ejected early in the encounter, the binary may be tidally disrupted before envelope ejection, or, in the case of unsuccessful envelope ejections, one or both of the binary components could merge with the core of the donor star.   The revised formalism presented in this paper should be used when examining  the various outcomes expected from these complex interactions.
For example, \citet{2015MNRAS.450.1716S} considered a formation scenario for \psr in which the outer binary system  remains intact as its CM spirals into the envelope of the donor star, reaching high-enough densities to be tidally disrupted. Our results indicate this channel is indeed plausible, as the timescale for the drag-induced inspiral of the CM of the embedded binary is shorter than the timescale for the embedded binary's finite semi-major axis to decay.


\section{Conclusions}\label{sec:summary}
This paper has studied the evolution of binaries moving through gaseous media as a function of the binary system's initial semi-major axis, $a_0$, relative to its gravitational focusing radius, $R_a$.  We extended the analytical and numerical formalisms of single-particle BHL accretion to the case of a binary system moving perpendicular to the orbital angular momentum vector of the pair.  We performed 3D hydrodynamical simulations to study the effects of accretion and drag on the translational and orbital motion of the system.  Below we summarize the main conclusions of this work:

\begin{enumerate}
    \item The introduction of a finite binary separation softens the gravitational potential of the pair relative to a single object and introduces time-dependent structures into the flow (as described in Section \ref{sec:flow} and seen in Figures \ref{fig:sma_series_density}-\ref{fig:flow-timeseries-041}).  We find that the ratio of the binary separation to the gravitational focusing scale is crucial in describing the flow structure. When $a/\rbh<1$, the binary orbits within a bow shock that is otherwise similar to that from a single mass. In these cases, the orbital motion is supersonic relative to the postshock gas and the pair launches spiral shocks into its surroundings (Figure \ref{fig:sma_series_divergence}).  By contrast, when $a/\rbh>1$, the binary members are each trailed by their own shock structure. 
    \item The orbiting binary's motion introduces periodic variation in the accretion rate onto the binary components (see Section \ref{sec:accretion} and Figure \ref{fig:mdot-time-series}). The overall accretion rate onto the binary is again modulated by the ratio of the binary separation to the gravitational focusing scale. When $a/\rbh \ll 1$, the accretion rate onto the binary approaches that of a single particle, with scaling $\dot M \propto M^2$. When $a/\rbh > 1$, the accretion rate approaches the sum of two independent particles with scaling $\dot M \propto m_1^2 + m_2^2$ (Figure \ref{fig:mdot-versus-a}).
    \item Gaseous DF and the accretion of momentum give rise to forces that modify the CM and orbital motion of the pair. In Section \ref{sec:forces}, we discuss the projection of these forces onto the CM and orbital velocities. CM deceleration is dominated by DF forces from the large-scale wake that trails the binary (Figure \ref{fig:forces_a041_cm}). Gas structures and flow on the scale of the binary orbit give rise to forces that drive orbital tightening \citep[e.g.,][]{2008ApJ...679L..33K}. In this case, momentum accretion and DF become similarly important (Figure \ref{fig:forces_a041_orb}). 
    \item We measure orbit-averaged timescales for accretion, $\ave\taccr$, slowdown of the CM, $\ave\tstop$, and orbital inspiral of the binary, $\ave\tinsp$, with Equations \eqref{eq:tdouble_num}-\eqref{eq:tinsp_num}, respectively. For all values of $a_0$ that we simulated, the timescales exhibit a clear hierarchy with $\tstop < \tinsp < \taccr$ (Figure \ref{fig:timescales}).  Therefore, a binary moving through a gaseous medium slows before being driven to merger or gaining significant mass (Section \ref{sec:timescales}). 
    \item  BHL theory predicts that $\tinsp\propto a_0^{-3/2}$, which implies that as the orbit tightens, the time to coalescence lengthens. By contrast, in our simulations, we find a mild, positive scaling, $\ave\tinsp \propto a^{0.19}$ (given by Equation \eqref{eq:tinspfit} and shown in Figure \ref{fig:decay_fit}). Previous semianalytic work predicting the sign and scaling of this relation has arrived at a variety of conclusions, as described in Section \ref{sec:previouswork}. We show that the simulation results can be reproduced by considering the enhanced density in which the binary orbits that is due to the convergence of gravitationally focused gas (Equation \eqref{eq:tildeinsp}), a key effect that was neglected in previous analytic studies. 
    \item The magnitude of the forces we derive implies that binaries in environments from the ISM to star-forming clouds, AGN disks, and stellar envelopes during CE phases all may experience transformation due to interaction with their gaseous surroundings. We discuss particular systems and the implications of the observed hierarchy of timescales, and we derive the transition between systems whose orbital evolution is dominated by gaseous interaction (at large separations) and those whose orbital evolution is driven by gravitational radiation (at small separations) in Section \ref{sec:discussion}. 
\end{enumerate}

 There are many aspects of this problem left to explore. In this work, we have simulated equal-mass binary systems in circular orbits inclined such that the incoming wind hits the orbital plane edge-on. A caveat in the extension of our results to astrophysical binaries is the assumption of a uniform background medium, which is valid when velocity and density scale heights are larger than both $a$ and $R_a$ of the binary and when conditions change over timescales longer than the orbital period.
 
Realistic binary systems are often subject to nonuniform gases with gradients in density, pressure, and Mach number.  Previous work has shown that breaking the symmetry of canonical BHL has significant effects on the flow.  For example, \citet{2015ApJ...803...41M} and \citet{2017ApJ...838...56M} found that in single-object BHL, the presence of  density gradients typical of CE encounters impart a net rotation and reduce accretion from postshock flow. Recently, \citet{2019MNRAS.488.5162X} have examined the criteria that define how the inhomogeneity of the background gas affects the stability of postshock flow.  \citet{2006ApJ...638..369K} simulated BHL flows in cold, turbulent media of relevance to star-forming clouds and found that the vorticity imparted by turbulence can dominate the flow structure and lead to highly variable accretion rates.

Future simulation work could relax our current assumptions of equal mass, circular orbits, edge-on inclination, and a homogeneous background medium to extend the applicability of these results to a broader range of astrophysical systems and environments. 

\begin{acknowledgements}
We gratefully acknowledge the anonymous referee and I. El Mellah for helpful suggestions on an earlier draft of this manuscript. We thank Z. Haiman, M. Hance, D. Lee, D.N.C. Lin, J. Naiman, E. Ostriker, A. Rosen, and S. Schr{\o}der for helpful discussions. The calculations for this research were carried out in part on the UCSC supercomputer Hyades, which is supported by the National Science Foundation (award number AST-1229745) and UCSC. A.A. is grateful for support from the NSF REU program Lamat at UCSC and the California Space Grant Consortium.
M.M. is grateful for support for this work provided by NASA through Einstein Postdoctoral Fellowship grant No. PF6-170155 awarded by the Chandra X-ray Center, which is operated by the Smithsonian Astrophysical Observatory for NASA under contract NAS8-03060.
E.R.-R. acknowledges support from NASA ATP grant NNX14AH37G, NSF grant AST-1615881, the Julie Packard Foundation, the Radcliffe Institute for Advanced Study, the Heising-Simons Foundation, and the Danish National Research Foundation (DNRF132).

\end{acknowledgements}
\software{{\tt FLASH} \citep{2000ApJS..131..273F}; {\tt yt}  \citep{2011ApJS..192....9T}; {\tt astropy} \citep{2013A&A...558A..33A}. }

\appendix

\section{Derivation of Analytical Timescales for Orbital Inspiral}
\label{sec:timescale_derive}
A principal objective of this paper is to learn how the predictions of BHL must be modified to describe a binary system. In this section, we derive the analytic timescales for orbital inspiral given by equations \eqref{eq:tinspBH} and \eqref{eq:tinspHL}.  

To associate a timescale with the decay of the orbit, we imagine the CM at rest and calculate the Bondi-Hoyle energy dissipation rate for $m_1$ and $m_2$ as they orbit the CM. For an equal-mass binary in a circular orbit, $m_1 = m_2 = M/2$, and the orbital speed is $v_{\rm orb}^2 = GM/a$.   If $V_1$ and $V_2$ are the barycentric speeds of $m_1$ and $m_2$, respectively, then $V_1 = V_2 = v_{\rm orb}/2$.
The BH accretion radius, Equation \eqref{eq:rbh}, of $m_1$ is 
\beq
R_{\rm BH,1} = \frac{2 G m_1}{V_1^2 + \cs^2} 
\label{eq:rbhone}
\eeq
with an energy dissipation rate, Equation \eqref{eq:ebh}, of 
\beq
\dot{E}_{\rm BH,1} = \frac{1}{2}\pi R_{\rm BH,1}^2 \rhoinf (V_1^2 + \cs^2)^{3/2}.
\label{eq:ebhone}
\eeq
The total energy dissipation rate for $m_1$ and $m_2$ is just twice $\dot{E}_{\rm BH,1}$, that is, 
\beq
\dot{E}_{\rm BH, orb} = \frac{4\pi G^2 m_1^2\rhoinf}{(V_1^2 + \cs^2)^{1/2}}.
\eeq
Replacing $m_1 = M / 2$ and $V_1 = \vorb/2$ and factoring out a 1/4 from the denominator give
\beq
\dot{E}_{\rm BH, orb}= \frac{2\pi G^2M^2\rhoinf}{(\vorb^2 +4 \cs^2)^{1/2}}.
\eeq
The total orbital energy of the binary is $E_{\rm orb} = -G m_1 m_2/2a = -GM^2/8a$, so the timescale for orbital decay is
\begin{align}
    \tinspBH &\equiv \frac{|E_{\rm orb}|}{\dot{E}_{\rm BH, orb}} \nonumber\\
    &= \frac{GM^2 / 8a}{2\pi G^2 M^2 \rhoinf (\vorb^2 + 4\cs^2)^{-1/2}} \nonumber \\
    &= \frac{(\vorb^2 + 4\cs^2)^{1/2}}{16\pi G \rhoinf a}.
\end{align}
If we factor $\vorb^2$ out of the parentheses and substitute $\vorb = (GM / a)^{1/2}$ then
\begin{align}
\tinspBH &= \frac{1}{16\pi G \rhoinf a}\bigg(\frac{GM}{a}\bigg)^{1/2}\bigg[1 + 4a\frac{\cs^2}{GM}\bigg]^{1/2}  \nonumber \\
&= \frac{1}{16\pi\rhoinf a^{3/2}}\bigg(\frac{M}{G}\bigg)^{1/2}\bigg[1 + 4a\frac{\cs^2}{GM}\bigg]^{1/2}.
\end{align}
Now the term $\cs^2/GM$ can be expressed as 
\beq 
    \frac{\cs^2}{GM} = \frac{2\cs^2}{\vinf^2}\frac{\vinf^2}{2GM} = \frac{2}{\mach^2 R_a},
\eeq
so we can write
\beq
\tinspBH = \frac{1}{16\pi}\bigg(\frac{M}{G}\bigg)^{1/2} \rhoinf^{-1}a^{-3/2} \bigg[1 + \frac{8a}{\mach^2 R_a}\bigg]^{1/2}.
\eeq
It is convenient to introduce a factor of $\ra$ in order to obtain the quantity $(a / \ra)$, giving 
\beq
\tinspBH = \frac{1}{8\sqrt{2}}\bigg[\frac{\vinf^3}{4\pi G^2M \rhoinf}\bigg] \bigg(\frac{a}{R_a}\bigg)^{-3/2} \bigg[1 + \frac{8a}{\mach^2 R_a}\bigg]^{1/2}.
\eeq
The first term in brackets is simply $\tstopHL$, so we have arrived at Equation \eqref{eq:tinspBH}:
\beq
\tinspBH = 
\frac{\tstopHL}{8\sqrt{2}}\bigg(\frac{a}{R_a}\bigg)^{-3/2}\bigg[1 + \frac{8a}{\mach^2 R_a}\bigg]^{1/2}.
\eeq
In the Hoyle-Lyttleton formalism, the sound speed is omitted from the characteristic velocity in the expressions \eqref{eq:rbhone} and \eqref{eq:ebhone}, resulting in a total orbital energy dissipation rate of
$\dot{E}_{\rm HL, orb} = 2\pi G^2 M^2 \rhoinf \vorb^{-1}$
and a decay timescale of 
\beq
\tinspHL = \frac{\tstopHL}{8\sqrt{2}}\bigg(\frac{a}{R_a}\bigg)^{-3/2},
\eeq
which is Equation \eqref{eq:tinspHL}.  This quantity is the leading coefficient of $\tinspBH$.  That is, 
\beq
\tinspBH = \tinspHL \bigg[1 + \frac{8a}{\mach^2 R_a}\bigg]^{1/2}.
\eeq

\section{Validation of Numerical Method}
\label{sec:validation}
\begin{figure*}
\gridline{\fig{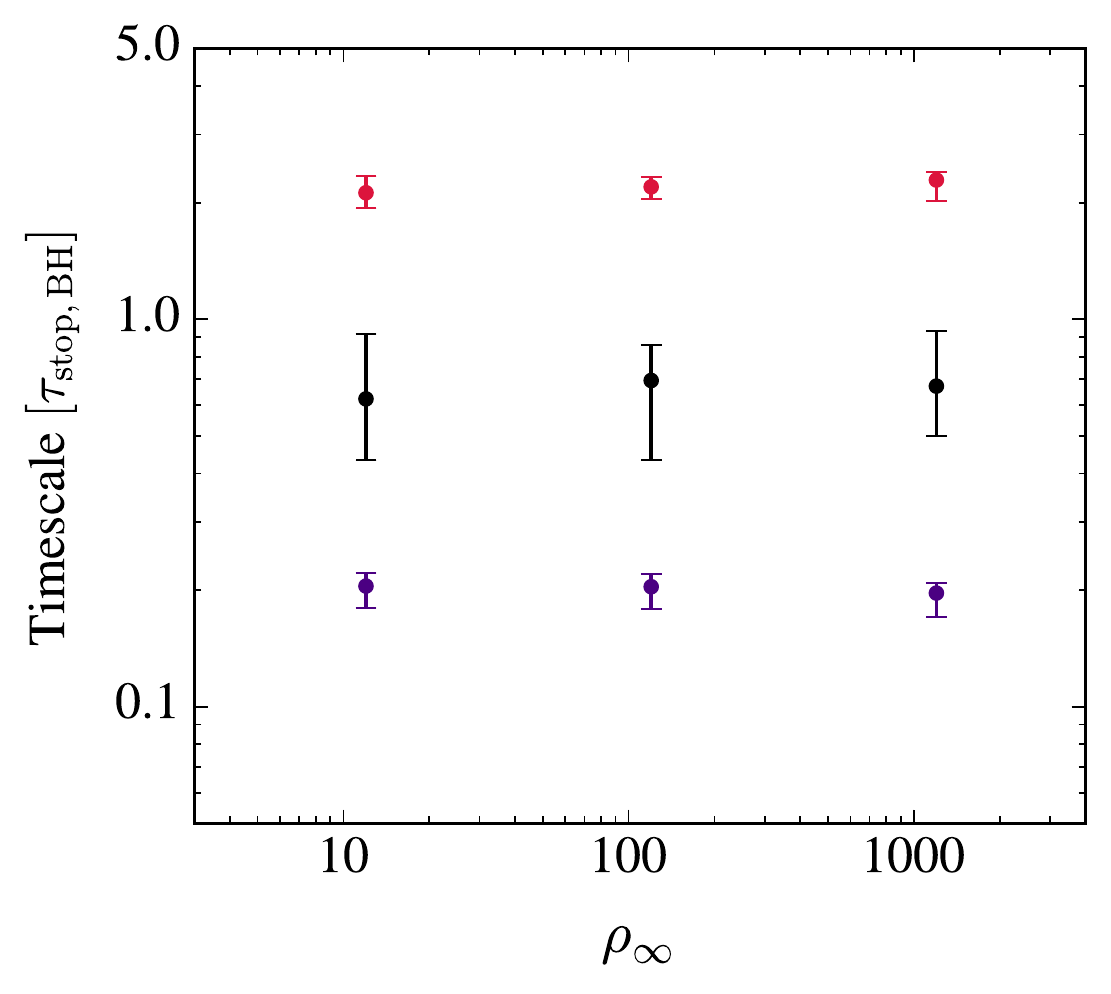}{0.32\textwidth}{(A)}
	      \fig{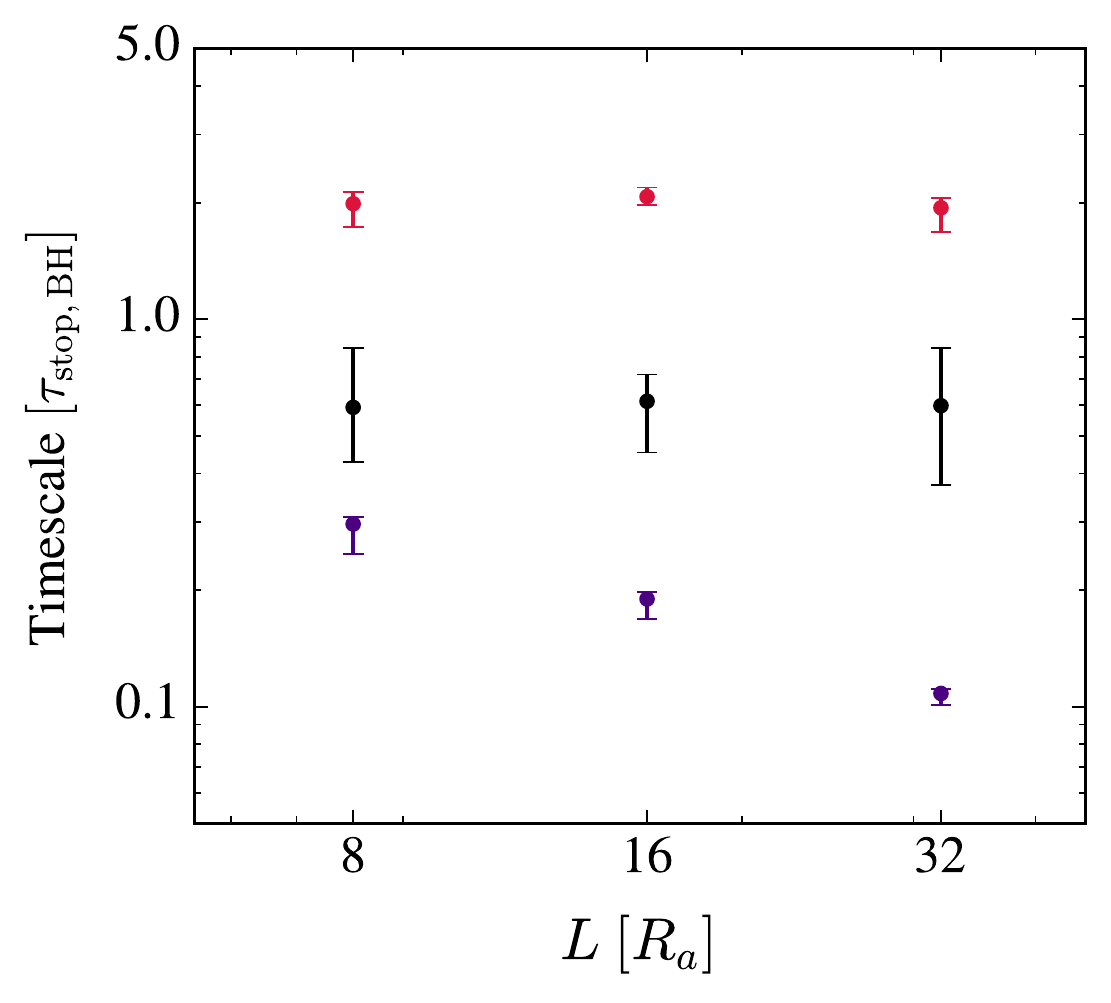}{0.32\textwidth}{(B)}
	      \fig{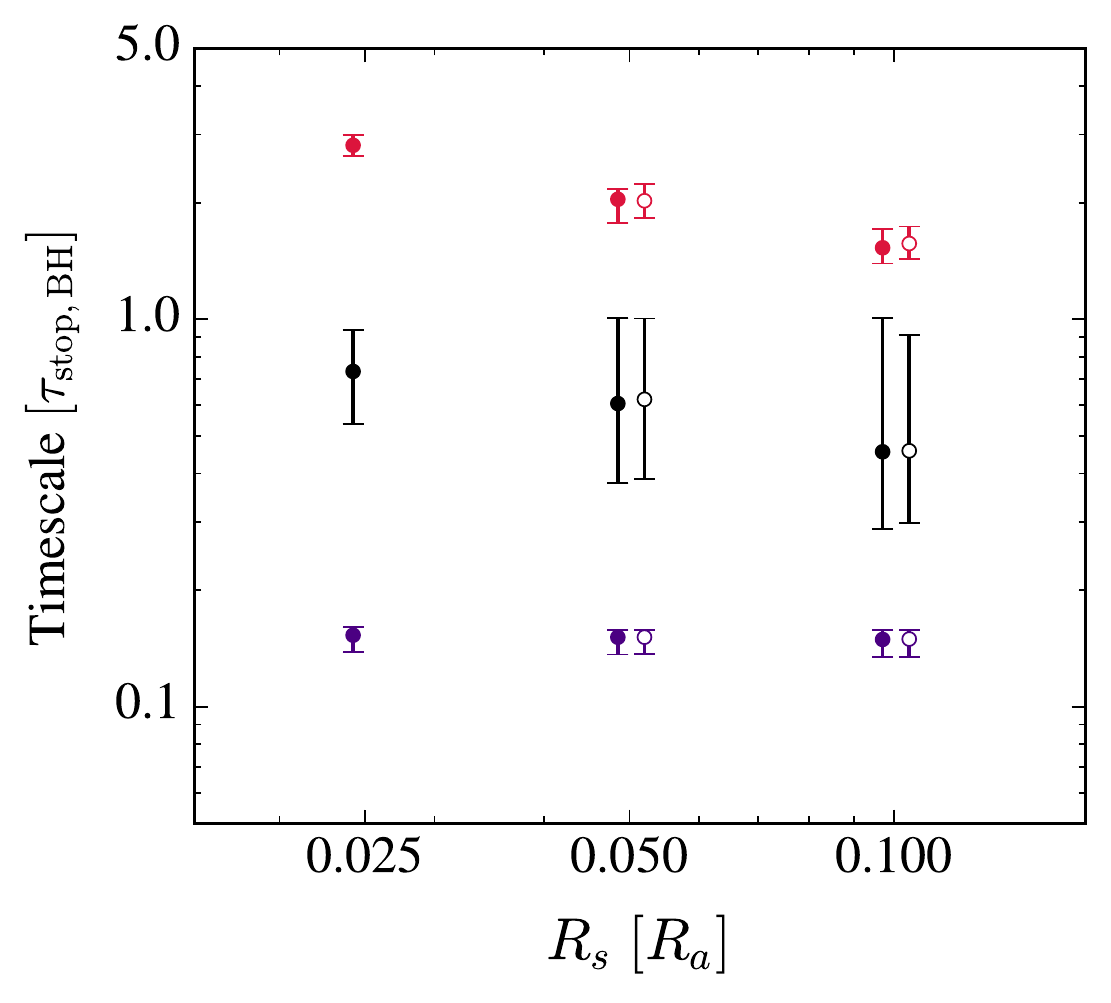}{0.32\textwidth}{(C)}
         }
\caption{Orbit-averaged timescales calculated from each of our numerical tests.  All of these tests were performed with $a_0 = 1.0R_a$ and $\gamma = 5/3$.  In each panel, $\ave\taccr$ is plotted in red, $\ave\tinsp$ in black, and $\ave\tstop$ in purple. The $y$ axis is in units of $\tstopBH$ and the $x$ axis shows the parameter under study. Each point gives the median over the simulation times used in the average, while the lower and upper error bars give the 15.87th to 87.13th percentile spread about the median.  Panel (A) shows the results of our studies of background density, in which we ran simulations with, $\rho_\infty = 10, 100,$ and $1000$. All three timescales show excellent convergence. Panel (B) shows our tests of the size of the computational domain. Here, $L$ is the width of the domain in the $x$, $y$, and $z$ directions, giving a computational volume of $L^3$. Panel (C) plays a dual role, showing tests of sink size, $R_s$, and finest grid resolution, $\delta_{\rm min}$.  The filled circles show our tests of $R_s = 0.025, 0.05,$ and $0.1 R_a$ with $\rs  / \delta_{\rm min} = 3$ cells across the sink radius.  The open circle plots (offset slightly to the right for clarity) show simulations with $R_s = 0.05$ and $1.0 R_a$, but at higher resolution ($R_s / \delta_{\rm min} = 6$ cells across the sink radius). }
\label{fig:studies}
\end{figure*}

In this section, we characterize the sensitivity of our results to 
background density, $\rhoinf$, the length of the computational domain in each Cartesian direction, $L$, the radius of the absorbing sinks used to represent the binary, $\rs$, and the length of each side of the finest grid cells,  $\delta_{\rm min}$.  The studies presented in this section adopt $\mach = 2$,  $\gamma = 5/3$, and $a_0 = 1.0 R_a$.  In the panels of  Figure \ref{fig:studies}, we plot $\ave\taccr$, $\ave\tinsp$, and $\ave\tstop$, normalized to $\tinspBH$, for each series of studies.
\subsection{Grid Refinement}
We maximally refine the region of our two sink particles so that sink cells have length of $\delta_{\rm min}$ in each Cartesian direction.   We test sensitivity to $\delta_{\rm min}$ by running simulations with $\rs / \delta_{\rm min} = $ 3 and 6 with sink sizes of $\rs = 0.05 \ra$ and $\rs = 0.1 \ra$. Comparing points with the same $\rs$ in panel (C) of Figure \ref{fig:studies} reveals that there is very little difference in the orbit-average timescales when the resolution is doubled from our fiducial value of $\rs / \delta_{\rm min} = $ 3.   When $\rs = 0.05 \ra$, $\ave\tstop$ and $\ave\taccr$ change by less than 1\% and $\ave\tinsp$ increases by 2.5\%.  Accretion rates are larger when $\rs = 0.1 \ra$ and, in this case, we find that a resolution of twice our fiducial value increases $\ave\taccr$ by $2.6\%$ and the other timescales change by less than 1\%.  We adopt $\rs / \delta_{\rm min} = 3$ in our production runs.

\subsection{Sink Radius}
 The influence of sink radius, $\rs$, on accretion rates in single-object BHL flows has been well studied \citep[e.g.][]{1994ApJ...427..351R,2012ApJ...752...30B,2015ApJ...803...41M,2017ApJ...838...56M}. For example, \citet{1994ApJ...427..351R} found that doubling the size of the sink increased accretion rates by $\approx 15\%-25\%$ in their simulations with $\mach = 3$ and $\gamma = 5/3$.  We test $\rs$ in our setup by running simulations with half and twice our fiducial value of $R_s  = 0.05 \ra$.  These simulations adopt $\rs / \delta_{\rm min} = 3$, $\rhoinf = 10$ and $L = 32 R_a$.   When $\rs / \ra$ is doubled from 0.05 to 0.1, $\ave\tstop$ decreases by 1\% and $\ave\tinsp$ and $\ave\taccr$ each decrease by 25\%.  A doubling of $\rs/\ra$ from 0.025 to 0.05 similarly decreases $\ave\tstop$ by 1\%, while $\ave\tinsp$ decreases by 17\% and $\ave\taccr$ decreases by 27\%.  
 
 The filled points in panel C of Figure \ref{fig:studies} show that the hierarchy of timescales $\ave\taccr \ge \ave\tinsp \ge \ave\tstop$ is  preserved across values of $\rs$.  However, because sink size influences  rates of mass accretion, the accretion timescale is sensitive to this choice. For an astrophysical accretor of size less than $\rs$, the mass accretion rate we derive therefore represents the upper limit defined by the flux of material into the object's vicinity, rather than the precise amount that is able to accrete. Because sink size affects the  momentum accreted, we note that the exact scaling of $\tinspfit$ as a function of $a_0$ will be influenced by the choice of sink radius.    We adopt $\rs = 0.05 \ra$ in our production runs.

\subsection{Background Density}
We performed tests of background density, $\rhoinf$, to ensure that the timescales we derive could be scaled to any astrophysical density.  We ran simulations with $\rhoinf =$ 10, 100, and 1000. These simulations were run with $\rs = 0.05 R_a$, $L = 16 R_a$, and $\rs / \delta_{\rm min} = 6$.  Panel (A) of Figure \ref{fig:studies}, shows that the background density scales out of the orbit-average timescales so that the values are the same when normalized to $\tstopBH$. In all cases, $\ave\taccr \gg \ave\tinsp \gg \ave\tstop$.  Increasing our fiducial value by a factor of 100 (to $\rhoinf=1000$) only  changes  $\ave\taccr / \tstopBH$ and  $\ave\tinsp/\tstopBH$ by $7.6\%$ each and decreases $\ave\tstop / \tstopBH$ by $4\%$.   We adopt $\rhoinf = 10$ for our production runs.
\begin{deluxetable}{c |ccc | c}
\tablecaption{Orbit-averaged Simulation Data \label{table:fittable}}
\tablehead{
$a_0$     &$\ave\tinsp$ &$\ave\tstop$ &$\ave\taccr$   & $\ave\mb$ \\
$[\ra]$     & $[\tstopBH]$          & $[\tstopBH]$           & $[\tstopBH]$ 
& $[\dot{M}_{\rm BH}]$
}
\startdata
0.0   & ---                       & $0.142_{-0.000}^{+0.000}$ & $1.265_{-0.007}^{+0.004}$ & $0.988_{-0.002}^{+0.007}$ \\
0.16  & $0.450_{-0.016}^{+0.026}$ & $0.143_{-0.003}^{+0.002}$ & $1.418_{-0.070}^{+0.070}$ & $0.883_{-0.006}^{+0.008}$ \\
0.3   & $0.503_{-0.035}^{+0.049}$ & $0.145_{-0.004}^{+0.003}$ & $1.595_{-0.085}^{+0.089}$ & $0.786_{-0.026}^{+0.022}$ \\
0.41  & $0.544_{-0.096}^{+0.088}$ & $0.145_{-0.002}^{+0.002}$ & $1.883_{-0.132}^{+0.145}$ & $0.663_{-0.036}^{+0.043}$ \\
0.55  & $0.545_{-0.088}^{+0.134}$ & $0.147_{-0.004}^{+0.003}$ & $2.166_{-0.181}^{+0.226}$ & $0.579_{-0.049}^{+0.041}$ \\
0.74  & $0.628_{-0.109}^{+0.121}$ & $0.148_{-0.009}^{+0.003}$ & $2.132_{-0.139}^{+0.153}$ & $0.588_{-0.035}^{+0.029}$ \\
1.0   & $0.603_{-0.219}^{+0.521}$ & $0.148_{-0.014}^{+0.006}$ & $2.040_{-0.243}^{+0.151}$ & $0.613_{-0.041}^{+0.079}$ \\
\enddata
\tablecomments{Here we give median values of the orbit-averaged timescales and accretion rates for our simulated binaries. The subscripts and superscripts give the 15.87th to 87.13th percentile spread, respectively, about the median. The timescales are given in units of $\tstopBH$. The accretion rates are in units of $\mbh$.}
\end{deluxetable}

\begin{figure*}
\gridline{\fig{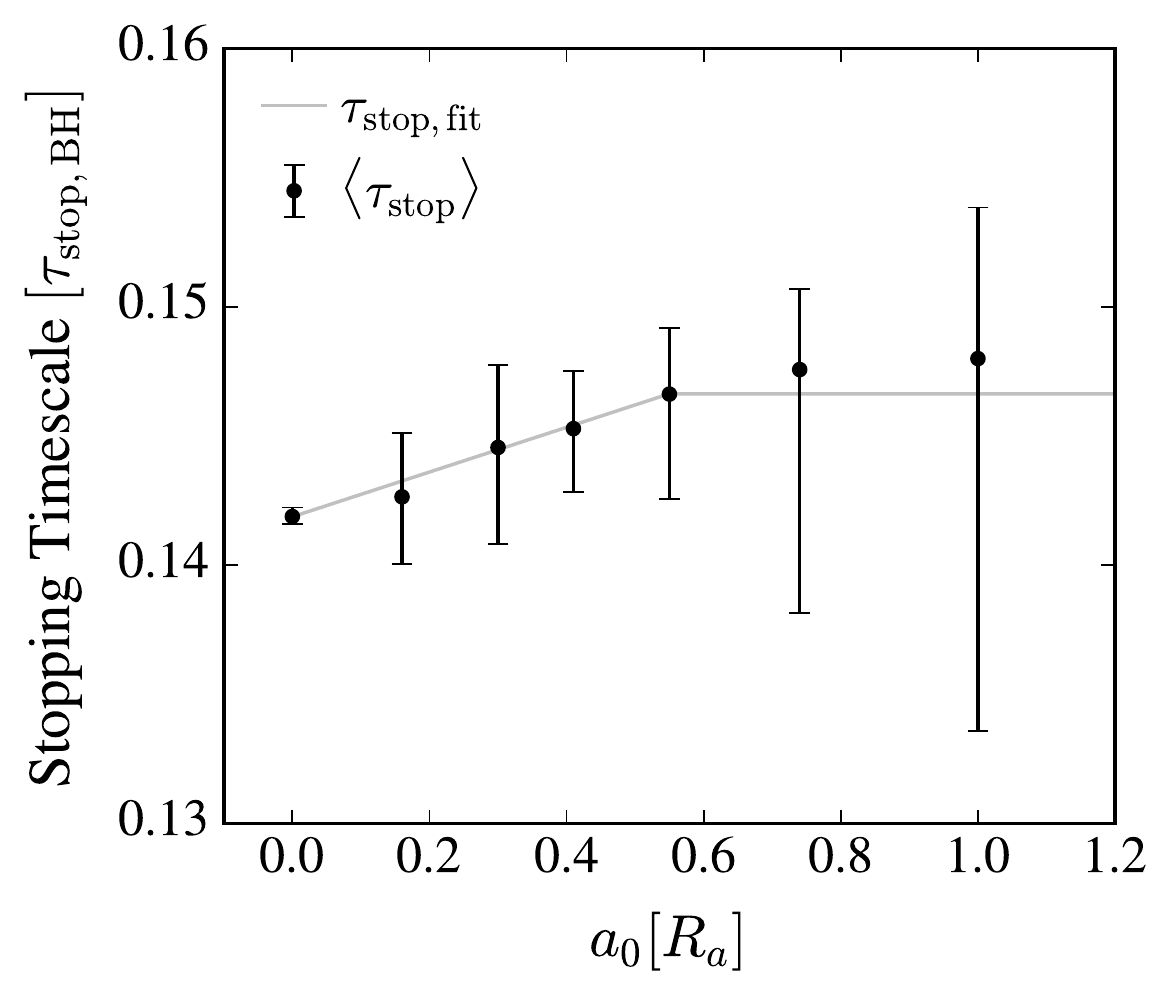}{0.32\textwidth}{(A)}
	      \fig{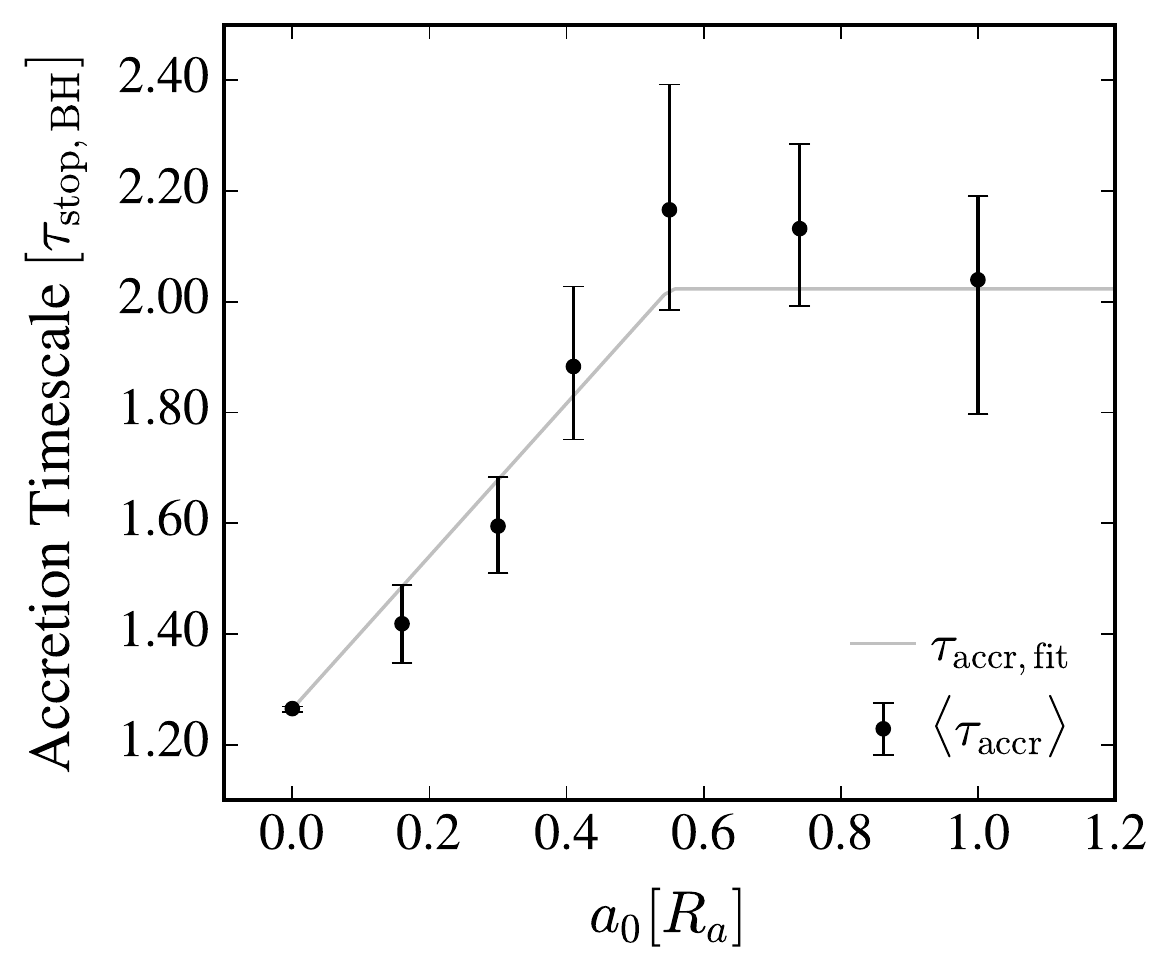}{0.32\textwidth}{(B)}
		  \fig{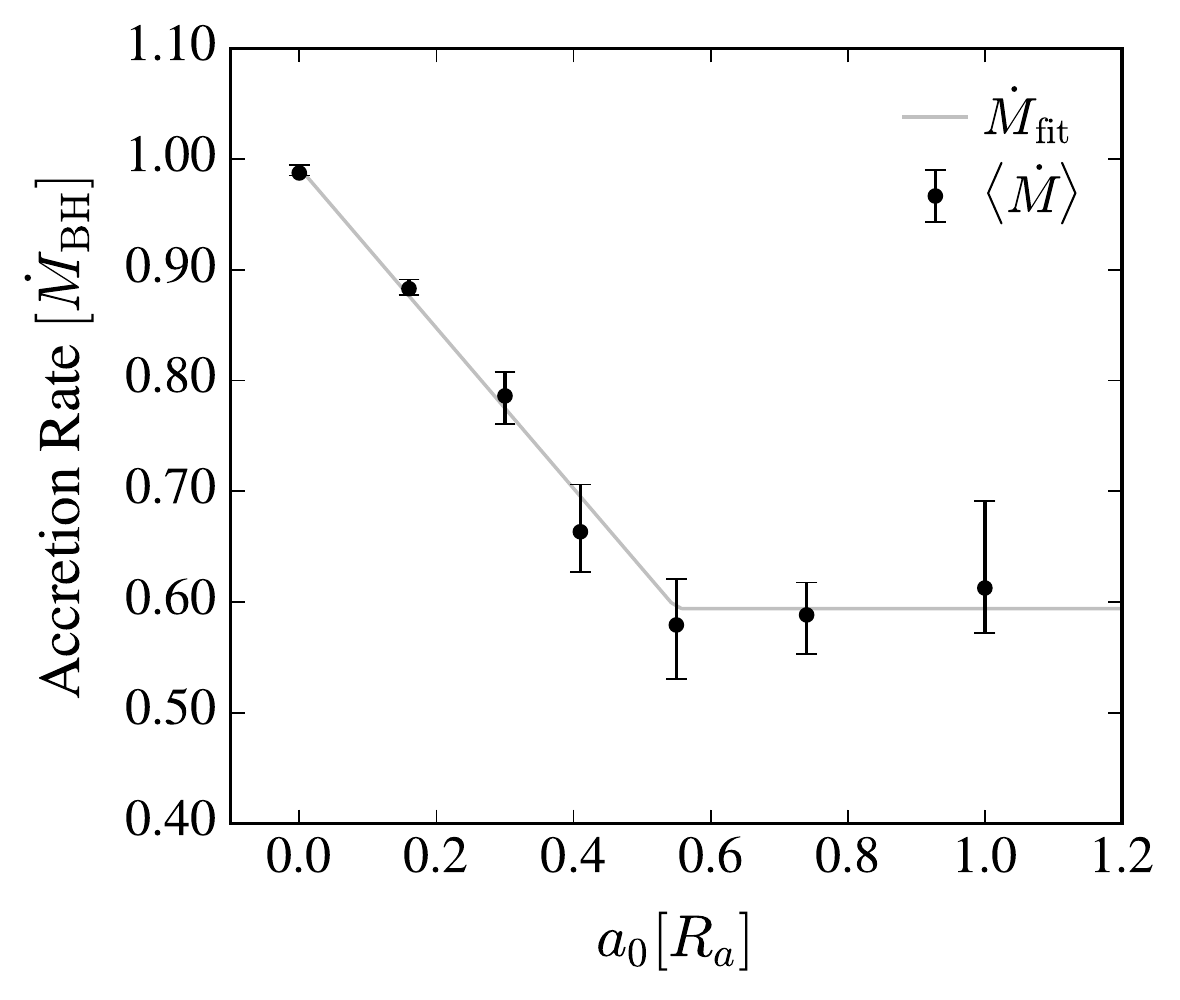}{0.32\textwidth}{(C)}
         }
\caption{Piecewise fits of $\ave\tstop$, $\ave\taccr$, and$\ave\mb$ (panels A, B, and C, respectively).  In each panel, the black points and error bars correspond to the data of Table \ref{table:fittable} and the gray line shows the fit, which are given by equations \eqref{eq:tstopfit}, \eqref{eq:taccrfit}, and \eqref{eq:mdotfit}, respectively.}
\label{fig:piecewisefits}
\end{figure*}

\subsection{Extent of the Computational Domain}
The DF drag forces exerted on an object depend on the size of the wake, with approximate scaling dictated by a Coulomb logarithm $\ln(r_{\rm max}/r_{\rm min})$, where $r_{\rm min}$ is the size scale outside of which a net force is contributed and $r_{\rm max}$ is the system size \citep{1999ApJ...513..252O}. While we are forced to choose a single system size in our numerical simulations, we measure the approximate scaling here. 

In our setup, the CM of the binary is located at the origin and we place all outer boundaries at a distance $L/2$ from the origin. We run simulations with half and double a fiducial value of $L = 16 \ra$.   Panel (B) of Figure \ref{fig:studies} shows that $\ave\taccr$ and $\ave\tinsp$ do not vary significantly with changing $L$.  Doubling $L$ from 16 to 32 reduces $\ave\taccr$ by 6.4\% and reduces $\ave\tinsp$ by 2.6\%. 

On the other hand, the same doubling of $L$ reduces $\ave\tstop$ by 43\%, which is expected because this timescale is dominated by DF due to gas at larger scales.  In otherwise similar simulations of single-object BHL, \citet{2016A&A...589A..10T} and \citet{2017ApJ...838...56M} found $r_{\rm min} \approx 0.5 \ra$ and $r_{\rm max}\approx L/2$.  A fit of the DF force for the simulations presented here reveals that $r_{\rm min} \sim 3\ra$ for the larger box sizes of $L=16\ra$ and $L=32\ra$ (adopting $r_{\rm max}\approx L/2$).  Since $\tstop \propto  1 / F_{\rm DF, CM}$, our measured stopping time can be approximately scaled to larger astrophysical systems by dividing by $\ln[(L/2)/ (3\ra) ]$.

\section{Fitting Formulae to Simulation Results}
In this section, we provide formulae for $\ave\tstop$, $\ave\taccr$, and $\ave\mb$ as a function of $a_0$.  The fit for $\ave\tinsp$ was given in Section \ref{sec:numtimescales}.  The remaining quantities $\ave\tstop$, $\ave\taccr$, and $\ave\mb$ do not follow a single power law, so we fit piecewise functions for $a_0 \le 0.55$ and $a_0 > 0.55$ of the functional form 
\beq
f(a_0) = \begin{cases}
		A (a_0/\ra) + B & a_0 \le 0.55 \ra\\[1em]
		A (0.55) + B     & a_0 > 0.55 \ra
		\end{cases}
\eeq
which goes to a constant value because the two particles become independent as $a_0 \to \infty$.  We perform least-squares fits of the median values given in Table \ref{table:fittable} to the model, $f(a_0)$, weighting the points according to one standard deviation of their time variability.  For$\ave\tstop$, we find $A = 0.009$ and $B = 0.142$, giving
\beq
\frac{\tau_{\rm stop,fit}}{\tstopBH} = \begin{cases}
		0.009(a_0/\ra) + 0.142   & a_0 \le 0.55 \ra\\[1em]
		0.147                    & a_0 > 0.55 \ra
		\end{cases}
\label{eq:tstopfit}
\eeq
in units of $\tstopBH$. For $\ave\taccr$, we obtain $A = 1.379$ and $B = 1.265$, or 
\beq
\frac{\tau_{\rm accr,fit}}{\tstopBH} = \begin{cases}
		1.379(a_0/\ra) + 1.265  & a_0 \le 0.55 \ra\\[1em]
		2.023                    & a_0 > 0.55 \ra
		\end{cases}
\label{eq:taccrfit}
\eeq
in units of $\tstopBH$. For the accretion rate, $\ave\mb$, in units of $\mbh$, we find $A = -0.724$ and $B = 0.992$, that is,
\beq
\frac{\dot{M}_{\rm fit}}{\mbh} = \begin{cases}
		-0.724(a_0/\ra) + 0.992  & a_0 \le 0.55 \ra\\[1em]
		0.594                    & a_0 > 0.55 \ra.
		\end{cases}
\label{eq:mdotfit}
\eeq
In panels (A), (B), and (C) of Figure \ref{fig:piecewisefits}, we plot $\ave\tstop$, $\ave\taccr$, and $\ave\mb$, respectively, with points and error bars. The associated fit of each quantity is shown with a gray line in each panel.

\bibliographystyle{aasjournal}

\begin{thebibliography}{}
\expandafter\ifx\csname natexlab\endcsname\relax\def\natexlab#1{#1}\fi
\providecommand{\url}[1]{\href{#1}{#1}}
\providecommand{\dodoi}[1]{doi:~\href{http://doi.org/#1}{\nolinkurl{#1}}}
\providecommand{\doeprint}[1]{\href{http://ascl.net/#1}{\nolinkurl{http://ascl.net/#1}}}
\providecommand{\doarXiv}[1]{\href{https://arxiv.org/abs/#1}{\nolinkurl{https://arxiv.org/abs/#1}}}

\bibitem[{{Abbate} {et~al.}(2018){Abbate}, {Possenti}, {Ridolfi}, {Freire},
  {Camilo}, {Manchester}, \& {D'Amico}}]{2018MNRAS.481..627A}
{Abbate}, F., {Possenti}, A., {Ridolfi}, A., {et~al.} 2018, \mnras, 481, 627,
  \dodoi{10.1093/mnras/sty2298}

\bibitem[{{Artymowicz} {et~al.}(1993){Artymowicz}, {Lin}, \&
  {Wampler}}]{1993ApJ...409..592A}
{Artymowicz}, P., {Lin}, D.~N.~C., \& {Wampler}, E.~J. 1993, \apj, 409, 592,
  \dodoi{10.1086/172690}

\bibitem[{{Astropy Collaboration} {et~al.}(2013){Astropy Collaboration},
  {Robitaille}, {Tollerud}, {Greenfield}, {Droettboom}, {Bray}, {Aldcroft},
  {Davis}, {Ginsburg}, {Price-Whelan}, {Kerzendorf}, {Conley}, {Crighton},
  {Barbary}, {Muna}, {Ferguson}, {Grollier}, {Parikh}, {Nair}, {Unther},
  {Deil}, {Woillez}, {Conseil}, {Kramer}, {Turner}, {Singer}, {Fox}, {Weaver},
  {Zabalza}, {Edwards}, {Azalee Bostroem}, {Burke}, {Casey}, {Crawford},
  {Dencheva}, {Ely}, {Jenness}, {Labrie}, {Lim}, {Pierfederici}, {Pontzen},
  {Ptak}, {Refsdal}, {Servillat}, \& {Streicher}}]{2013A&A...558A..33A}
{Astropy Collaboration}, {Robitaille}, T.~P., {Tollerud}, E.~J., {et~al.} 2013,
  \aap, 558, A33, \dodoi{10.1051/0004-6361/201322068}

\bibitem[{{Bartos} {et~al.}(2017){Bartos}, {Kocsis}, {Haiman}, \&
  {M{\'a}rka}}]{2017ApJ...835..165B}
{Bartos}, I., {Kocsis}, B., {Haiman}, Z., \& {M{\'a}rka}, S. 2017, \apj, 835,
  165, \dodoi{10.3847/1538-4357/835/2/165}

\bibitem[{{Baruteau} {et~al.}(2011){Baruteau}, {Cuadra}, \&
  {Lin}}]{2011ApJ...726...28B}
{Baruteau}, C., {Cuadra}, J., \& {Lin}, D.~N.~C. 2011, \apj, 726, 28,
  \dodoi{10.1088/0004-637X/726/1/28}

\bibitem[{{Blondin} \& {Pope}(2009)}]{2009ApJ...700...95B}
{Blondin}, J.~M., \& {Pope}, T.~C. 2009, \apj, 700, 95,
  \dodoi{10.1088/0004-637X/700/1/95}

\bibitem[{{Blondin} \& {Raymer}(2012)}]{2012ApJ...752...30B}
{Blondin}, J.~M., \& {Raymer}, E. 2012, \apj, 752, 30,
  \dodoi{10.1088/0004-637X/752/1/30}

\bibitem[{{Bondi}(1952)}]{1952MNRAS.112..195B}
{Bondi}, H. 1952, \mnras, 112, 195, \dodoi{10.1093/mnras/112.2.195}

\bibitem[{{Bondi} \& {Hoyle}(1944)}]{1944MNRAS.104..273B}
{Bondi}, H., \& {Hoyle}, F. 1944, \mnras, 104, 273,
  \dodoi{10.1093/mnras/104.5.273}

\bibitem[{{Bovy} {et~al.}(2012){Bovy}, {Rix}, {Hogg}, {Beers}, {Lee}, \&
  {Zhang}}]{2012ApJ...755..115B}
{Bovy}, J., {Rix}, H.-W., {Hogg}, D.~W., {et~al.} 2012, \apj, 755, 115,
  \dodoi{10.1088/0004-637X/755/2/115}

\bibitem[{{Brown} {et~al.}(2010){Brown}, {Geller}, {Kenyon}, \&
  {Diaferio}}]{2010AJ....139...59B}
{Brown}, W.~R., {Geller}, M.~J., {Kenyon}, S.~J., \& {Diaferio}, A. 2010, \aj,
  139, 59, \dodoi{10.1088/0004-6256/139/1/59}

\bibitem[{{Chanam{\'e}} \& {Gould}(2004)}]{2004ApJ...601..289C}
{Chanam{\'e}}, J., \& {Gould}, A. 2004, \apj, 601, 289, \dodoi{10.1086/380442}

\bibitem[{{Chandrasekhar}(1943)}]{1943ApJ....97..255C}
{Chandrasekhar}, S. 1943, \apj, 97, 255, \dodoi{10.1086/144517}

\bibitem[{Cherepashchuk(2014)}]{1063-7869-57-4-359}
Cherepashchuk, A.~M. 2014, Physics-Uspekhi, 57, 359

\bibitem[{{Colella} \& {Woodward}(1984)}]{1984JCoPh..54..174C}
{Colella}, P., \& {Woodward}, P.~R. 1984, Journal of Computational Physics, 54,
  174, \dodoi{10.1016/0021-9991(84)90143-8}

\bibitem[{{Cowley}(1992)}]{1992ARA&A..30..287C}
{Cowley}, A.~P. 1992, \araa, 30, 287,
  \dodoi{10.1146/annurev.aa.30.090192.001443}

\bibitem[{{del Valle} \& {Escala}(2014)}]{2014ApJ...780...84D}
{del Valle}, L., \& {Escala}, A. 2014, \apj, 780, 84,
  \dodoi{10.1088/0004-637X/780/1/84}

\bibitem[{{Di Stefano}(2018)}]{2018arXiv180509338D}
{Di Stefano}, R. 2018, ArXiv e-prints.
\newblock \doarXiv{1805.09338}

\bibitem[{{Duch{\^e}ne} \& {Kraus}(2013)}]{2013ARA&A..51..269D}
{Duch{\^e}ne}, G., \& {Kraus}, A. 2013, \araa, 51, 269,
  \dodoi{10.1146/annurev-astro-081710-102602}

\bibitem[{{Edgar}(2004)}]{2004NewAR..48..843E}
{Edgar}, R. 2004, \nar, 48, 843, \dodoi{10.1016/j.newar.2004.06.001}

\bibitem[{{Edgar}(2005)}]{2005A&A...434...41E}
{Edgar}, R.~G. 2005, \aap, 434, 41, \dodoi{10.1051/0004-6361:20042067}

\bibitem[{{Farris} {et~al.}(2015){Farris}, {Duffell}, {MacFadyen}, \&
  {Haiman}}]{2015MNRAS.446L..36F}
{Farris}, B.~D., {Duffell}, P., {MacFadyen}, A.~I., \& {Haiman}, Z. 2015,
  \mnras, 446, L36, \dodoi{10.1093/mnrasl/slu160}

\bibitem[{{Farris} {et~al.}(2010){Farris}, {Liu}, \&
  {Shapiro}}]{2010PhRvD..81h4008F}
{Farris}, B.~D., {Liu}, Y.~T., \& {Shapiro}, S.~L. 2010, \prd, 81, 084008,
  \dodoi{10.1103/PhysRevD.81.084008}

\bibitem[{{Federrath} {et~al.}(2010){Federrath}, {Banerjee}, {Clark}, \&
  {Klessen}}]{2010ApJ...713..269F}
{Federrath}, C., {Banerjee}, R., {Clark}, P.~C., \& {Klessen}, R.~S. 2010,
  \apj, 713, 269, \dodoi{10.1088/0004-637X/713/1/269}

\bibitem[{{Foglizzo} {et~al.}(2005){Foglizzo}, {Galletti}, \&
  {Ruffert}}]{2005A&A...435..397F}
{Foglizzo}, T., {Galletti}, P., \& {Ruffert}, M. 2005, \aap, 435, 397,
  \dodoi{10.1051/0004-6361:20042201}

\bibitem[{{Fryxell} {et~al.}(2000){Fryxell}, {Olson}, {Ricker}, {Timmes},
  {Zingale}, {Lamb}, {MacNeice}, {Rosner}, {Truran}, \&
  {Tufo}}]{2000ApJS..131..273F}
{Fryxell}, B., {Olson}, K., {Ricker}, P., {et~al.} 2000, \apjs, 131, 273,
  \dodoi{10.1086/317361}

\bibitem[{{Hilditch}(2001)}]{2001icbs.book.....H}
{Hilditch}, R.~W. 2001, {An Introduction to Close Binary Stars} (Cambridge
  University Press), 392

\bibitem[{{Hoyle} \& {Lyttleton}(1939)}]{1939PCPS...35..405H}
{Hoyle}, F., \& {Lyttleton}, R.~A. 1939, Proceedings of the Cambridge
  Philosophical Society, 35, 405, \dodoi{10.1017/S0305004100021150}

\bibitem[{{Hunt}(1971)}]{1971MNRAS.154..141H}
{Hunt}, R. 1971, \mnras, 154, 141, \dodoi{10.1093/mnras/154.2.141}

\bibitem[{{Ivanov} {et~al.}(1998){Ivanov}, {Igumenshchev}, \&
  {Novikov}}]{1998ApJ...507..131I}
{Ivanov}, P.~B., {Igumenshchev}, I.~V., \& {Novikov}, I.~D. 1998, \apj, 507,
  131, \dodoi{10.1086/306324}

\bibitem[{{Just} {et~al.}(2012){Just}, {Yurin}, {Makukov}, {Berczik}, {Omarov},
  {Spurzem}, \& {Vilkoviskij}}]{2012ApJ...758...51J}
{Just}, A., {Yurin}, D., {Makukov}, M., {et~al.} 2012, \apj, 758, 51,
  \dodoi{10.1088/0004-637X/758/1/51}

\bibitem[{{Kaaz} {et~al.}(2019){Kaaz}, {Antoni}, \&
  {Ramirez-Ruiz}}]{2019ApJ...876..142K}
{Kaaz}, N., {Antoni}, A., \& {Ramirez-Ruiz}, E. 2019, \apj, 876, 142,
  \dodoi{10.3847/1538-4357/ab158b}

\bibitem[{{Kafle} {et~al.}(2014){Kafle}, {Sharma}, {Lewis}, \&
  {Bland-Hawthorn}}]{2014ApJ...794...59K}
{Kafle}, P.~R., {Sharma}, S., {Lewis}, G.~F., \& {Bland-Hawthorn}, J. 2014,
  \apj, 794, 59, \dodoi{10.1088/0004-637X/794/1/59}

\bibitem[{{Karas} \& {{\v S}ubr}(2007)}]{2007A&A...470...11K}
{Karas}, V., \& {{\v S}ubr}, L. 2007, \aap, 470, 11,
  \dodoi{10.1051/0004-6361:20066068}

\bibitem[{{Kelly} {et~al.}(2017){Kelly}, {Baker}, {Etienne}, {Giacomazzo}, \&
  {Schnittman}}]{2017PhRvD..96l3003K}
{Kelly}, B.~J., {Baker}, J.~G., {Etienne}, Z.~B., {Giacomazzo}, B., \&
  {Schnittman}, J. 2017, \prd, 96, 123003, \dodoi{10.1103/PhysRevD.96.123003}

\bibitem[{{Kennedy} {et~al.}(2016){Kennedy}, {Meiron}, {Shukirgaliyev},
  {Panamarev}, {Berczik}, {Just}, \& {Spurzem}}]{2016MNRAS.460..240K}
{Kennedy}, G.~F., {Meiron}, Y., {Shukirgaliyev}, B., {et~al.} 2016, \mnras,
  460, 240, \dodoi{10.1093/mnras/stw908}

\bibitem[{{Kim} \& {Ostriker}(2017)}]{2017ApJ...846..133K}
{Kim}, C.-G., \& {Ostriker}, E.~C. 2017, \apj, 846, 133,
  \dodoi{10.3847/1538-4357/aa8599}

\bibitem[{{Kim} \& {Kim}(2007)}]{2007ApJ...665..432K}
{Kim}, H., \& {Kim}, W.-T. 2007, \apj, 665, 432, \dodoi{10.1086/519302}

\bibitem[{{Kim} {et~al.}(2008){Kim}, {Kim}, \&
  {S{\'a}nchez-Salcedo}}]{2008ApJ...679L..33K}
{Kim}, H., {Kim}, W.-T., \& {S{\'a}nchez-Salcedo}, F.~J. 2008, \apjl, 679, L33,
  \dodoi{10.1086/589149}

\bibitem[{{Krumholz} {et~al.}(2006){Krumholz}, {McKee}, \&
  {Klein}}]{2006ApJ...638..369K}
{Krumholz}, M.~R., {McKee}, C.~F., \& {Klein}, R.~I. 2006, \apj, 638, 369,
  \dodoi{10.1086/498844}

\bibitem[{{MacLeod} {et~al.}(2017){MacLeod}, {Antoni}, {Murguia-Berthier},
  {Macias}, \& {Ramirez-Ruiz}}]{2017ApJ...838...56M}
{MacLeod}, M., {Antoni}, A., {Murguia-Berthier}, A., {Macias}, P., \&
  {Ramirez-Ruiz}, E. 2017, \apj, 838, 56, \dodoi{10.3847/1538-4357/aa6117}

\bibitem[{{MacLeod} \& {Ramirez-Ruiz}(2015)}]{2015ApJ...803...41M}
{MacLeod}, M., \& {Ramirez-Ruiz}, E. 2015, \apj, 803, 41,
  \dodoi{10.1088/0004-637X/803/1/41}

\bibitem[{{McKernan} {et~al.}(2012){McKernan}, {Ford}, {Lyra}, \&
  {Perets}}]{2012MNRAS.425..460M}
{McKernan}, B., {Ford}, K.~E.~S., {Lyra}, W., \& {Perets}, H.~B. 2012, \mnras,
  425, 460, \dodoi{10.1111/j.1365-2966.2012.21486.x}

\bibitem[{{McKernan} {et~al.}(2011){McKernan}, {Ford}, {Lyra}, {Perets},
  {Winter}, \& {Yaqoob}}]{2011MNRAS.417L.103M}
{McKernan}, B., {Ford}, K.~E.~S., {Lyra}, W., {et~al.} 2011, \mnras, 417, L103,
  \dodoi{10.1111/j.1745-3933.2011.01132.x}

\bibitem[{{Miralda-Escud{\'e}} \& {Kollmeier}(2005)}]{2005ApJ...619...30M}
{Miralda-Escud{\'e}}, J., \& {Kollmeier}, J.~A. 2005, \apj, 619, 30,
  \dodoi{10.1086/426467}

\bibitem[{{Moe} \& {Di Stefano}(2017)}]{2017ApJS..230...15M}
{Moe}, M., \& {Di Stefano}, R. 2017, \apjs, 230, 15,
  \dodoi{10.3847/1538-4365/aa6fb6}

\bibitem[{{Murguia-Berthier} {et~al.}(2017){Murguia-Berthier}, {MacLeod},
  {Ramirez-Ruiz}, {Antoni}, \& {Macias}}]{2017ApJ...845..173M}
{Murguia-Berthier}, A., {MacLeod}, M., {Ramirez-Ruiz}, E., {Antoni}, A., \&
  {Macias}, P. 2017, \apj, 845, 173, \dodoi{10.3847/1538-4357/aa8140}

\bibitem[{{Naiman} {et~al.}(2011){Naiman}, {Ramirez-Ruiz}, \&
  {Lin}}]{2011ApJ...735...25N}
{Naiman}, J.~P., {Ramirez-Ruiz}, E., \& {Lin}, D.~N.~C. 2011, \apj, 735, 25,
  \dodoi{10.1088/0004-637X/735/1/25}

\bibitem[{{Naiman} {et~al.}(2018){Naiman}, {Ramirez-Ruiz}, \&
  {Lin}}]{2018MNRAS.478.2794N}
---. 2018, \mnras, 478, 2794, \dodoi{10.1093/mnras/sty1198}

\bibitem[{{Naoz} \& {Fabrycky}(2014)}]{2014ApJ...793..137N}
{Naoz}, S., \& {Fabrycky}, D.~C. 2014, \apj, 793, 137,
  \dodoi{10.1088/0004-637X/793/2/137}

\bibitem[{{Nice} {et~al.}(2008){Nice}, {Stairs}, \&
  {Kasian}}]{2008AIPC..983..453N}
{Nice}, D.~J., {Stairs}, I.~H., \& {Kasian}, L.~E. 2008, in American Institute
  of Physics Conference Series, Vol. 983, 40 Years of Pulsars: Millisecond
  Pulsars, Magnetars and More, ed. C.~{Bassa}, Z.~{Wang}, A.~{Cumming}, \&
  V.~M. {Kaspi}, 453--458

\bibitem[{{Ostriker}(1999)}]{1999ApJ...513..252O}
{Ostriker}, E.~C. 1999, \apj, 513, 252, \dodoi{10.1086/306858}

\bibitem[{{Panamarev} {et~al.}(2018){Panamarev}, {Shukirgaliyev}, {Meiron},
  {Berczik}, {Just}, {Spurzem}, {Omarov}, \&
  {Vilkoviskij}}]{2018MNRAS.476.4224P}
{Panamarev}, T., {Shukirgaliyev}, B., {Meiron}, Y., {et~al.} 2018, \mnras, 476,
  4224, \dodoi{10.1093/mnras/sty459}

\bibitem[{{Perets} {et~al.}(2007){Perets}, {Hopman}, \&
  {Alexander}}]{2007ApJ...656..709P}
{Perets}, H.~B., {Hopman}, C., \& {Alexander}, T. 2007, \apj, 656, 709,
  \dodoi{10.1086/510377}

\bibitem[{{Peters}(1964)}]{1964PhRv..136.1224P}
{Peters}, P.~C. 1964, Physical Review, 136, 1224,
  \dodoi{10.1103/PhysRev.136.B1224}

\bibitem[{{Postnov} \& {Yungelson}(2014)}]{2014LRR....17....3P}
{Postnov}, K.~A., \& {Yungelson}, L.~R. 2014, Living Reviews in Relativity, 17,
  3, \dodoi{10.12942/lrr-2014-3}

\bibitem[{{Roupas} \& {Kazanas}(2019)}]{2019A&A...621L...1R}
{Roupas}, Z., \& {Kazanas}, D. 2019, \aap, 621, L1,
  \dodoi{10.1051/0004-6361/201834609}

\bibitem[{{Ruffert}(1994)}]{1994A&AS..106..505R}
{Ruffert}, M. 1994, \aaps, 106, 505

\bibitem[{{Ruffert}(1995)}]{1995A&AS..113..133R}
---. 1995, \aaps, 113, 133

\bibitem[{{Ruffert}(1996)}]{1996A&A...311..817R}
---. 1996, \aap, 311, 817

\bibitem[{{Ruffert}(1999)}]{1999A&A...346..861R}
---. 1999, \aap, 346, 861

\bibitem[{{Ruffert} \& {Arnett}(1994)}]{1994ApJ...427..351R}
{Ruffert}, M., \& {Arnett}, D. 1994, \apj, 427, 351, \dodoi{10.1086/174145}

\bibitem[{{Sabach} \& {Soker}(2015)}]{2015MNRAS.450.1716S}
{Sabach}, E., \& {Soker}, N. 2015, \mnras, 450, 1716,
  \dodoi{10.1093/mnras/stv717}

\bibitem[{{S{\'a}nchez-Salcedo} \& {Chametla}(2014)}]{2014ApJ...794..167S}
{S{\'a}nchez-Salcedo}, F.~J., \& {Chametla}, R.~O. 2014, \apj, 794, 167,
  \dodoi{10.1088/0004-637X/794/2/167}

\bibitem[{{Shapiro} \& {Teukolsky}(1983)}]{1983bhwd.book.....S}
{Shapiro}, S.~L., \& {Teukolsky}, S.~A. 1983, {Black holes, white dwarfs, and
  neutron stars: The physics of compact objects} (Wiley-VHC)

\bibitem[{{Shen} {et~al.}(2017){Shen}, {Kulkarni}, {Madau}, \&
  {Mayer}}]{2017MNRAS.469.4012S}
{Shen}, S., {Kulkarni}, G., {Madau}, P., \& {Mayer}, L. 2017, \mnras, 469,
  4012, \dodoi{10.1093/mnras/stx1094}

\bibitem[{{Shima} {et~al.}(1985){Shima}, {Matsuda}, {Takeda}, \&
  {Sawada}}]{1985MNRAS.217..367S}
{Shima}, E., {Matsuda}, T., {Takeda}, H., \& {Sawada}, K. 1985, \mnras, 217,
  367, \dodoi{10.1093/mnras/217.2.367}

\bibitem[{{Stahler}(2010)}]{2010MNRAS.402.1758S}
{Stahler}, S.~W. 2010, \mnras, 402, 1758,
  \dodoi{10.1111/j.1365-2966.2009.15994.x}

\bibitem[{{Stone} {et~al.}(2017){Stone}, {Metzger}, \&
  {Haiman}}]{2017MNRAS.464..946S}
{Stone}, N.~C., {Metzger}, B.~D., \& {Haiman}, Z. 2017, \mnras, 464, 946,
  \dodoi{10.1093/mnras/stw2260}

\bibitem[{{Syer} {et~al.}(1991){Syer}, {Clarke}, \&
  {Rees}}]{1991MNRAS.250..505S}
{Syer}, D., {Clarke}, C.~J., \& {Rees}, M.~J. 1991, \mnras, 250, 505,
  \dodoi{10.1093/mnras/250.3.505}

\bibitem[{{Tauris} \& {van den Heuvel}(2014)}]{2014ApJ...781L..13T}
{Tauris}, T.~M., \& {van den Heuvel}, E.~P.~J. 2014, \apjl, 781, L13,
  \dodoi{10.1088/2041-8205/781/1/L13}

\bibitem[{{Thun} {et~al.}(2016){Thun}, {Kuiper}, {Schmidt}, \&
  {Kley}}]{2016A&A...589A..10T}
{Thun}, D., {Kuiper}, R., {Schmidt}, F., \& {Kley}, W. 2016, \aap, 589, A10,
  \dodoi{10.1051/0004-6361/201527629}

\bibitem[{{Toonen} {et~al.}(2016){Toonen}, {Hamers}, \& {Portegies
  Zwart}}]{2016ComAC...3....6T}
{Toonen}, S., {Hamers}, A., \& {Portegies Zwart}, S. 2016, Computational
  Astrophysics and Cosmology, 3, 6, \dodoi{10.1186/s40668-016-0019-0}

\bibitem[{{Turk} {et~al.}(2011){Turk}, {Smith}, {Oishi}, {Skory}, {Skillman},
  {Abel}, \& {Norman}}]{2011ApJS..192....9T}
{Turk}, M.~J., {Smith}, B.~D., {Oishi}, J.~S., {et~al.} 2011, \apjs, 192, 9,
  \dodoi{10.1088/0067-0049/192/1/9}

\bibitem[{{{\v S}ubr} \& {Karas}(2005)}]{2005A&A...433..405S}
{{\v S}ubr}, L., \& {Karas}, V. 2005, \aap, 433, 405,
  \dodoi{10.1051/0004-6361:20042089}

\bibitem[{{Xu} \& {Stone}(2019)}]{2019MNRAS.488.5162X}
{Xu}, W., \& {Stone}, J.~M. 2019, \mnras, 488, 5162,
  \dodoi{10.1093/mnras/stz2002}

\end{thebibliography}

\end{document}